\documentclass[a4paper,fleqn,usenatbib]{mnras}

\usepackage{newtxtext,newtxmath}

\usepackage[T1]{fontenc}
\usepackage{ae,aecompl}
\usepackage{pdflscape}
\usepackage{threeparttable}
\usepackage{color}

\usepackage{graphicx}	
\usepackage{amsmath}	
\usepackage{amssymb}	

\title[CO, H$_{2}$CO, HCO$^{+}$, CS lines in proto-BDs]{Chemical tracers in proto-brown dwarfs: CO, ortho-H$_{2}$CO, para-H$_{2}$CO, HCO$^{+}$, CS observations}

\author[Riaz et al.]{
Riaz, B.,$^{1}$\thanks{E-mail: briaz@usm.lmu.de}
Thi, W.-F.,$^{2}$
Caselli, P.$^{2}$
\\
$^{1}$  Universit\"ats-Sternwarte M\"unchen, Ludwig-Maximilians-Universit\"at, Scheinerstr.~1, 81679 M\"unchen, Germany
\\
$^{2}$  Max-Planck-Institut f\"{u}r Extraterrestrische Physik, Giessenbachstrasse 1, D-85748 Garching, Germany
\\
}

\date{Accepted XXX. Received YYY; in original form ZZZ}

\pubyear{2018}

\begin{document}
\label{firstpage}
\pagerange{\pageref{firstpage}--\pageref{lastpage}}
\maketitle

\begin{abstract}

We present a study of the CO isotopologues and the high-density tracers H$_{2}$CO, HCO$^{+}$, and CS in Class 0/I proto-brown dwarfs (proto-BDs). We have used the IRAM 30m telescope to observe the $^{12}$CO (2-1), $^{13}$CO (2-1), C$^{18}$O (2-1), C$^{17}$O (2-1), H$_{2}$CO (3-2), HCO$^{+}$ (3-2), and CS (5-4) lines in 7 proto-BDs. The hydrogen column density for the proto-BDs derived from the CO gas emission is $\sim$2-15 times lower than that derived from the dust continuum emission, indicating CO depletion from the gas-phase. The mean H$_{2}$CO ortho-to-para ratio is $\sim$3 for the proto-BDs and indicates gas-phase formation for H$_{2}$CO. We have investigated the correlations in the molecular abundances between the proto-BDs and protostars. Proto-BDs on average show a factor of $\sim$2 higher ortho-to-para H$_{2}$CO ratio than the protostars. Possible explanations include a difference in the H$_{2}$CO formation mechanism, spin-selective photo-dissociation, self-shielding effects, or different emitting regions for the ortho and para species. There is a tentative trend of a decline in the HCO$^{+}$ and H$_{2}$CO abundances with decreasing bolometric luminosity, while the CS and CO abundances show no particular difference between the proto-BDs and protostars. These trends reflect the scaled-down physical structures for the proto-BDs compared to protostars and differences in the peak emitting regions for these species. The C$^{17}$O isotopologue is detected in all of the proto-BDs as well as the more evolved Class Flat/Class II BDs in our sample, and can probe the quiescent gas at both early and late evolutionary stages.

\end{abstract}




\section{Introduction}

The formation and evolution of sub-stellar mass objects have been a subject of much theoretical and observational work. Numerical simulation studies have provided an understanding of the various phenomenon of the brown dwarf formation processes, whether it is a star-like formation via gravitational core collapse (Reipurth and Clarke 2001; Machida et al. 2009), or alternative mechanisms of formation via disk fragmentation and as ejected embryos (Stamatellos \& Whitworth  2008; Goodwin \& Whitworth  2007; Bate 2012; Basu \& Vorobyov 2012). The past few years have seen a rapid development of a new generation of highly sensitive, cutting-edge instruments across a wide range in wavelengths. This has led to the confirmed detection of brown dwarfs in their early Class 0/I proto-brown dwarf stage in various star-forming regions (e.g., Palau et al. 2014; Riaz et al. 2015; 2016; Huelamo et al. 2017; Barrado et al. 2009; 2018). While the proto-brown dwarf studies are sparse in number, they have already provided hints of a similar range in the accretion and outflow activity rates and a scaled-down physical structure compared to low-mass protostars (Riaz et al. 2016; Whelan et al. 2018). A notable result is the discovery of a parsec length jet driven by a proto-brown dwarf in Orion, with a variety of features that are commonly seen in protostellar jets (Riaz et al. 2017). Most observational results for both the early-stage proto-brown dwarfs and the more evolved T Tauri analog brown dwarfs indicate that their diverse properties are similar to low-mass stars (e.g., Luhman et al. 2008; Riaz et al. 2012; 2016; Whelan et al. 2009; 2018), and point towards a star-like formation for sub-stellar mass objects. 

Low-mass star formation takes place in dense cores of molecular clouds (e.g., Benson \& Myers 1989). Previous studies on molecular line observations of low-mass Class 0/I protostars have revealed a number of molecular species, such as HCO$^{+}$, CS, and H$_{2}$CO, which can trace the dense interiors of the cores and provide an insight into the chemical composition, kinematics, and dynamics of the cores (e.g., Mardones et al. 1997; Lee et al. 1999; Dickens \& Irvine 1999; Maret et al. 2004; Joergensen et al. 2004; 2005; Larson \& Liseau 2017; Vidal et al. 2017). The HCO$^{+}$ and CS molecules are known tracers of high-density structure, and have been used to search for signatures of infall and rotation in dense cores and protostars (e.g., Mardones et al. 1997; Lee et al. 1999; Dickens \& Irvine 1999; Maret et al. 2004; Joergensen et al. 2004; Doty et al. 2002). HCO$^{+}$ is also a good tracer of gas ionization and plays an important role in the destruction of other molecules (e.g., Bergin \& Langer 1997; Sternberg \& Dalgarno 1995). H$_{2}$CO can be formed in the gas-phase and on the surface of dust grains, and is used for studying the physical conditions of the gas in astrophysical sources (e.g., Charnley et al. 1992; Maret et al. 2004; Lindberg et al. 2017). An interconversion between the ortho and para H$_{2}$CO species has a longer lifetime than destruction by grains, therefore the ratio retains its initial information on the environment in which these species formed (e.g., Dickens \& Irvine 1999; Guzman et al. 2011; Mangum \& Wooten 1993; Tudorie et al. 2006). 

Observations of the CO isotopes with a low critical density can trace the gas density in different components, with $^{13}$CO probing the intermediate optical depth envelope, C$^{17}$O and C$^{18}$O probing the high optical depth core, and the highly abundant $^{12}$CO tracing the optically thin bipolar outflows and the outer envelope layers (e.g., Lee et al. 2004; Redman et al. 2002; Hatchell et al. 1999; Carolan et al. 2008). The HCO$^{+}$, CS, and H$_{2}$CO molecules have also been used as an evolutionary clock, with the fractional abundance of CS, H$_{2}$CO, CO reaching a peak at an earlier time than HCO$^{+}$ (e.g., Aikawa et al. 2005; Maret et al. 2007; Wakelam et al. 2004). A well established result from various studies on dense cores is that the formation of ice grain mantles leads to the removal of chemical reservoirs like CO, H$_{2}$CO, CS, and other abundant species from the gas phase (e.g., Caselli et al. 1999; Lee et al. 2004; Bacmann et al. 2002; Joergensen et al. 2002; Redman et al. 2002). The mantles are chemically active despite their low temperature. An understanding of the inhomogeneties in the chemical composition due to the depletion of molecules from the gas phase in the pre-stellar material is important in order to correctly interpret the molecular line observations.

As molecular lines are an important aid to study the internal structure of dense cores, we have conducted an exploratory study to identify the best and easily detectable chemical tracers in Class 0/I proto-BDs. Chemistry can also be a tool to identify such faint, embedded objects, particularly if the dust mass is too low and the thermal continuum emission is too weak to detect the proto-BDs. Our sample covers a range in bolometric luminosities from $\sim$0.1 down to 0.008 L$_{\sun}$, and extends well below the luminosities of protostars studied in previous molecular line surveys. The aim of this exploratory study is to probe the chemical composition of brown dwarfs during their early evolutionary stages, and to investigate if the molecular abundances of some of the commonly detected species in low-mass protostars show a different trend for bolometric luminosities below the sub-stellar limit. Our deep molecular line survey has revealed emission in 13 molecular species in the proto-BDs. The first paper in this series presented the CN, HCN, and HNC observations (Riaz et al. 2018). This paper presents the results from the CO, H$_{2}$CO, HCO$^{+}$, and CS observations. The sample and observations are described in Sect.~\ref{obs}. In Sect.~\ref{analysis}, we present an analysis of the line profiles and the derivation of the line parameters, column densities, and the molecular abundances. Section~\ref{discussion} investigates the extent of CO freeze-out in the proto-BDs and the trends in their molecular abundances in comparison with protostellar objects.

\section{Targets, Observations and Data Reduction}
\label{obs}

\subsection{Sample}

\begin{table*}
\centering
\caption{Sample}
\label{sample}
\begin{threeparttable}
\begin{tabular}{lllp{0.8cm}p{0.8cm}lp{1.2cm}} 
\hline
Object & RA (J2000) & Dec (J2000) & L$_{\textrm{bol}}$ (L$_{\sun}$)\tnote{a} & M$^{d+g}_{\textrm{total}}$ (M$_{\textrm{Jup}}$) & Classification\tnote{b} & Region \\
\hline
\multicolumn{7}{c}{Stage 0+I} \\
SSTc2d J182854.9+001833 (J182854) & 18h28m54.90s &  00d18m32.68s &  0.05 & 48$\pm$11 & Stage 0+I, Stage I, Class 0/I & Serpens \\ 
SSTc2d J182844.8+005126 (J182844) & 18h28m44.78s  & 00d51m25.79s &  0.04 & 70$\pm$20 & Stage 0+I, Stage 0, Class 0/I & Serpens \\ 
SSTc2d J183002.1+011359 (J183002) &  18h30m02.09s  & 01d13m58.98s &  0.09 & 74$\pm$18 & Stage 0+I, Stage 0, Class 0/I & Serpens \\ 
SSTc2d J182959.4+011041 (J182959) & 18h29m59.38s  & 01d10m41.08s &  0.008 & 62$\pm$20 & Stage 0+I, Stage II, Class 0/I & Serpens \\ 
SSTc2d J163143.8-245525 (J163143) & 16h31m43.75s & -24d55m24.61s &  0.16 & 65$\pm$14 &  Stage 0+I, Stage I, Class Flat & Ophiuchus \\ 
SSTc2d J182953.0+003607 (J182953) & 18h29m53.05s  & 00d36m06.72s &  0.15 & $<$25 		& Stage 0+I, Stage 0, Class Flat & Serpens \\ 
SSTc2d J163136.8-240420 (J163136) & 16h31m36.77s & -24d04m19.77s &  0.09 & 56$\pm$28 &  Stage 0+I, Stage I, Class Flat & Ophiuchus \\ 
 \multicolumn{7}{c}{Stage I-T/II} \\
SSTc2d J182940.2+001513 (J182940) & 18h29m40.20s &  00d15m13.11s &  0.074 & 28$\pm$12 & Stage II, Stage I-T, Class 0/I & Serpens \\ 
SSTc2d J182927.4+003850 (J182927) & 18h29m27.35s &  00d38m49.75s &  0.012 & $<$12 	& Stage II, Stage I-T, Class 0/I & Serpens \\ 
SSTc2d J182952.1+003644 (J182952) & 18h29m52.06s &  00d36m43.63s &  0.016 & $<$28 	& Stage II, Stage II, Class Flat & Serpens \\ 
\hline
\end{tabular}
\begin{tablenotes}
\item[a] Errors on L$_{\textrm{bol}}$ are estimated to be $\sim$30\%. 
\item[b] The first, second, and third values are using the classification criteria based on the integrated intensity in the HCO$^{+}$ (3-2) line, the physical characteristics, and the SED slope, respectively.
\end{tablenotes}
\end{threeparttable}
\end{table*}

Our sample consists of 10 proto-BD candidates, 8 of which were identified in Serpens and 2 in the Ophiuchus region. Table~\ref{sample} lists the properties for the targets in the sample. We applied the same target selection and identification criteria as described in Riaz et al. (2015; 2016), which is based on a cross-correlation of infrared data from the UKIDSS and/or 2MASS, Spitzer, and Herschel archives, followed by deep sub-millimeter 850$\mu$m continuum observations obtained with the JCMT/SCUBA-2 bolocam. Out of the 10 targets, 7 are detected at a $>$5-$\sigma$ level in the 850$\mu$m sub-millimeter continuum observations, while the rest are detected at a marginal ($\sim$2-$\sigma$) level. We have derived the total (dust+gas) mass, M$^{d+g}_{\textrm{total}}$, for the proto-BD candidates using their 850 $\mu$m flux density or the $\sim$2-$\sigma$ upper limits for the marginal detections, assuming a dust temperature of 10 K, a gas to dust mass ratio of 100, and a dust mass opacity coefficient at 850 $\mu$m of 0.0175 cm$^{2}$ gm$^{-1}$ (Ossenkopf \& Henning 1994). The masses thus derived are listed in Table~\ref{sample}, and are found to be in the range of $\sim$0.01--0.1 M$_{\sun}$. The uncertainties on the masses have been estimated using the flux errors. The bolometric luminosity, L$_{\textrm{bol}}$, for the targets as measured from integrating the observed infrared to sub-millimeter spectral energy distribution (SED) is in the range of $\sim$0.008--0.1 L$_{\sun}$. We assumed a distance to the Serpens region of 436$\pm$9 pc (Ortiz-Leon et al. 2017; Dzib et al. 2010) based on astrometric observations, and 140$\pm$6 pc to Ophiuchus (Mamajek 2008; Schlafly et al. 2014). We note that since these objects are still accreting, the total mass may not represent the final mass for these systems. However, based on the accretion models by Baraffe et al. (2012), the L$_{\textrm{bol}}$ for these candidates is below the luminosity threshold considered between very low-mass stars and brown dwarfs ($<$0.2 L$_{\sun}$), and considering the sub-stellar mass reservoir in the (envelope+disk) for these systems, the final mass is also expected to stay within the sub-stellar limit.

We have determined the evolutionary stage of the candidates using the Stage 0+I/II criteria based on the strength in the HCO$^{+}$ (3-2) line emission, and the Stage 0/I/I-T/II classification criteria based on the physical characteristics of the system. A detailed description of these criteria is presented in Riaz et al. (2016). Briefly, a young stellar object (YSO) with an integrated intensity of $>$0.4 K km s$^{-1}$ in the HCO$^{+}$ (3-2) line is classified as a Stage 0+I object (e.g., van Kempen et al. 2009; Heiderman \& Evans 2015). Based on the physical characteristics estimated from the radiative transfer modelling of the observed spectral energy distribution (SED), a Stage 0 object is expected to have a disk-to-envelope mass ratio of $<<$1, while for a Stage I YSO, this mass ratio would be in the range of 0.1--2 (e.g., Whitney et al. 2003). A Stage 0 object is also expected to have a total circumstellar (disk+envelope) mass to stellar mass ratio of $\sim$1. The Stage thus determined has been compared to the Class 0/I/Flat classification based on the 2-24 $\mu$m slope of the observed SED. Stage I-T and Class Flat are considered to be intermediate between the Stage I and Stage II, or Class I and Class II evolutionary phases. As listed in Table~\ref{sample}, we find a good match between the Stage of the system based on the physical characteristics and the strength in the molecular line emission, whereas the classification based on the SED slope does not relate well to this evolutionary stage. This is particularly for the Class Flat cases, three of which are found to be Stage 0/I objects. The physical structure from modelling of these objects indicates an edge-on inclination, which could result in a flatter spectral slope in the SED compared to genuine Stage 0/I systems (e.g., Whitney et al. 2003; Riaz et al. 2016). Thus, 7 proto-BD candidates in our sample are both in the early Stage 0/I evolutionary phase and sub-stellar in nature, and can be considered as bona fide proto-BDs, whereas the sources J182940, J182927, and J182952 are sub-stellar objects that are in the intermediate Stage I-T/Stage II phase (Table~\ref{sample}).  A detailed analysis on the physical and chemical structure of the targets based on continuum and line modelling will be presented in a forthcoming paper.

\subsection{Molecular Line Observations and Data Reduction}

\begin{table*}
\centering
\caption{IRAM Molecular line observations}
\label{line-obs}
\begin{threeparttable}
\begin{tabular}{cccccc} 
\hline
Molecule & Line & Frequency (GHz) 	& $E_{j}$/k (K) & $A_{jk}$ (s$^{-1}$) & $n_{crit}^{thin}$(10 K) (cm$^{-3}$)   	\\
\hline

p-H$_{2}$CO \tnote{a}	& 3(0,3)-2(0,2)	& 218.2222	& 21.0 & 2.8$\times 10^{-4}$ & 9.7$\times 10^{5}$	\\
C$^{18}$O 			& 2-1 		& 219.5603	& 15.8 & 6.0$\times 10^{-7}$ & 1.8$\times 10^{4}$	\\
$^{13}$CO			& 2-1		& 220.3986	& 15.8 & 6.0$\times 10^{-7}$ & 1.8$\times 10^{4}$	\\
C$^{17}$O 			& 2-1 		& 224.7142 	& 16.2 & 6.4$\times 10^{-7}$ & 1.9$\times 10^{4}$	\\
o-H$_{2}$CO \tnote{a}	& 3(1,2)-2(1,1) 	& 225.6977	& 33.4 & 2.8$\times 10^{-4}$ & 8.4$\times 10^{5}$	\\
$^{12}$CO			& 2-1 		& 230.5380	& 16.6 & 6.9$\times 10^{-7}$ & 2.1$\times 10^{4}$	\\
$^{13}$CS			& 5-4 		& 231.2207 	& 35.3 & 3.0$\times 10^{-4}$ & 2.0$\times 10^{6}$	\\
C$^{34}$S			& 5-4 		& 241.0161 	& 35.3 & 3.0$\times 10^{-4}$ & 2.0$\times 10^{6}$	\\
CS 					& 5-4 		& 244.9355	& 35.3 & 3.0$\times 10^{-4}$ & 2.0$\times 10^{6}$	\\
H$^{13}$CO$^{+}$		& 3-2		& 260.2553	& 24.9 & 1.3$\times 10^{-3}$ & 3.5$\times 10^{6}$	\\
HCO$^{+}$			& 3-2 		& 267.5576 	& 25.7 & 1.5$\times 10^{-3}$ & 3.8$\times 10^{6}$	\\
				
\hline
\end{tabular}
\begin{tablenotes}
  \item[a] These are the only H$_{2}$CO lines detected for the targets. 
\end{tablenotes}
\end{threeparttable}
\end{table*}

We observed the 10 targets with the IRAM 30m telescope using the EMIR heterodyne receiver in May, 2016. We used two tuning setups in the E230 band to observe the main $^{12}$CO (2-1) and the C$^{18}$O (2-1), $^{13}$CO (2-1), C$^{17}$O (2-1) isotopologues, the H$_{2}$CO (3-2), CS (5-4), and HCO$^{+}$ (3-2) lines. An overview of the molecular lines observed within these frequency ranges are listed in Table~\ref{line-obs}. We used the FTS backend in the wide mode, with a spectral resolution of 200 kHz ($\sim$0.3 km s$^{-1}$ at 224 and 267 GHz). The large 32 GHz bandwidth (4 sidebands of 4 GHz each) available with FTS allowed observing several additional molecular lines in the same tuning setups. Follow-up observations in the H$^{13}$CO$^{+}$ (3-2) isotopologue were obtained for the proto-BDs J182844 and J183002 that show the strongest emission in HCO$^{+}$ (3-2). Due to time limitations, this isotopologue could not be observed for the whole sample. The observations were taken in the frequency switching mode with a frequency throw of approximately 7 MHz. The source integration times ranged from 3 to 4 hours per source per tuning reaching a typical RMS (on T$_{\textrm{A}}^{*}$ scale) of $\sim$0.02--0.05 K. The telescope absolute RMS pointing accuracy is better than 3$\arcsec$ (Greve et al. 1996). All observations were taken under good weather conditions (0.08 $<$$\tau$ $<$0.12; PWV $<$2.5 mm). 

The spectra were calibrated at the telescope onto the antenna temperature T$_{\textrm{A}}^{*}$ scale using the standard chopper wheel method. The absolute calibration accuracy for the EMIR receiver is around 10\% (Carter et al. 2012). The telescope intensity scale was converted into the main beam brightness temperature (T$_{\textrm{mb}}$) using standard beam efficiencies that vary between 59\% at 210 GHz to 49\% at 280 GHz. The half power beam width of the telescope beam varies between 11$\arcsec$ at 210 GHz and 9$\arcsec$ at 260 GHz. The spectral reduction was conducted using the CLASS software (Hily-Blant et al. 2005) of the GILDAS package\footnote{http://www.iram.fr/IRAMFR/GILDAS}. The standard data reduction process consisted of averaging multiple observations for each transition line, extracting a subset around the line rest frequency, folding the spectrum to correct for the frequency switching effects, and finally a low-order polynomial baseline was subtracted for each spectrum. In case of a line detection, a single or multiple Gaussians were fitted, as described further in Sect.~\ref{analysis}.


\section{Data Analysis and Results}
\label{analysis}

\subsection{Line Profiles and Parameters}
\label{lineFits}

The spectra for the sources show a variety in line shapes and strengths, from narrow, weak detection at T$_{\textrm{mb}}$ $<$1 K, to broad, multi-component profiles (Figs.~A1~-~A4;~B1~-~B4 in the Supplementary Material). We have measured the parameters of the line center, line width, the peak and integrated intensities for all spectra. A single-peaked fit was used for typical Gaussian shaped profiles. For all spectra that show non-Gaussian line shapes with multiple peaks, we have used the IDL routine $MPFIT$ in an attempt to decompose the observed profile into multiple Gaussians. The line fits are shown in Figs.~A1~-~A4;~B1~-~B4, and the parameters derived from the Gaussian fits are listed in Tables~C1~-~C8 (Supplementary Material). In case of a non-detection, a 3-$\sigma$ upper limit was calculated by integrating over the velocity range of $\pm$2 km s$^{-1}$ from the cloud systemic velocity of $\sim$8 km s$^{-1}$ for Serpens (e.g., Burleigh et al. 2013) and $\sim$4.4 km s$^{-1}$ for Ophiuchus (e.g., White et al. 2015). We estimate uncertainties of $\sim$15\%-20\% for the integrated intensity, $\sim$0.01-0.02 km s$^{-1}$ in V$_{lsr}$, $\sim$5\%-10\% on T$_{\textrm{mb}}$ and $\Delta$v. For J163136, J182853, and J182952, all of the lines are shifted from the cloud systemic velocity. This suggests that the emission for these objects is tracing molecular gas at a different bulk velocity than the others, which is possibly due to differences in the dynamical processes such as collapse versus rotation or expansion that can lead to slight Doppler shifts of the gas components (e.g., Evans 1999).


\subsubsection{C$^{17}$O hyperfine lines}

 \begin{figure*}
  \centering              
     \includegraphics[width=3in]{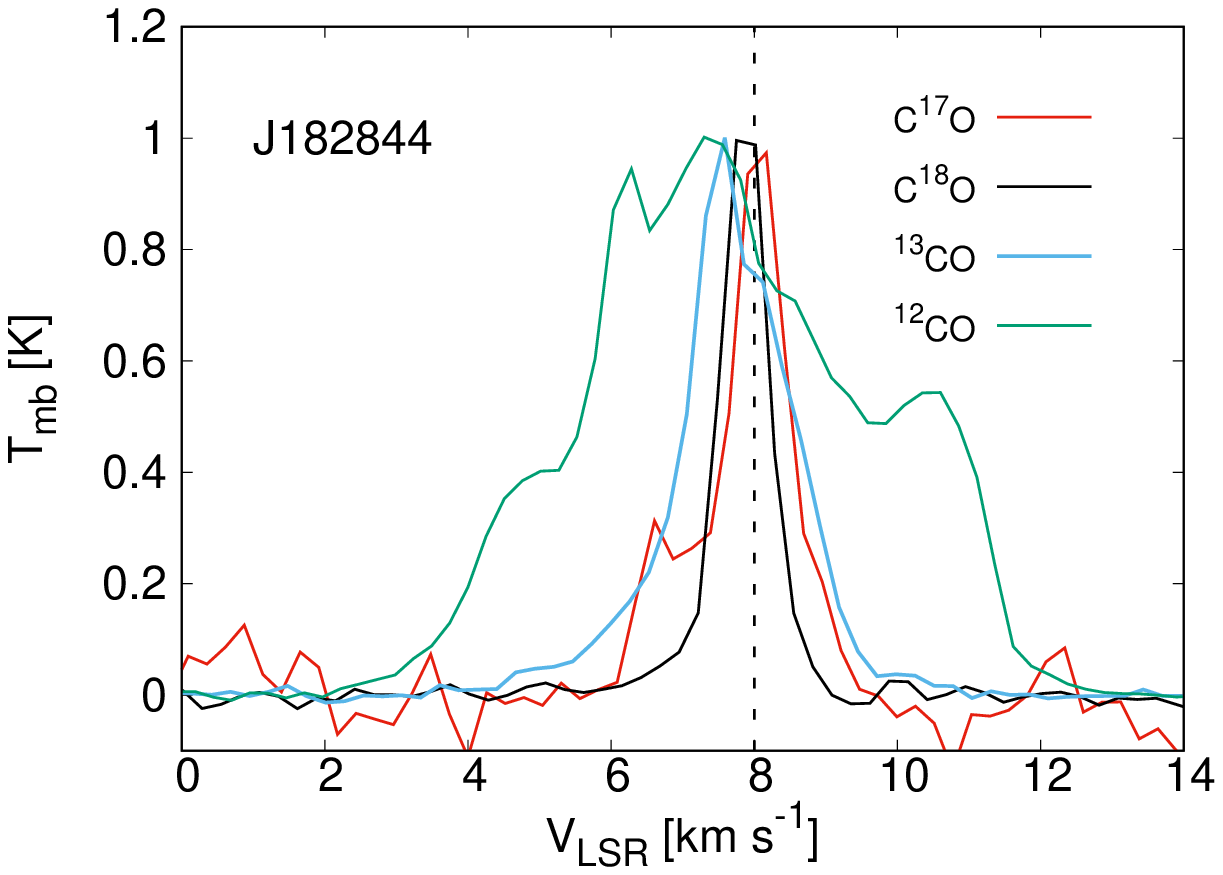}
     \includegraphics[width=3in]{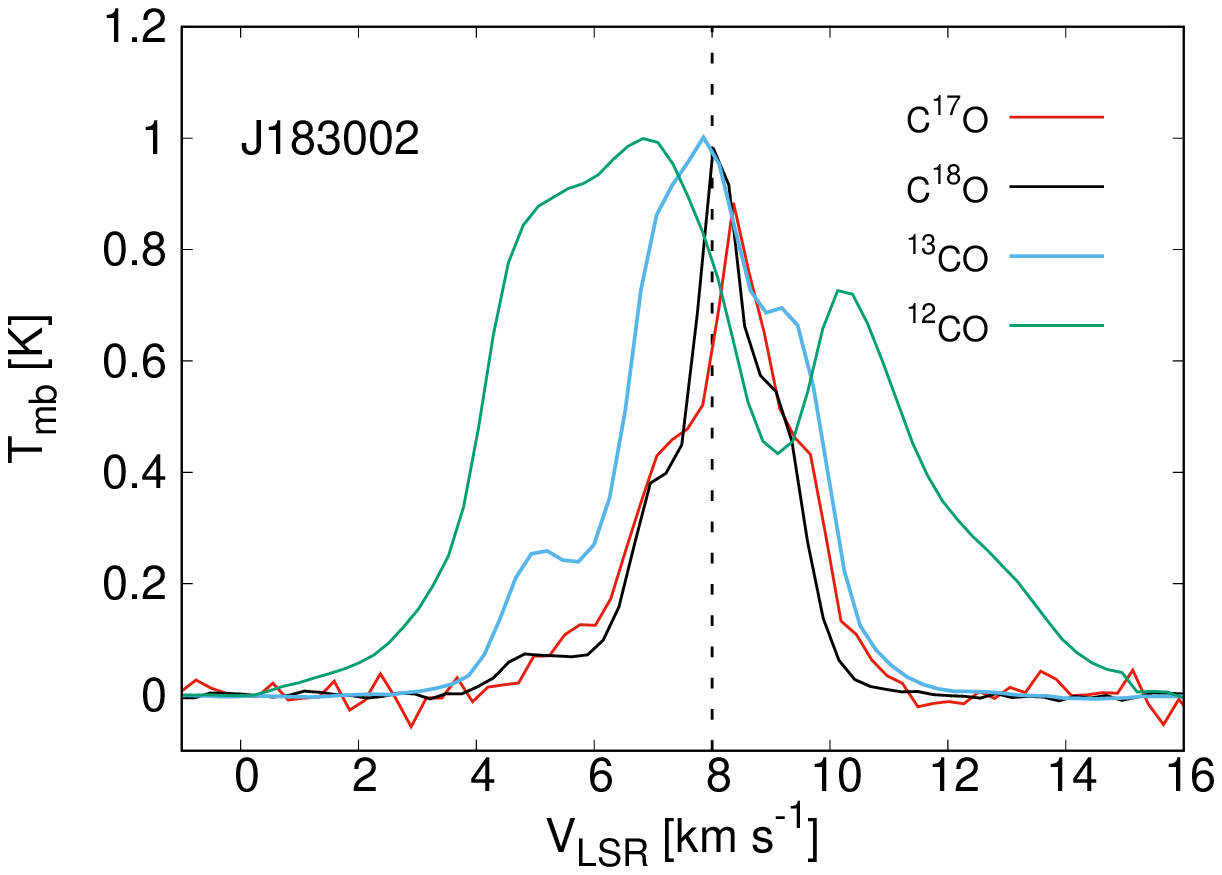}
     \caption{An overplot of the $^{12}$CO (2-1), $^{13}$CO (2-1), C$^{18}$O (2-1), and C$^{17}$O (2-1) spectra for J182844 (left) and J183002 (right). Black dashed line marks the cloud systemic velocity of $\sim$8 km/s in Serpens. The spectra have been normalized to unity. The normalization factors in the C$^{17}$O, C$^{18}$O, $^{13}$CO, and $^{12}$CO lines, respectively, are 4.3, 1.1, 0.7, and 0.6 for J182844, and 1.5, 0.4, 0.1, and 0.08 for J183002.   }
     \label{co-overplot}
  \end{figure*}


The C$^{17}$O (2-1) line is detected in all of the proto-BDs as well as the more evolved Stage I-T/II objects (Fig.~A1), even though it is the least abundant CO isotopomer and thus difficult to detect. This tracer has a low optical depth and can trace deep into the quiescent envelope. C$^{17}$O is known to be a species with non-anomalous hyperfine components (e.g., Loughane et al. 2012). It is robust against the line overlap effect due to an even number of routes down to the lower state hyperfine energy levels and the hyperfine lines are excited equally in the lower rotational state. Being the least abundant observable CO isotopologue, the C$^{17}$O emission is expected to agree with the optically thin LTE structure irrespective of the brightness temperature or optical depth (e.g., Redman et al. 2002). The C$^{17}$O (2-1) consists of 9 hyperfine components within a narrow frequency range of $\sim$1.7 MHz that cannot be spectrally resolved. The hyperfine structure under optically thin LTE conditions is plotted in the top, left panel in Fig.~A1. The observed C$^{17}$O (2-1) emission is a composite of all 9 hyperfine components, and the spectrum appears in most sources as a bright, narrowly peaked central component with a weaker, satellite component (Fig.~A1). Due to the close blending in the hyperfine lines seen in C$^{17}$O (2-1), the multiple Gaussian fitting method cannot be employed. We have used the hyperfine structure fitting routine in the GILDAS/CLASS package. The close fits to the observed lines do not show any saturation effects. J183002 shows a broad, single-peaked profile that is a blend of all of the hyperfine components. The satellite peak is nearly as strong as the central peak in J182959, which suggests that the emission is affected by a large optical depth. Results from line radiative transfer modelling indicate an opacity $\tau \sim$1.7 for this object, which is comparatively higher than $\leq$0.5 for the other objects (Sect.~\ref{col-den}).





Figure~\ref{co-overplot} compares the profiles of the main CO and isotopomers for the proto-BDs J182844 and J183002. The spectra are plotted as normalized colored curves for a clear comparison. The narrowest profiles are seen for the C$^{17}$O and C$^{18}$O isotopes with the lowest abundance. As mentioned, these optically thin lines can trace deep into the proto-BD core, and can be used as reliable tracers to estimate the total CO abundance for the proto-BDs. From the line overlap analysis, we can estimate that roughly one-third of the emission in the optically thick $^{12}$CO and $^{13}$CO lines arises from the proto-BD, while the rest is likely tracing the cloud material surrounding the central source. The $^{12}$CO (2-1), $^{13}$CO (2-1), and C$^{18}$O (2-1) spectra are discussed in the Appendix A (Supplementary Material).

\subsubsection{HCO$^{+}$, CS, H$_{2}$CO}
\label{profiles}

The HCO$^{+}$ (3-2) spectra for J182854, J182844, and J163143 show a blue- or red-dominated asymmetry with a central dip that is suggestive of self-absorption (Fig.~B1). A blue-dominated asymmetry is typically indicative of an infalling envelope and is considered a signature of a collapsing core, whereas the red asymmetry is indicative of outflowing material or rotating collapse that can cause the line profiles to shift from blue- to red-skewed depending on the rotation axis (e.g., Evans 1999). All of these objects show extended wings in the HCO$^{+}$ spectra, which is indicative of a contribution from an outflow. The HCO$^{+}$ spectrum for J183002 shows a spectrally resolved blue-shifted component at $\sim$5 km s$^{-1}$ that is much weaker in strength than the main profile and possibly arises from a blue-shifted outflow. Likewise, the HCO$^{+}$ profile for J182959 suggests that the observed emission arises from different physical components, with the narrow peak tracing the inner, dense region and the broad base could be tracing the blue- and red-shifted outflow material. J163136 and J182952 are the only two sources in our sample that show a narrow, Gaussian-like HCO$^{+}$ profile, which suggests that the observed emission is tracing the high-density region towards the inner envelope/disk in these sources.


 \begin{figure*}
  \centering              
     \includegraphics[width=2.5in]{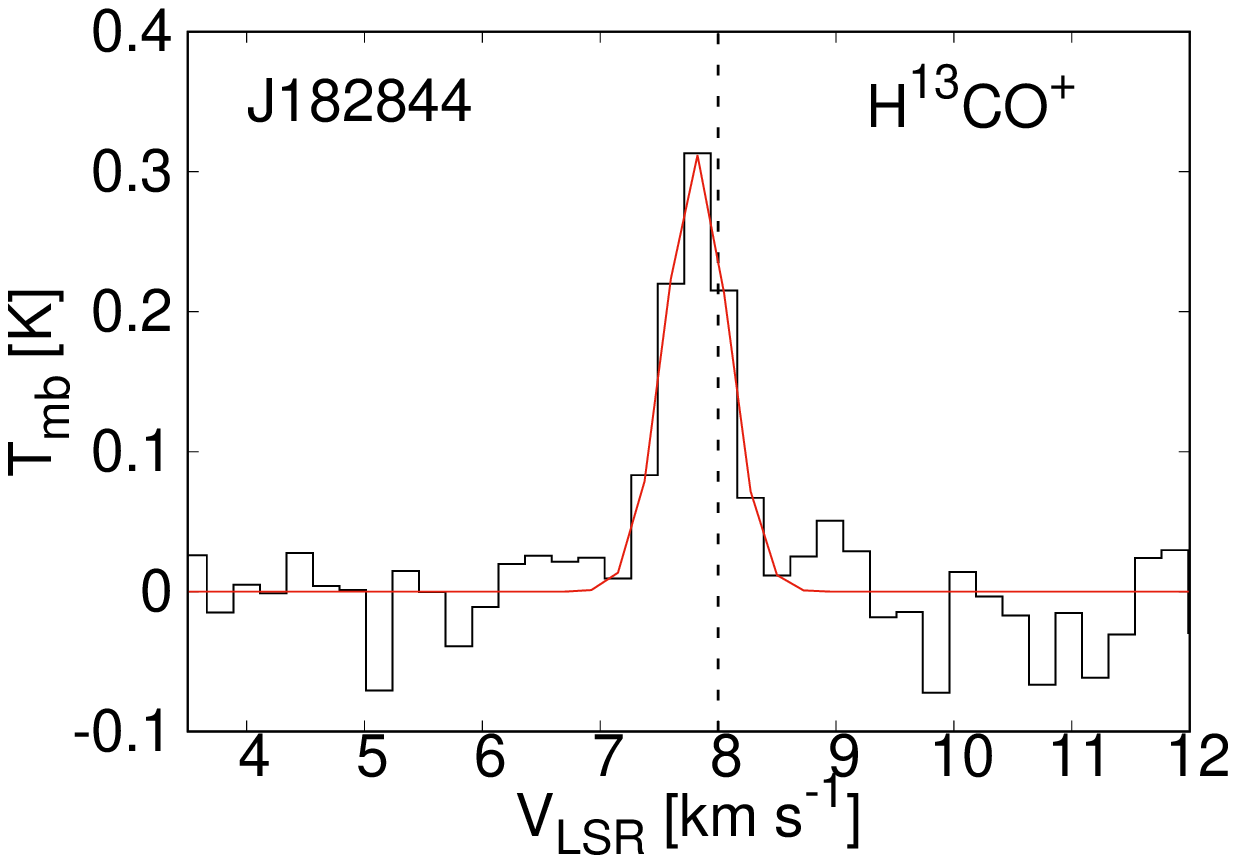}
     \includegraphics[width=2.5in]{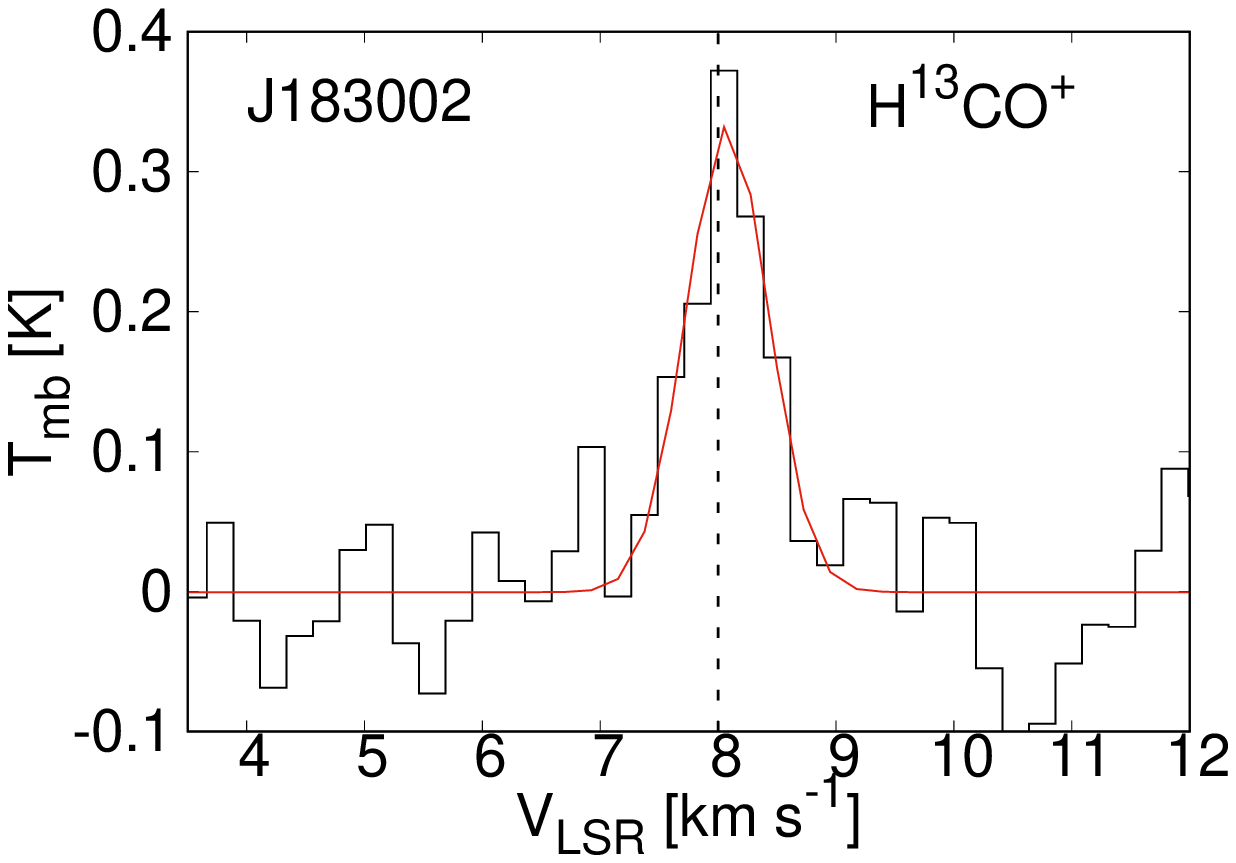}    \\
     \includegraphics[width=2.5in]{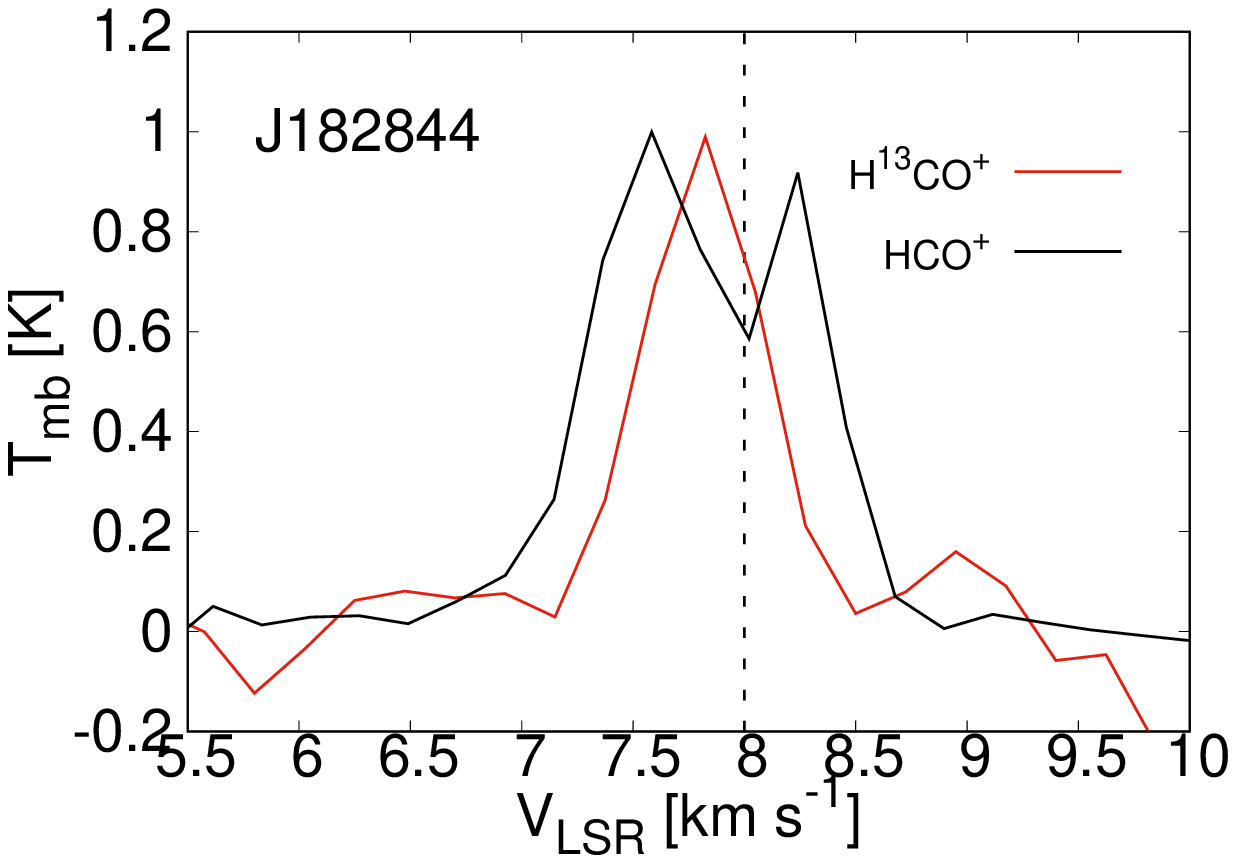}         
     \includegraphics[width=2.5in]{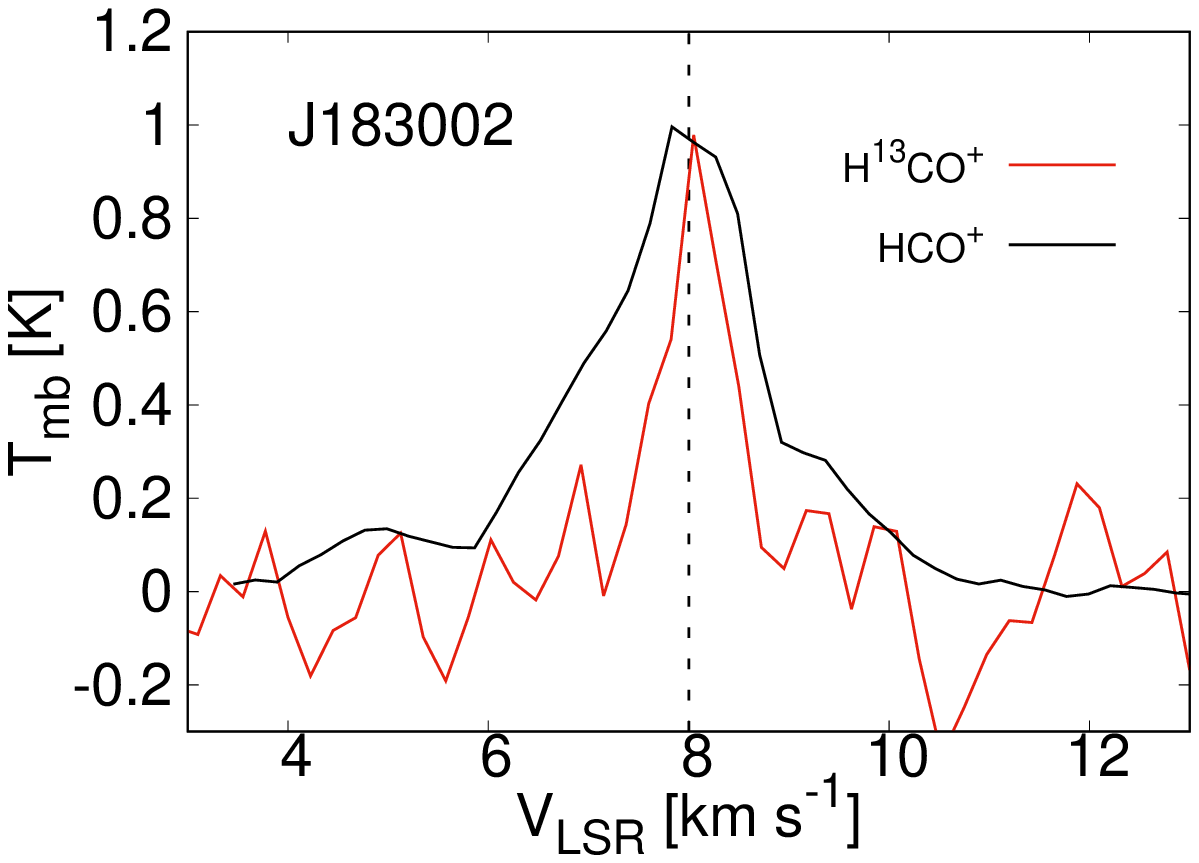}     
     \caption{Top panels show the observed H$^{13}$CO$^{+}$ (3-2) spectra (black) with the Gaussian fits (red) for J182844 and J183002. Bottom panels show an overplot of the HCO$^{+}$ (3-2) spectra (black) and H$^{13}$CO$^{+}$ (3-2) spectra (red) for the same objects. The line intensities in the overplots have been normalized to unity. The normalization factors in the H$^{13}$CO$^{+}$ and HCO$^{+}$ lines, respectively, are 3.2 and 1.0 for J182844, and 2.6 and 0.3 for J183002. Black dashed line marks the cloud systemic velocity of $\sim$8 km/s and $\sim$4.4 km/s in Serpens and Ophiuchus, respectively.  }
     \label{h13cop-figs}     
  \end{figure*}

It is important to observe both optically thick and optically thin lines in order to distinguish double-peak profiles produced by optical depth effects from those due to two velocity components along the line of sight. The optically thin isotopologue of H$^{13}$CO$^{+}$ can provide a better assessment of the abundance of the HCO$^{+}$ molecule in the proto-BDs. Due to time limitations, we could only obtain H$^{13}$CO$^{+}$ observations for the proto-BDs J182844 and J183002. Both proto-BDs show a $>$5-$\sigma$ detection in H$^{13}$CO$^{+}$ (Fig.~\ref{h13cop-figs}). The right panel in Fig.~\ref{h13cop-figs} shows an overplot of the HCO$^{+}$ and H$^{13}$CO$^{+}$ lines. For J182844, the H$^{13}$CO$^{+}$ peak is nearly at the centre of the self-absorption feature seen in the HCO$^{+}$ line. This is consistent with the interpretation that the self-absorption is caused by the low excitation outlying layer in the outer envelope ($>$1000 AU) regions that surround the proto-BD core, while the H$^{13}$CO$^{+}$ emission arises from the dense core itself. For J183002, both HCO$^{+}$ and H$^{13}$CO$^{+}$ are centered at the same velocity, and we can say that the bulk of the HCO$^{+}$ emission in this proto-BD arises from the proto-BD core. HCO$^{+}$ shows excess emission on the blue side, which may be associated with an outflow.

 \begin{figure*}
  \centering              
     \includegraphics[width=3in]{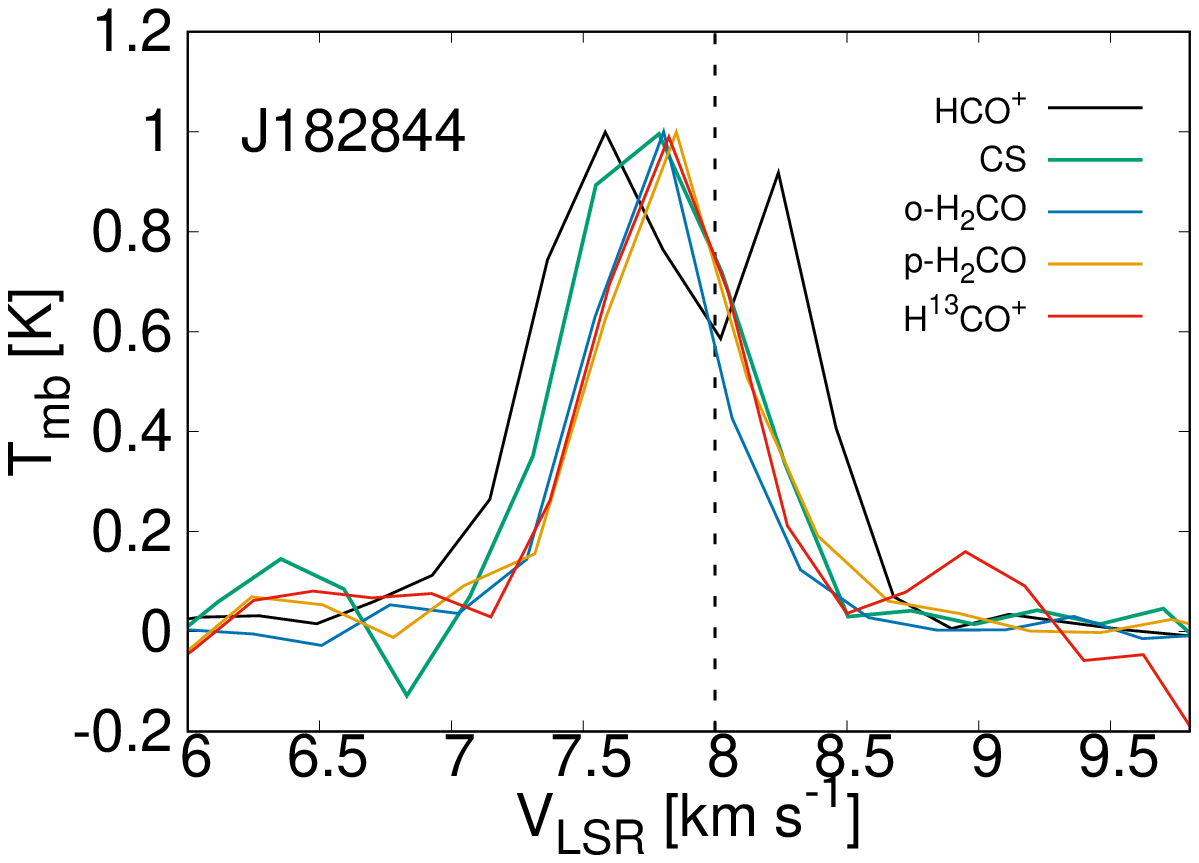}                           
     \includegraphics[width=3in]{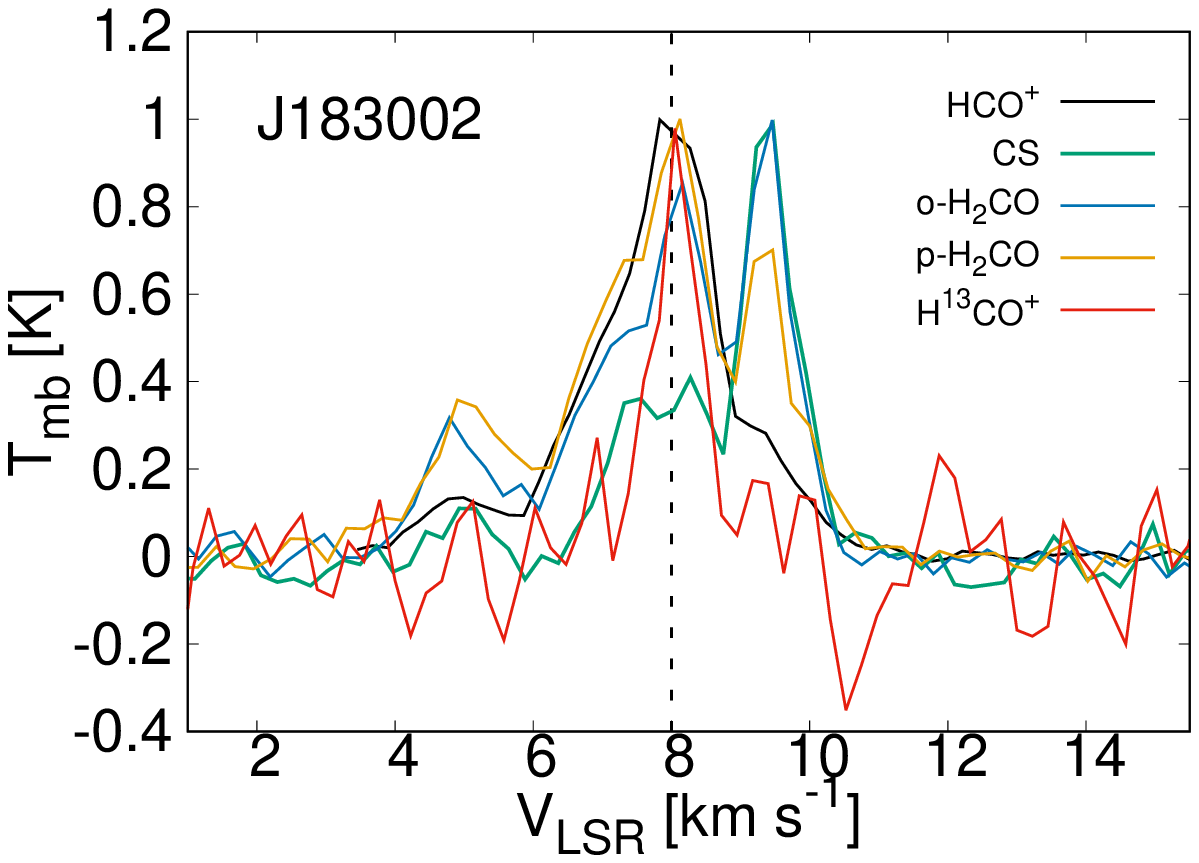}                           
     \caption{An overplot of the HCO$^{+}$, H$^{13}$CO$^{+}$, o-H$_{2}$CO, p-H$_{2}$CO, and CS spectra for J182844 (left) and J183002 (right). Black dashed line marks the cloud systemic velocity of $\sim$8 km/s in Serpens. The spectra have been normalized to unity. The normalization factors in the HCO$^{+}$, H$^{13}$CO$^{+}$, o-H$_{2}$CO, p-H$_{2}$CO, and CS lines, respectively, are 1.0, 3.2, 1.2, 2.0, and 3.2 for J182844, and 0.3, 2.6, 0.9, 1.7, and 1.2 for J183002.   }
  \label{overplots}       
\end{figure*}     
			

The ortho-H$_{2}$CO 3(1,2)-2(1,1) and para-H$_{2}$CO 3(0,3)-2(0,2) lines are detected in the five proto-BDs J182844, J183002, J182959, J163143, J182953, and one Stage I-T/II object J182952. The proto-BD J182854 shows weak emission in the para-H$_{2}$CO but none in ortho-H$_{2}$CO (Figs.~B2;~B3). The CS (5-4) line is detected in the four proto-BDs J182844, J183002, J182959, and J163143, with a weak detection in J182953 (Fig.~B4). All of these objects show single peaked Gaussian shaped profiles in H$_{2}$CO and CS, except J183002 that shows clear contamination with multiple peaks arising from different velocity components along the line of sight. None of the sources show emission in the $^{13}$CS and C$^{34}$S isotopologues.

HCO$^{+}$, H$_{2}$CO, and CS are all high-density tracers but may have different origins. In Fig.~\ref{overplots}, we have over-plotted the HCO$^{+}$, H$^{13}$CO$^{+}$, o-H$_{2}$CO, p-H$_{2}$CO, and CS spectra for J182844, which shows a blue-dominated asymmetry in the HCO$^{+}$ profile, and J183002, which shows multi-peaked line profiles. For J182844, the CS, o-H$_{2}$CO and p-H$_{2}$CO line profiles are similar to H$^{13}$CO$^{+}$, indicating an origin from the proto-BD core. For J183002, the central broad component in the o-H$_{2}$CO and p-H$_{2}$CO lines is coincident with HCO$^{+}$ and H$^{13}$CO$^{+}$, indicating a common origin from the central core. As mentioned, the HCO$^{+}$ and H$_{2}$CO lines show excess emission on the blue side along with a weak blue-shifted peak in H$_{2}$CO at $\sim$5 km s$^{-1}$, which may be associated with an outflow or another velocity component along the line of sight. On the other hand, the CS profile for J183002 shows a narrow peak at $\sim$9.5 km s$^{-1}$, which is also seen in the o-H$_{2}$CO and p-H$_{2}$CO lines but not in HCO$^{+}$ and H$^{13}$CO$^{+}$ (Fig.~\ref{overplots}). The differences in the density and velocity structure of the physical components probed by these high-density tracers could explain the differences in their observed profiles. None of the Stage I-T/Stage II objects show emission in any of these high-density tracers, which appears consistent with the relatively advanced evolutionary stage.


\subsection{Column Densities and Abundances}
\label{col-den}

We have used the non-LTE radiative transfer code RADEX (van der Tak et al. 2007) to estimate the column densities for all species detected towards the proto-BDs. RADEX considers a static, uniform density sphere, with the CMB as the only background radiation field, and with H$_{2}$ as the dominant collision partner for all calculations. The online version of RADEX\footnote{http://var.sron.nl/radex/radex.php} uses as input the kinetic temperature, $T_{kin}$, of the gas, the H$_{2}$ number density, $n_{H_{2}}$, the turbulent line width, and the column density for a given species. The kinetic temperature is expected to be relatively low ($\sim$10 K) throughout the outer and inner envelope layers in the proto-BDs (Machida et al. 2009). The H$_{2}$ number density and column density, $N (H_{2})$, were calculated from the total mass M$_{total}^{d+g}$ (Sect.~\ref{sample}) and size of the source. Tables~\ref{column1} lists the values for $n_{H_{2}}$ and $N (H_{2})$ derived for all targets.



\begin{table*}
\caption{Column Densities and Abundances}
\label{column1}
\begin{threeparttable}
\begin{tabular}{lp{0.9cm}p{0.75cm}p{0.9cm}p{0.7cm}p{1cm}p{0.9cm}p{0.6cm}p{0.6cm}p{1.55cm}p{1.3cm}p{1.6cm}p{1.3cm}} 
\hline
Object & $n_{H_{2}}$   & $N_{H_{2}}$   & $N$(C$^{17}$O) & [C$^{17}$O] & $N$(HCO$^{+}$)  & [HCO$^{+}$] & $N$(CS)  & [CS] & $N$(o-H$_{2}$CO) & [o-H$_{2}$CO]  & $N$(p-H$_{2}$CO)  & [p-H$_{2}$CO]    \\ 

	   & x10$^{6}$ & x10$^{22}$ &  x10$^{14}$ &   x10$^{-9}$ &  x10$^{12}$ &   x10$^{-10}$ &  x10$^{13}$ 			      &   x10$^{-10}$ &  x10$^{13}$ &   x10$^{-10}$ &  x10$^{13}$ &   x10$^{-10}$ 		\\

	   & (cm$^{-3}$)	& (cm$^{-2}$) &  (cm$^{-2}$) &   &  (cm$^{-2}$) &  & (cm$^{-2}$) &  &  (cm$^{-2}$) &  & (cm$^{-2}$) 			      &  		\\
\hline
\multicolumn{13}{c}{Stage 0+I} \\ \hline
J182854 & 1.0$\pm$0.4	& 3.7$\pm$0.6 & 3.2 & 8.6    & 2.2        & 0.6         & $<$1.3       & $<$3.5    & $<$0.15     & $<$0.1 &  0.024 & 0.065 \\
J182844 & 0.7$\pm$0.3	& 3.6$\pm$0.5 & 3.8 & 10.5    & 2.4      & 0.66        & 1.1  & 3.0 & 0.3 & 0.88 & 0.15 & 0.42  \\
J183002 & 2.1$\pm$0.2	& 7.6$\pm$0.5 & 42.0    & 55.0     & 34.5  & 4.5   & 3.8        & 5.0           & 3.0       & 3.9  & 0.65 & 0.85  \\
J182959 & 0.8$\pm$0.5	& 4.5$\pm$0.5 & 98.0 & 210.0 & 7.0       & 1.5        & 2.5 & 5.5   & 2.0   & 4.4 & 0.65 & 1.4 \\
J163143 & 0.3$\pm$0.2	& 1.8$\pm$0.2  & 4.2  & 23.3  & 25.0 & 13.8 & 11.0 & 61.1 & 5.4 & 30.0 & 1.0 & 5.5 \\
J182953  & $<$0.5	& $<$2.0  &  12.0     &   60.0  & 20.0 & 10.0 & 2.5 & 12.5 & 0.25 & 1.25  & 0.16 & 0.8 \\
J163136 & 0.6$\pm$0.1	& 2.9$\pm$0.9  & 10.5  & 36.2 & 1.5 & 0.5 & $<$0.25 & $<$0.9 & $<$0.06 & $<$0.2  & $<$0.03 & $<$0.1 \\
\hline \multicolumn{13}{c}{Stage I-T/II} \\ \hline
J182940 & 0.05$\pm$0.04 & 0.4$\pm$0.07 & 0.4  & 10.0   & $<$6.0 & $<$15.0  &  $<$2.5     & $<$42.0     & $<$0.3       & $<$0.8   & $<$0.3 & $<$7.5  \\
J182927 & $<$0.1	& $<$0.6  & 2.0 & 33.8  & $<$1.0 & $<$2 &  $<$1.5     & $<$25     & $<$0.25       & $<$4.2  & $<$0.1 & $<$2.0    \\ 
J182952  & $<$0.9	& $<$3.0  &  4.3    &   14.3  & 0.6 & 0.2 & $<$0.4 & $<$1.0 &  0.035 & 0.12  & 0.035 & 0.12 \\
\hline
\end{tabular}
\begin{tablenotes}
  \item The uncertainty is estimated to be $\sim$40\%-50\% for the column densities and molecular abundances. 
\end{tablenotes}
\end{threeparttable}
\end{table*}

\begin{table*}
\centering
\caption{H$^{13}$CO$^{+}$ (3-2) Line Parameters\tnote{a}}
\label{h13cop}
\begin{threeparttable}
\begin{tabular}{lcccccccc} 
\hline
Object   & V$_{lsr}$ & T$_{\textrm{mb}}$ & $\int{T_{mb} dv}$ &  $\Delta$v & $N$ (H$^{13}$CO$^{+}$)   & [H$^{13}$CO$^{+}$]  & $T_{ex}$ & $\tau$  \\
	     & (km s$^{-1}$) & (K)     & (K km s$^{-1}$) &  (km s$^{-1}$) & x10$^{11}$ (cm$^{-2}$)  & x10$^{-12}$ & (K) & \\
\hline
\multicolumn{9}{c}{Stage 0+I} \\ \hline
J182844 & 7.8 & 0.3 & 0.2 & 0.6 & 5.5 & 15.3 & 6.0 & 0.2 \\
J183002 & 8.0 & 0.4 & 0.3 & 0.7 & 4.0 & 5.3 & 7.6 &	0.1 \\
\hline
\end{tabular}
\begin{tablenotes}
  \item The uncertainty is estimated to be $\sim$15\%-20\% for the integrated intensity, $\sim$0.01-0.02 km s$^{-1}$ on V$_{lsr}$, $\sim$5\%-10\% on T$_{\textrm{mb}}$ and $\Delta$v, $\sim$20\% on $T_{ex}$ and $\tau$. 
\end{tablenotes}
\end{threeparttable}
\end{table*}

\begin{table*}
\centering
\caption{Excitation temperature and optical depth}
\label{tex-tau}
\begin{threeparttable}
\begin{tabular}{lcccccccccc} 
\hline
Object & \multicolumn{2}{c}{C$^{17}$O (2-1)} & \multicolumn{2}{c}{HCO$^{+}$ (3-2)} & \multicolumn{2}{c}{CS (5-4)} & \multicolumn{2}{c}{o-H$_{2}$CO (3-2)}  & \multicolumn{2}{c}{p-H$_{2}$CO (3-2)}     \\
\hline
	   & $T_{ex}$ (K)  & $\tau$  & $T_{ex}$ (K) & $\tau$ & $T_{ex}$ (K) & $\tau$ & $T_{ex}$ (K) & $\tau$ & $T_{ex}$ (K) & $\tau$  \\
	  
\hline
\multicolumn{11}{c}{Stage 0+I} \\ \hline
J182854 & 10.0 & 0.08   & 6.6 	& 0.3 	& -- 	   & --      & --   & -- & 6.9 & 0.04	    	\\
J182844 & 10.0 & 0.07   & 6.2 	& 0.5 	& 5.9 & 0.2 & 6.3   & 0.3 & 6.5 & 0.2  \\
J183002 & 10.0  & 0.3    & 8.5 	& 2.5 	&  7.5	   &   0.3   & 8.1    & 0.5	   & 8.3 &  0.4	\\
J182959 & 10.0 & 1.7     & 6.4 	& 0.9	& 6.1 & 0.3 & 6.7 & 0.8  & 6.9 & 0.6 \\
J163143 & 10.0 & 0.1     & 6.3 	& 4.8 	& 5.0 & 0.8 & 5.5 & 1.5  & 5.5 & 0.8 \\
J182953 & 10.0 &  0.2    & 6.1 	& 1.7	& 5.6 	   & 0.1      & 5.6   	& 0.2	  & 5.9 & 0.1	 \\
J163136 & 10.0 & 0.3     & 5.9 	& 0.5 	& -- 	   & --      & --   & -- & --   & --   	 \\
\hline \multicolumn{11}{c}{Stage I-T/II} \\ \hline
J182940 & 9.5 & 0.01   & -- 	& -- & -- 	   & --      & --   & -- 	& --   & -- 	\\
J182927 & 9.7 & 0.03   & -- & --	& -- 	   & --      & --   & -- 	& --   & --  	\\
J182952 & 10.0 & 0.06       & 6.3 	& 0.2 	& -- 	   & --      & 6.5     & 0.03	& 6.7 & 0.04  \\
\hline
\end{tabular}
\begin{tablenotes}
  \item The uncertainty is estimated to be $\sim$20\% on $T_{ex}$ and $\tau$ estimates. 
\end{tablenotes}
\end{threeparttable}
\end{table*}

For a fixed $T_{kin}$, $n_{H_{2}}$, and line width in RADEX, the input column density is varied to match the output peak intensity from RADEX with the observed peak line intensity (T$_{mb}$). The output from RADEX is the radiation temperature T$_{R}$, which was converted to T$_{mb}$ by making correction for the beam filling factor. This is the ratio of the source size to the beam size and is estimated to be $\sim$0.6-0.8. RADEX assumes a Gaussian profile that is convolved by the line width. Thus, in effect, the observed integrated intensity is matched with the model value. 

For HCO$^{+}$, o-H$_{2}$CO, p-H$_{2}$CO, and CS lines, we have used the total integrated intensity (sum of all Gaussian components) to estimate the column densities (Table~\ref{column1}). The C$^{17}$O column density has been estimated using the integrated intensity measured from the hyperfine fit. Due to the heavy contamination to the $^{12}$CO and $^{13}$CO emission, we have not derived the column densities for the main isotope line, and have instead derived the CO abundance from the C$^{17}$O abundance using the isotopic ratio of [$^{16}$O/$^{17}$O] = 2044 (Wilson \& Rood 1994; Penzias 1984). Since C$^{17}$O is a rare CO isotopologue, the derived abundance should be a more reliable estimate on the CO emission.

The column densities thus derived are listed in Tables~\ref{column1};~\ref{h13cop}. Also listed are the molecular abundances relative to H$_{2}$, [N(X)/N(H$_{2}$)]. We estimate an uncertainty of $\sim$20\% on the column densities. The uncertainty was estimated by varying the input column density to match the 1-$\sigma$ range in the observed integrated line intensity. The uncertainty on the derived molecular abundances is estimated to be $\sim$40\% and is propagated from the 1-$\sigma$ error on the H$_{2}$ density and the column density. For the non-detections, we have derived the upper limits on the column density and abundance using the 3-$\sigma$ upper limit on the integrated intensity, assuming a line width of 1 km s$^{-1}$ and the same kinetic temperature.


Table~\ref{h13cop} lists the H$^{13}$CO$^{+}$ (3-2) column density and abundance for J182844 and J183002. The  measured abundance ratio [HCO$^{+}$/H$^{13}$CO$^{+}$] is 4.4 for J182844 and 80 for J183002. In comparison with the ISM isotopic ratio of [$^{12}$C/$^{13}$C] = 77, the abundance ratio measured for J183002 is close to the ISM value while the one measured for J182844 indicates HCO$^{+}$ to be under-abundant. This can be expected due to the the self-absorption signature seen in the HCO$^{+}$ line profile for J182844 (Fig.~\ref{h13cop-figs}). Since RADEX does not take into account the effects of such features, this could result in an under-estimate of the HCO$^{+}$ column density and the derived abundance. No such self-absorption is seen in HCO$^{+}$ for J183002, and therefore its HCO$^{+}$ abundance derived from the main isotope should be a reliable estimate.

The output from RADEX also includes the excitation temperature $T_{ex}$ and the optical depth $\tau$. These are listed in Table~\ref{tex-tau}. Other than C$^{17}$O for which $T_{ex}$$\approx$$T_{kin}$, the $T_{ex}$ for all other tracers is in the range of $\sim$6-8 K and lower than the assumed $T_{kin}$$\sim$10 K. This indicates sub-thermal non-LTE conditions. The effect of the slightly lower dust temperature than 10 K on the M$^{d+g}_{\textrm{total}}$ estimates are within the uncertainties listed in Table~\ref{sample}. The RADEX estimates on line opacities are the largest for HCO$^{+}$ ($\sim$0.5-4.8) and $\leq$2 for the other tracers, suggesting that these are marginally optically thick (Table~\ref{tex-tau}). Table~\ref{line-obs} lists the critical density under optically thin LTE conditions, $n_{crit}^{thin}$, which can be used to indicate the density at which a transition would appear in emission. A comparison of $n_{crit}^{thin}$ with the number density of the sources (Table~\ref{column1}) indicates that other than the CO lines, none of the dense gas tracers should be detected. However, the number density of the sources have been derived from single-dish observations integrated over a beam size of $\sim$14.5$\arcsec$. Due to this, the inner, dense regions ($<$300 AU) in the proto-BDs where $n_{H_{2}}$ is expected to reach values of $>$10$^{7}$ cm$^{-3}$ (e.g., Machida et al. 2009) are beam-diluted in the dust continuum observations, but can be probed with these dense gas tracers.

\section{Results and Discussion}
\label{discussion}

\subsection{Depletion of CO}
\label{depletion}

 \begin{figure*}
  \centering              
     \includegraphics[width=3in]{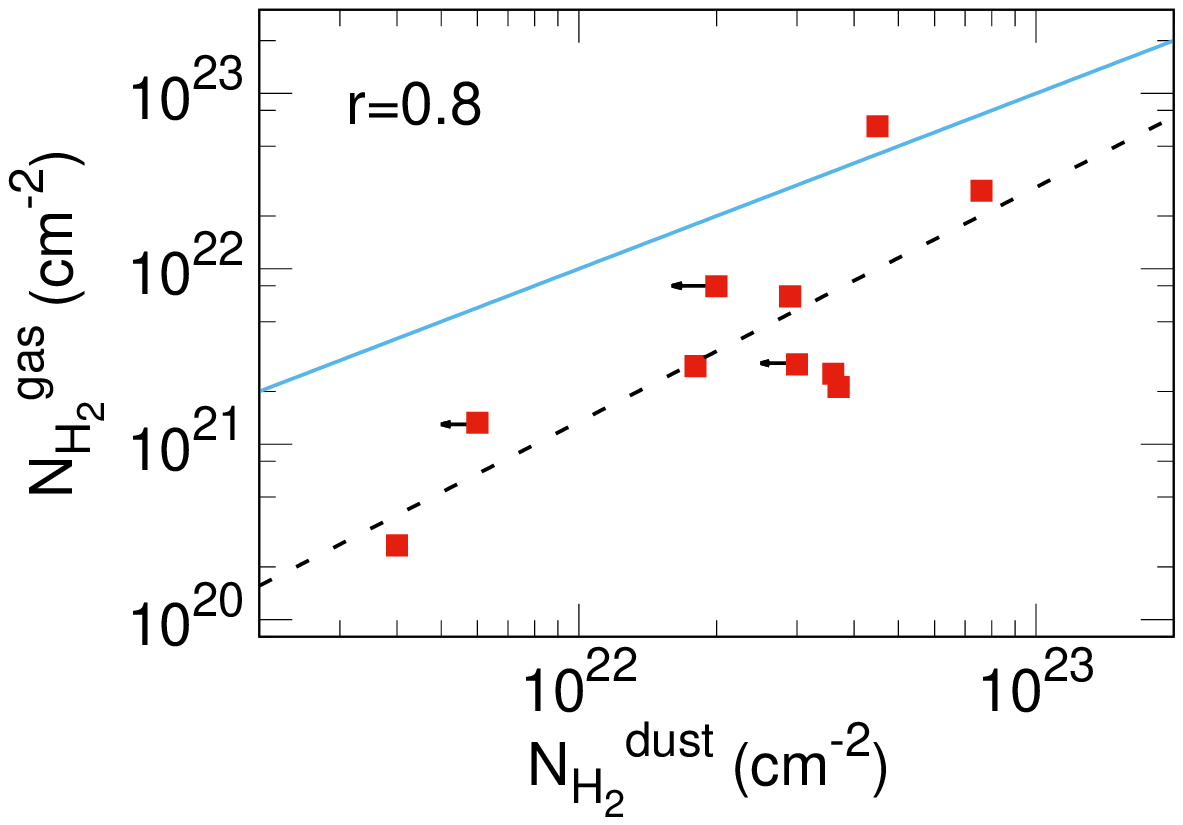}       
     \includegraphics[width=3in]{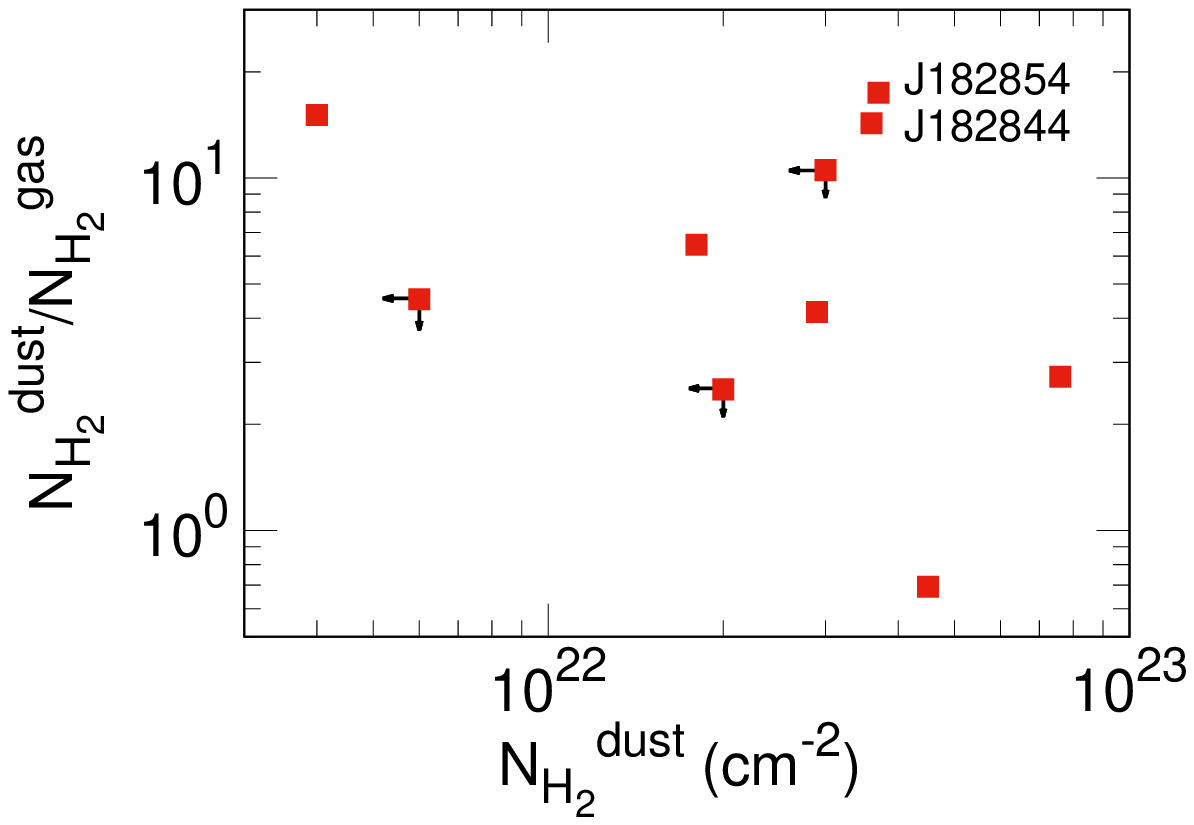}
     \caption{Left panel compares the H$_{2}$ column density derived from the sub-millimeter 850$\mu$m continuum observations, $N_{H_{2}}^{dust}$, with the H$_{2}$ column density derived from the C$^{17}$O column density, $N_{H_{2}}^{gas}$. The black dashed line is the best-fit to the data points. The blue line represents a 1:1 relation if there was no CO gas depletion. In the right panel, the ratio $N_{H_{2}}^{dust}$/$N_{H_{2}}^{gas}$ represents the extent to which CO is depleted from the gas phase, with J182844 and J182854 showing severe depletion. }
  \label{co-dep}       
\end{figure*}

The CO molecules have traditionally been used as a proxy tracer of mass in molecular clouds. In dense and cold cores, a comparison of the hydrogen column density derived from the CO gas emission with that derived from the dust continuum emission can provide an estimate on the fraction of CO that has frozen onto the surfaces of dust grains and thus depleted from the gas phase. Using this method, CO is found to be depleted by an order of magnitude in low-mass Class 0/I protostars, by factors of $\sim$4-15 in pre-stellar cores, and in extreme cases such as the L1544 and L1689B dense cores, more than 90\% of the CO is frozen onto grains (e.g., Caselli et al. 1999; Joergensen et al. 2002; Bergin et al. 2002; Bacmann et al. 2002; Redman et al. 2002). C$^{17}$O emission in our proto-BDs is optically thin and traces the quiescent gas deep into the core (Sect.~\ref{analysis}), and thus it allows an estimate of the total column density as traced by this molecule. To calculate the hydrogen column density from the C$^{17}$O emission, we have used the H$_{2}$/CO abundance ratio of 3700 (Lacy et al. 1994), and the oxygen isotopic ratio of [$^{16}$O/$^{17}$O] = 2044 (Wilson \& Rood 1994; Penzias 1981). The final conversion factor from C$^{17}$O to H$_{2}$ is 7.6 $\times$ 10$^{6}$. 

Figure~\ref{co-dep} compares the H$_{2}$ column density derived from the sub-millimeter 850$\mu$m continuum observations, $N_{H_{2}}^{dust}$, with the H$_{2}$ column density derived from the C$^{17}$O column density, $N_{H_{2}}^{gas}$ for our sample. The proto-BDs J182844 and J182854 have $N_{H_{2}}^{gas}$ that is $\sim$15 times lower than $N_{H_{2}}^{dust}$, indicating that a significant fraction ($\sim$90\%) of CO is depleted from the gas phase and frozen onto the surfaces of grains in these objects. For the other objects, $N_{H_{2}}^{dust}$ is moderately correlated with $N_{H_{2}}^{gas}$ (correlation coefficient $\sim$0.8), with a $N_{H_{2}}^{dust}$/$N_{H_{2}}^{gas}$ ratio of $\sim$0.7--6. The timescale and spatial extent of the CO freeze-out would depend on the local density and temperature, since the freeze-out is expected to take a longer time ($>$10,000 yr) in the outer low-density regions than the denser, cooler regions towards the central core (e.g., Aikawa et al. 2005). Another object J182940, which is a Stage I-T/II object, also shows a high $N_{H_{2}}^{dust}$/$N_{H_{2}}^{gas}$ ratio of $\sim$15, but its C$^{17}$O spectrum has a low signal to noise ratio of $\sim$2 due to which the ratio has a high 50\% uncertainty. As noted, both $N_{H_{2}}^{dust}$ and $N_{H_{2}}^{gas}$ are beam-averaged measurements. Alternative factors such as the effects of beam dilution of the central dense ($n_{H_{2}}$ $>$ 10$^{7}$ cm$^{-3}$) regions in the dust continuum observations, and optical depth on the line emission can also result in differences in the H$_{2}$ dust and gas column densities. 

     
We can use the ratios of the measured abundances of the optically thin (C$^{17}$O and C$^{18}$O) and optically thick ($^{13}$CO) CO isotopologues to compare with the corresponding ISM isotopic ratio, which is usually adopted for predicting CO isotopologue abundances. We consider the proto-BD J182844 that shows a single peaked profile in all CO isotopologues. The expected [$^{13}$CO/C$^{17}$O] and [$^{13}$CO/C$^{18}$O] isotopic ratios are 26.5 and 7.3, respectively. For J182844, we have measured [$^{13}$CO/C$^{17}$O] = 7.5 and [$^{13}$CO/C$^{18}$O] = 3.6, indicating that C$^{17}$O and C$^{18}$O are $\sim$2--3 times more abundant than the expected isotopic ratios. The discrepancy could be caused by significant optical depths of the observed $^{13}$CO emission.

\subsection{Trends in molecular abundances}
\label{trends}

\begin{figure}
  \centering              
     \includegraphics[width=3in]{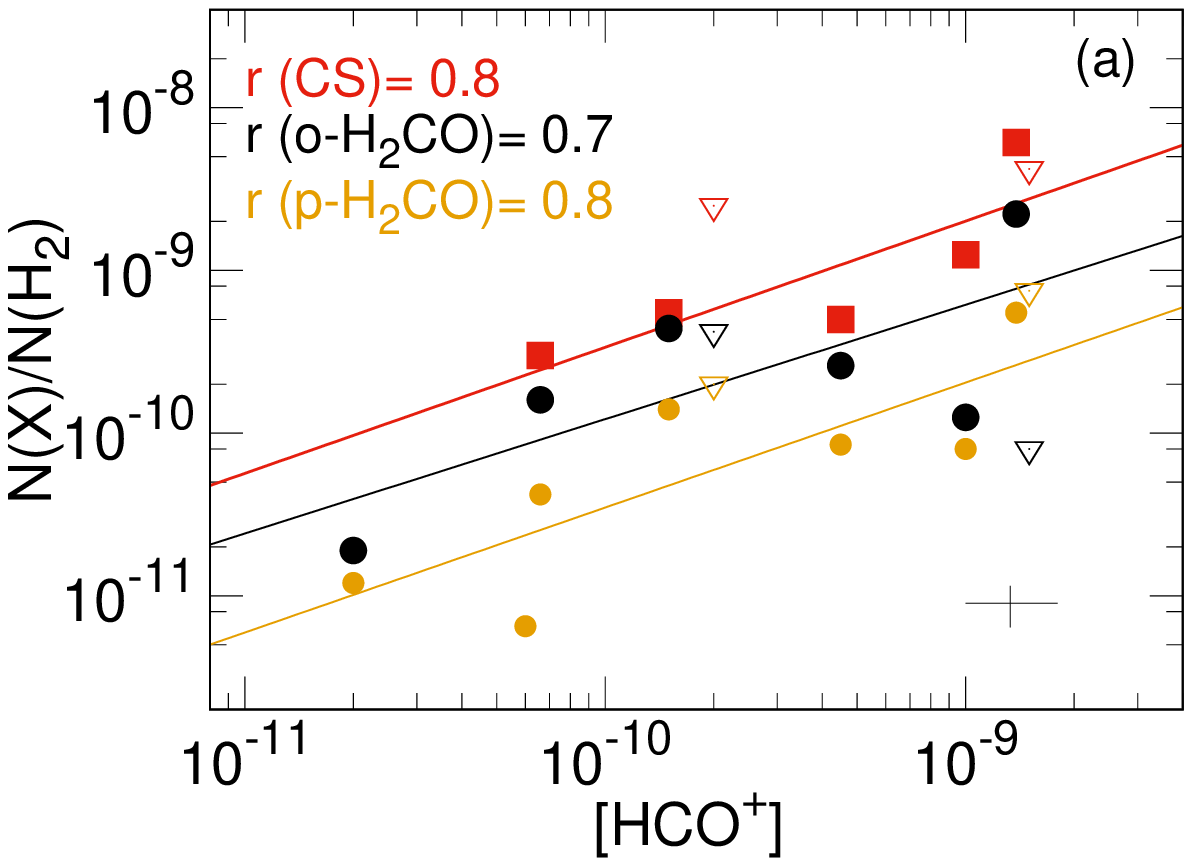}     
     \includegraphics[width=3in]{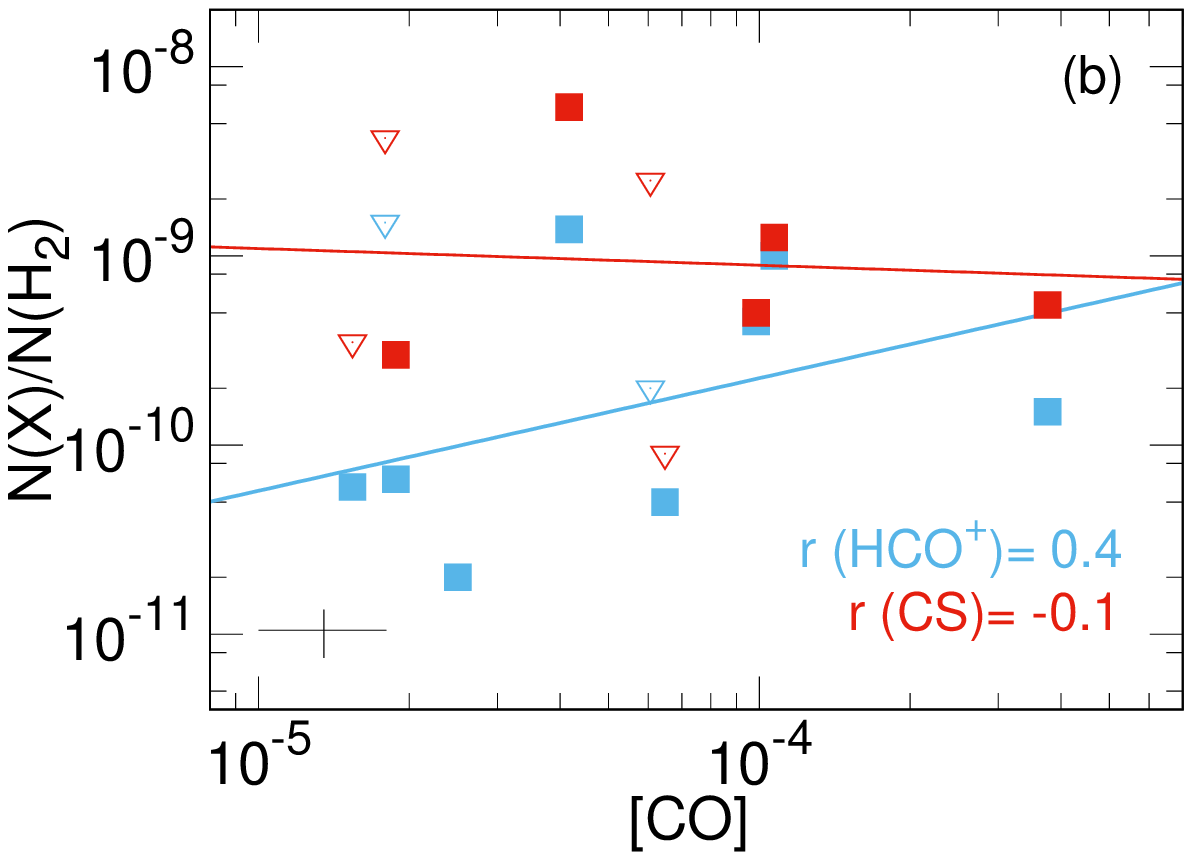}
     \includegraphics[width=3in]{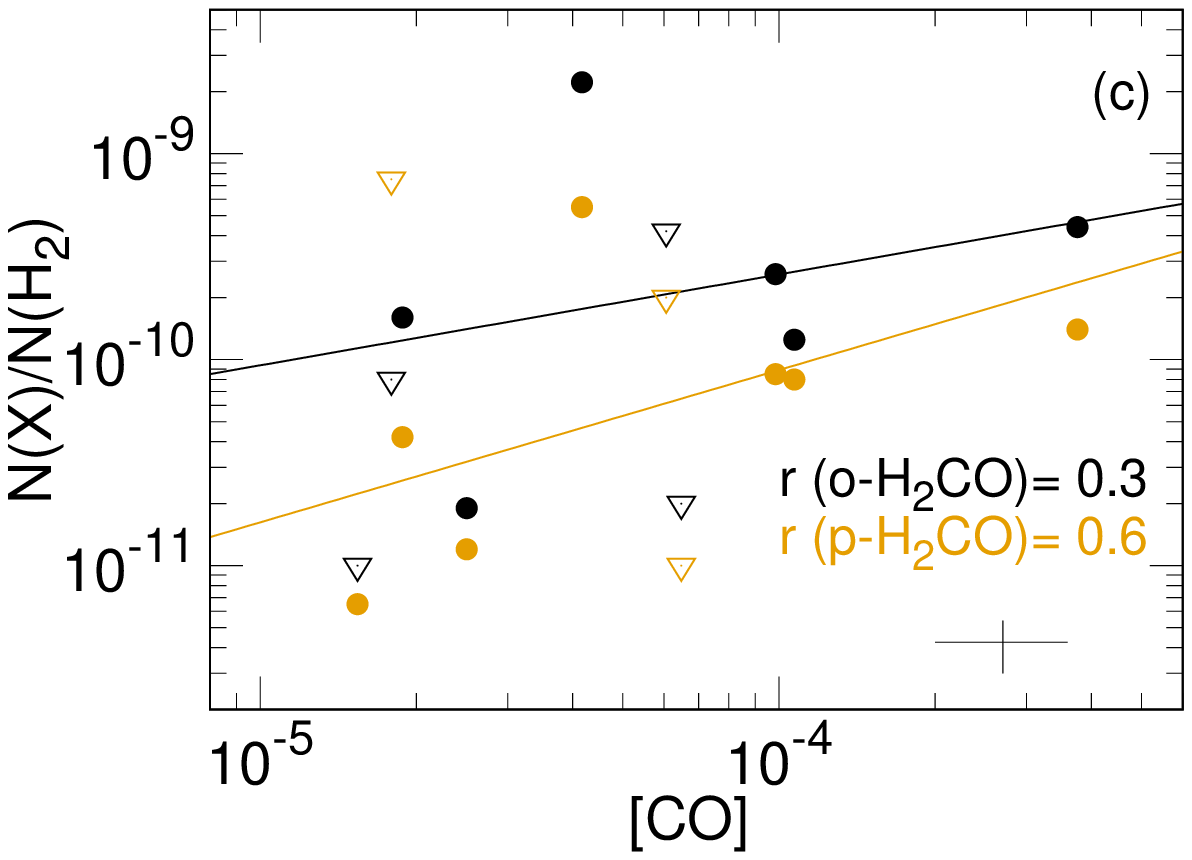}   
     \includegraphics[width=3in]{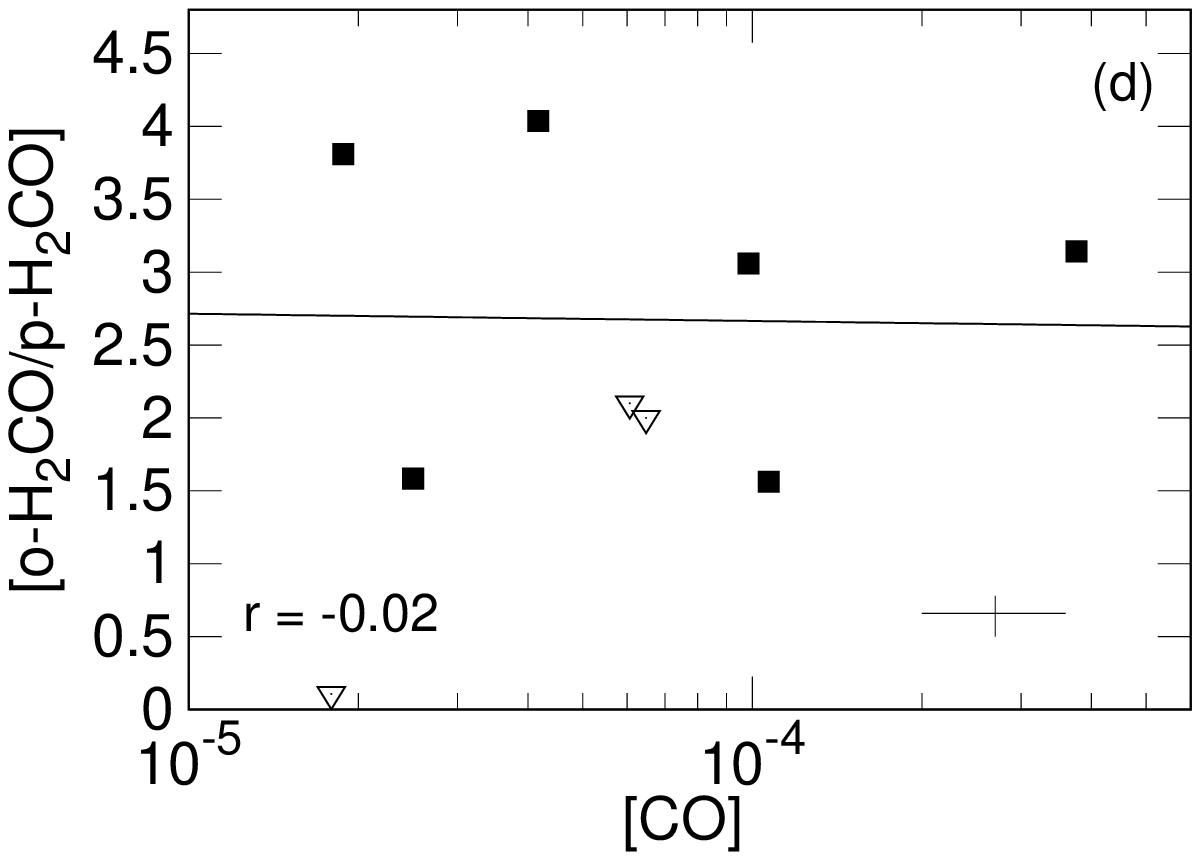}             
     \caption{Correlations in the abundances (with respect to H$_{2}$) among the HCO$^{+}$, CS, and H$_{2}$CO tracers, and with CO. The correlation coefficients are noted in each plot. Diamonds denote the non-detections, which have not been included in the correlations. Typical uncertainties in the abundances are estimated to be $\sim$40\% and are plotted in the bottom, right corner of each plot.}
  \label{hcop-cs-h2co}       
\end{figure}

Figure~\ref{hcop-cs-h2co} explores the correlations in the abundances relative to H$_{2}$ in the various tracers for the sample. The correlation coefficient $r$ from the best-fit to the observed data points is noted in the plots; a value of $\pm$1 for $r$ indicates a tight correlation (100\%) between the two parameters, depending on the slope of the best-fit. The upper limits have not been included when fitting the data points. We estimate an uncertainty of 20\% on the coefficients.


As discussed in Sect.~\ref{profiles}, most objects show complex profiles in the optically thick HCO$^{+}$ (3-2) line, with blue or red asymmetry indicative of infall or rotation signatures. The asymmetries in HCO$^{+}$ can also be caused by outflows (e.g., Rawlings et al. 2004). In comparison, CS and H$_{2}$CO profiles are narrowly peaked and Gaussian-shaped, suggesting a different origin than HCO$^{+}$. It is likely that the HCO$^{+}$ emission is dominated by the outer envelope layers and the CS and H$_{2}$CO emission represents the inner, dense regions of the proto-BDs. Figure~\ref{hcop-cs-h2co}a shows that the beam-averaged abundances for these high-density tracers are well-correlated ($r\sim$70-80\%) with each other, despite the differences in their emitting regions. 

Figures~\ref{hcop-cs-h2co}b,c probe the dependence of the HCO$^{+}$, CS, and H$_{2}$CO abundances on the depletion in CO. The CO abundances have been derived from the C$^{17}$O abundances (Sect.~\ref{col-den}). CS shows a notably flat distribution with no dependence on CO ($r\sim$10\%), HCO$^{+}$ and o-H$_{2}$CO are weakly correlated ($r\sim$30\%--40\%), while p-H$_{2}$CO shows a comparatively moderate correlation with CO ($r\sim$60\%). The fact that high-density tracers do not follow CO is probably related to the fact that the inner structure of the proto-BDs is complex, perhaps with clumpy dense regions traced by the high-density tracers, embedded in a smoother lower density envelope traced by the CO isotopologues. This has to be tested with high angular resolution observations.

The weak correlation with CO could be related to the formation/destruction pathways for these species. One pathway for CS formation is via neutral-neutral reaction of S with CH$_{2}$ or CH or C$_{2}$ intermediates to form CS, but these reactions have a lower rate of 10$^{-11}$ cm$^{3}$ s$^{-1}$. A faster reaction is S + O$_{2}$ $\rightarrow$ SO + C $\rightarrow$ CS + O. There is no SO detection for any proto-BDs even though there are several SO lines in the frequency range covered by our observations. This suggests that most of the atomic sulphur is incorporated into forming CS, although SO may also be mainly on the surface of dust grains. The lack of correlation between CS and CO suggests that CS may be forming in the inner, denser regions via these neutral-neutral reactions where O and S abundances are expected to be high.






The main formation mechanism of HCO$^{+}$ is via the reaction H$_{3}^{+}$ + CO $\rightarrow$ HCO$^{+}$ + H$_{2}$, and the main destruction mechanism is via dissociative recombination HCO$^{+}$ + e$^{-}$ $\rightarrow$ CO + H. So both the formation and destruction are dependent on the CO abundance. The lack of a strong correlation between HCO$^{+}$ and CO suggests that alternative pathways via CH$_{2}$$^{+}$ + O $\rightarrow$ HCO$^{+}$ + H and C$^{+}$ + H$_{2}$O $\rightarrow$ HCO$^{+}$ + H are also active in HCO$^{+}$ formation though the reactions rates will be slower than the main formation pathway via CO. On the other hand, C$^{+}$ abundances cannot be large in dense gas unless in the outflowing gas where molecules are dissociated. HCO$^{+}$ can efficiently be destroyed by H$_{2}$O, forming H$_{3}$O$^{+}$, and this could be important in regions where ices are desorbed, i.e. next to the proto-BDs or along the molecular outflow. 


H$_{2}$CO can form via gas phase as well as through ice surface reactions. The dominant gas-phase formation mechanism is via repeated reactions of C$^{+}$ with H$_{2}$ to form CH$_{4}^{+}$ followed by dissociative recombination to form CH$_{3}$, which then reacts with atomic oxygen to form H$_{2}$CO (e.g., Kahane et al. 1984). H$_{2}$CO formation on grains requires the freeze-out of CO and is only efficient at low temperature of $\sim$10--15 K. Basically, ice surface reactions undergoing two successive hydrogenation can produce H$_{2}$CO, i.e. CO + H $\rightarrow$ HCO and HCO + H $\rightarrow$ H$_{2}$CO. For the proto-BDs, no dependence is seen between the o-H$_{2}$CO and CO abundances (correlation coefficient = 0.2), while [p-H$_{2}$CO] declines with decreasing CO abundance (Fig.~\ref{hcop-cs-h2co}c). If both o-H$_{2}$CO and p-H$_{2}$CO were forming via ice surface reactions in the proto-BDs, their abundances would show an increase as CO is depleted from the gas phase. This suggests gas-phase formation for H$_{2}$CO in these objects. 



\subsection{H$_{2}$CO ortho-to-para ratio}
\label{ratio}

The ratio between the column densities of ortho-H$_{2}$CO and para-H$_{2}$CO can provide information about the formation of the molecule (e.g., Kahane et al. 1984; Tudorie et al. 2006). When the molecule forms in the gas-phase via reaction with CH$_{3}$, a ratio of 3 is expected, which corresponds to the statistical weight ratio between the ground states of the ortho and para species. An alternative gas-phase reaction to produce H$_{2}$CO is via H$_{3}$CO$^{+}$ (H$_{3}$CO$^{+}$ + e$^{-}$ $\rightarrow$ H$_{2}$CO + H), which results in an ortho to para H$_{2}$CO ratio of $\sim$3-5 at 10 K and $\sim$3 at 70 K (Kahane et al. 1984). This formation pathway, however, is uncertain due to the lack of information on the interconversion process via formation of H$_{3}$CO$^{+}$. A ratio lower than 3 is expected when the molecule is formed on the surface of cold (T$_{dust} \sim$ 10-15 K) dust grains (e.g., Kahane et al. 1984; Dickens \& Irvine 1999; Guzman et al. 2011). Previous studies have measured H$_{2}$CO ortho-to-para ratios of $\sim$3 towards dense cores, and in the range of $\sim$1.5--2 in the envelopes around low-mass protostars and towards star-forming cores with outflows (Dickens \& Irvine 1999; Jørgensen et al. 2005; Guzman et al. 2011).

\begin{table}
\centering
\caption{H$_{2}$CO abundance ratios}
\label{ratios}
\begin{threeparttable}
\begin{tabular}{ccc} 
\hline
Object & [o-H$_{2}$CO/p-H$_{2}$CO] \tnote{a}   \\
\hline
\multicolumn{2}{c}{Stage 0+I} \\ \hline
J182854 & $<$1.5 \\ 
J182844 & 3.8 \\ 
J183002 & 3.0 \\ 
J182959 & 3.1 \\ 
J163143 & 4.0 \\ 
J182953 & 1.6 \\ 
J163136 & $<$2.0 \\ 
\hline \multicolumn{2}{c}{Stage I-T/II} \\ \hline
J182940 & $<$0.1 \\ 
J182927 & $<$2.1 \\ 
J182952 & 1.6 \\ 
\hline
\end{tabular}
\begin{tablenotes}
  \item[a] The uncertainty is estimated to be $\sim$20\% on the abundance ratios. 
\end{tablenotes}
\end{threeparttable}
\end{table}

We find H$_{2}$CO ortho-to-para ratios in the range of $\sim$1.6--4.1 for our sample, with a mean value of 2.8$\pm$0.9 (Table~\ref{ratios}; Fig.~\ref{hcop-cs-h2co}d). The low ratio of 1.6 for J182953 and J182952 indicates H$_{2}$CO formation on dust grains. H$_{2}$CO molecules in the gas phase can be adsorbed onto grain mantles. Due to thermal equilibrium between the grains and the dense gas, the ortho/para ratio will reflect the equilibrium gas and grain temperature of $\sim$10-15 K, which corresponds to an ortho to para ratio of $\sim$1.5 (Kahane et al. 1984). Interestingly, J182844 does not show emission in ortho but in para-H$_{2}$CO, with an ortho-to-para ratio of $<$1.5. This suggests H$_{2}$CO formation on the surface of cold dust grains, and that the ortho species in this proto-BD may be frozen towards the central core, while para is detected from an intermediate layer.

In comparison, the H$_{2}$CO ortho-to-para ratio of $\sim$3 for the proto-BDs J182959 and J183002 indicates gas phase formation for H$_{2}$CO. The proto-BDs J182844 and J163143 are two unusual cases with H$_{2}$CO ortho/para ratio of 3.8-4.0. High ratios of $>$3 typically imply high-temperature ($>$50 K) formation (Kahane et al. 1984; Guzman et al. 2011). Note that the proto-BDs show high ortho/para ratios but low H$_{2}$CO abundances of the order of 10$^{-9}$--10$^{-11}$. This is unlike the high H$_{2}$CO abundances (10$^{-7}$--10$^{-9}$) typically measured in PDRs such as Orion where the ortho/para ratios of $\sim$3-4 have been explained by the presence of an external ISRF and a warm ($>$50-70 K) environment (e.g., Leurini et al. 2010).



There is evidence that J182844, J163143, and J182959 are driving outflows and show signs of strong accretion activity in the near-infrared observations, with accretion and outflow rates of the order of 10$^{-8}$ M$_{\sun}$ yr$^{-1}$ as typically measured in low-mass protostars (Whelan et al. 2018; Riaz et al. 2018). Considering the H$_{2}$CO desorption energy of 2050 K, the evaporation temperature would be $\sim$40 K using the formula in Hollenbach et al. (2009). In the innermost regions ($<$5 AU) of the proto-BDs close to where the jet/outflow interacts with the inner envelope layers, the temperature may be raised locally to $\sim$40-50 K and such thermal processes could lead to the H$_{2}$CO formed on the ice surface to be released into the gas phase. Non-thermal processes such as photo-desorption could also play a role. An active accretion zone can provide a source of stellar FUV flux that can be 100-10,000 times stronger than the interstellar UV radiation field (e.g., Bergin et al. 2003). A high flux of UV photons can impinge on the ice surface grains to desorb the H$_{2}$CO molecule and enhance the gas-phase abundance. Similar abundance enhancements in H$_{2}$CO due to the impact of outflows and shocks on the inner envelope have been previously reported in low-mass dense cores and protostars (e.g., Bachiller et al. 2001; Maret et al. 2004; Joergensen et al. 2005).


We may have a case of spin-selective photo-dissociation. The trend in Fig.~\ref{hcop-cs-h2co}c indicates an order of magnitude higher abundance for ortho compared to para in the dense core where CO is depleted, whereas the ortho and para abundances are nearly equal in the outer envelope layers where CO gas-phase abundance is high. This suggests that ortho can survive in the gas phase in the inner, dense regions whereas the para species shows some depletion. Considering the statistical weight ratio between the ground states of the ortho and para species of 3:1, this would imply 3 times higher abundance for ortho than para. H$_{2}$CO photo-dissociates by line absorptions, and the UV cross-sections for ortho and para-H$_{2}$CO are different (Heays et al. 2017; Meller et al. 2000; Dodson et al. 2015; Cooper et al. 1996). Lines are more attenuate for the more abundant of the two, which results in spin-selective photo-dissociation. The incident UV radiation would be optically thick to ortho but optically thin to para, resulting in destroying more of the para species and a high ortho to para ratio. Alternatively, ortho- and para-H$_{2}$CO may be predominately formed in regions exposed to different UV field strengths. The p-H$_{2}$CO emission in the proto-BDs likely originates from the less dense intermediate layers where it is shielded from the UV photons. A high ($>$ 3) ortho/para ratio comes then from a differential destruction rate, and this could be related to the proximity of the emitting region to the internal stellar UV radiation field.




The mean ortho/para ratio of $\sim$3 for the proto-BDs suggests gas-phase H$_{2}$CO formation but there appears to be a significant contribution from grain surface chemistry. If strong accretion activity and outflow effects are mainly responsible for the desorption then this can also happen at the low kinetic temperature prevalent in the proto-BD envelopes, and this may not necessarily be a case of high-temperature thermalization on the grain surface (e.g., Dickens \& Irvine 1999; Guzman et al. 2011). The relative contribution of the grains to the value of the ortho/para ratio in the gas phase would depend on multiple factors including the low-temperature desorption mechanisms, the likelihood of differential photo-dissociation that destroys para more than ortho, self-shielding effects, different emitting regions for the two species. The density versus temperature dependence of ortho versus para species could also explain the enhanced ratios, irrespective of differential photo-dissociation or desorption effects. 


\subsection{Trends in comparison with Class 0/I low-mass protostars}
\label{protostar}

\begin{figure*}
  \centering              
     \includegraphics[width=2.8in]{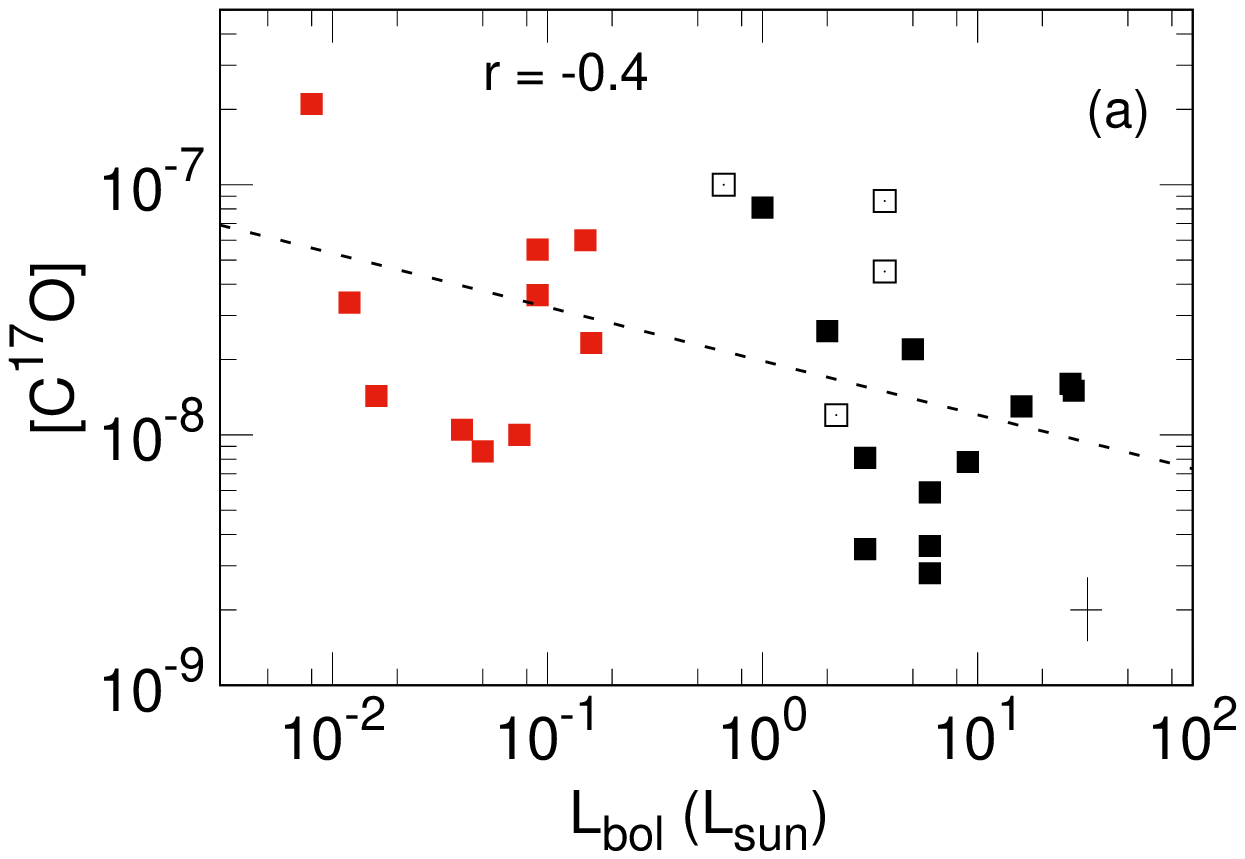}
     \includegraphics[width=2.8in]{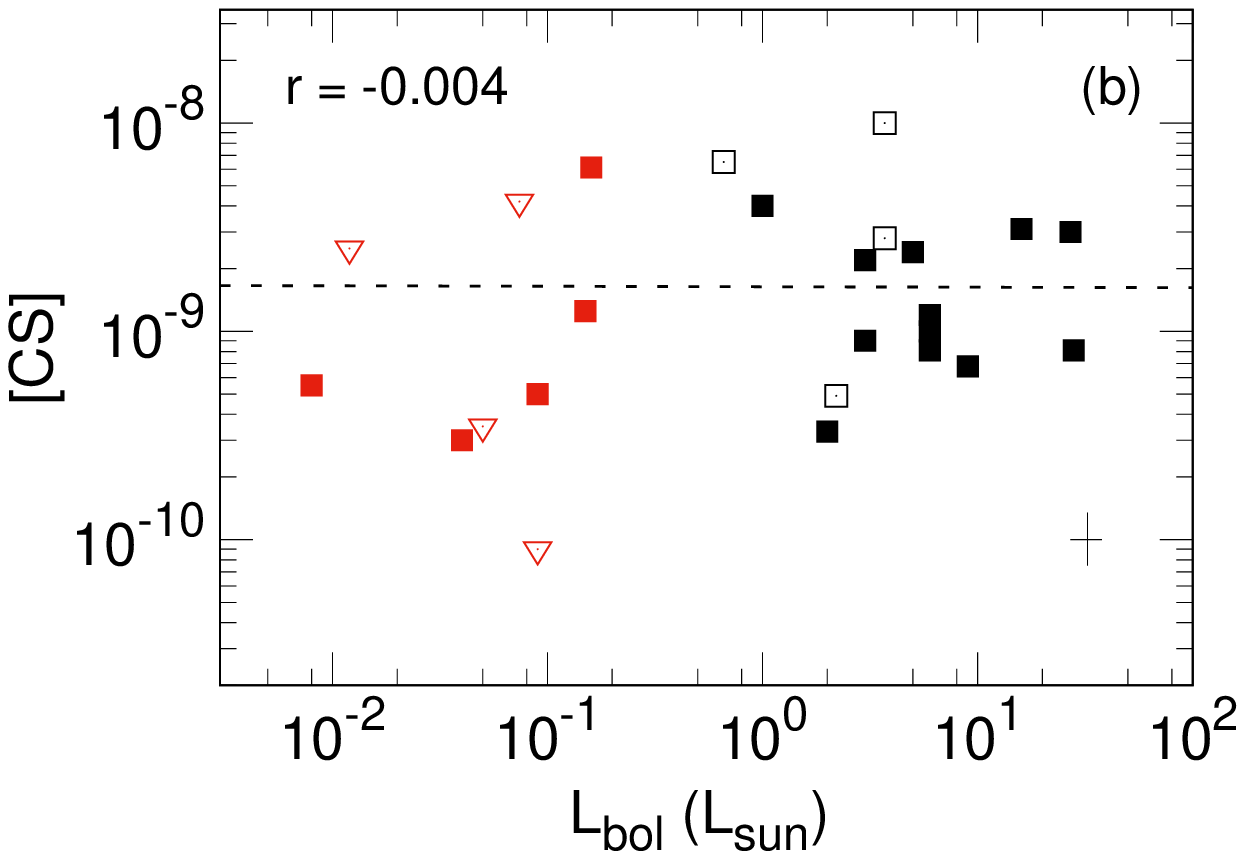}           
     \includegraphics[width=2.8in]{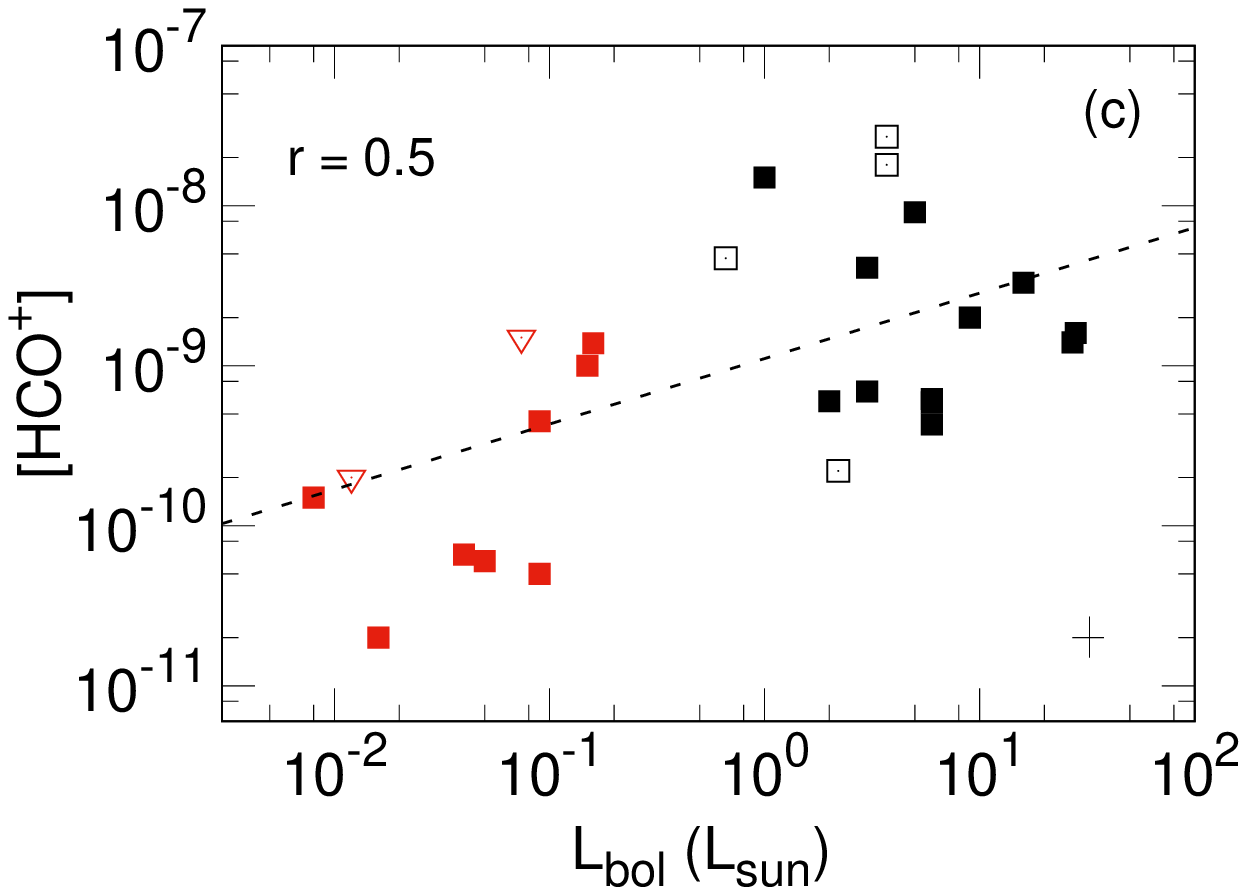}        
     \includegraphics[width=2.8in]{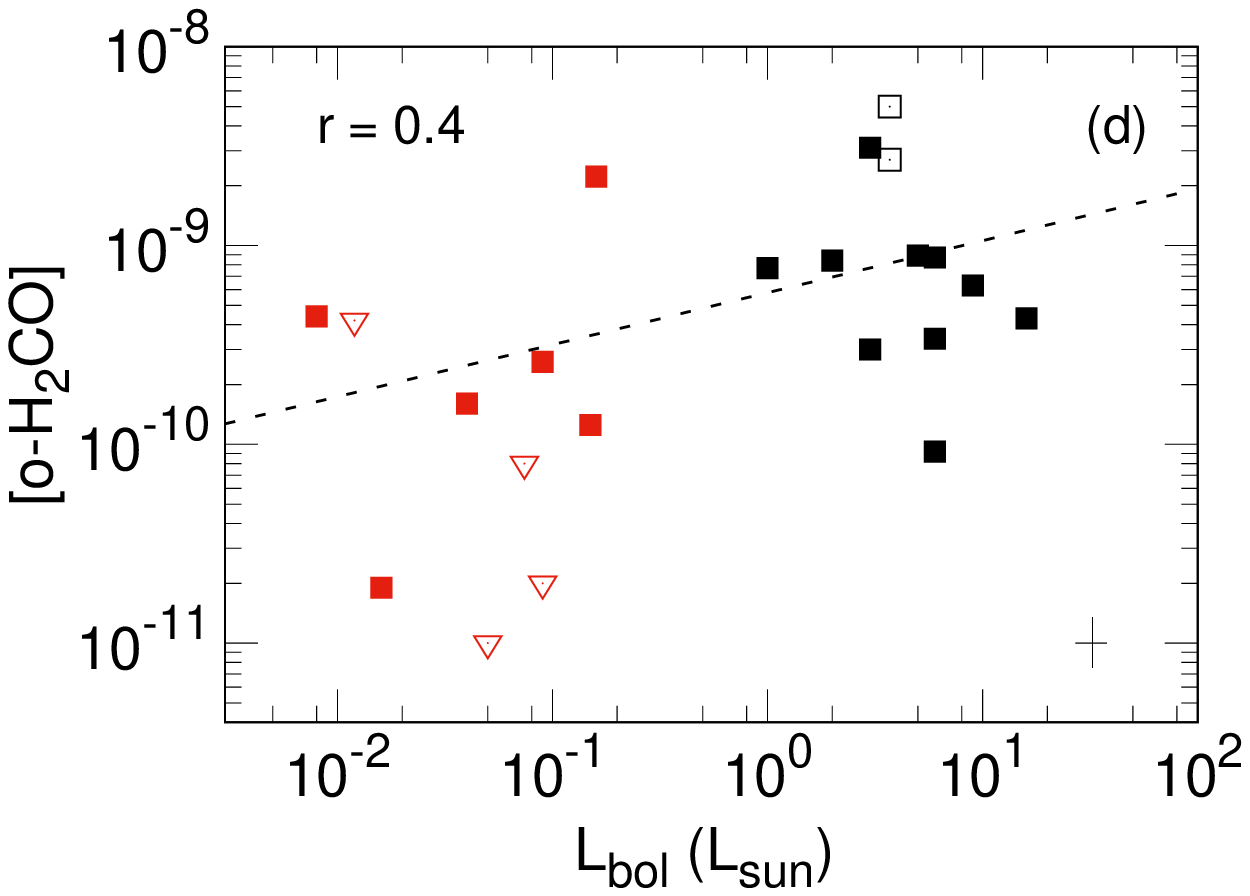}      
     \includegraphics[width=2.8in]{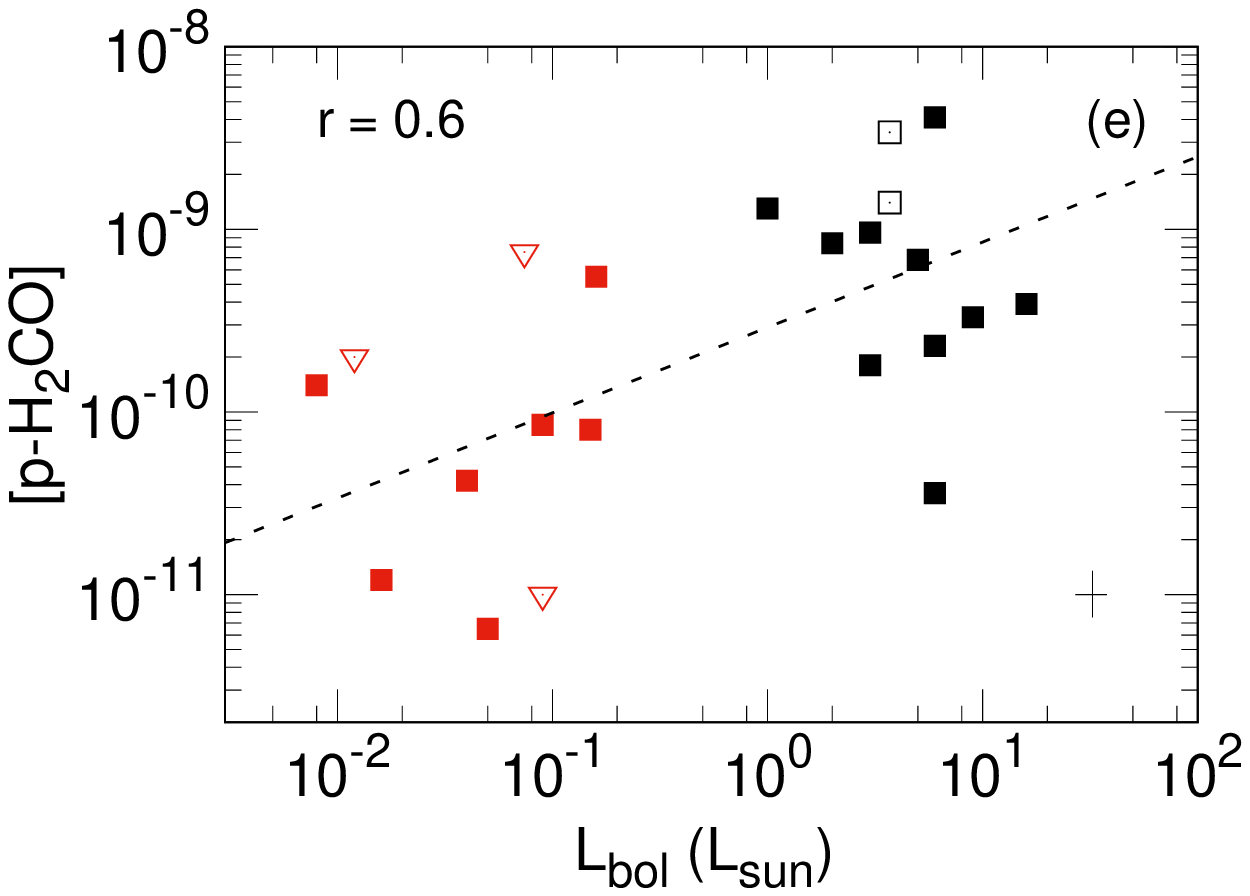}      
     \includegraphics[width=2.8in]{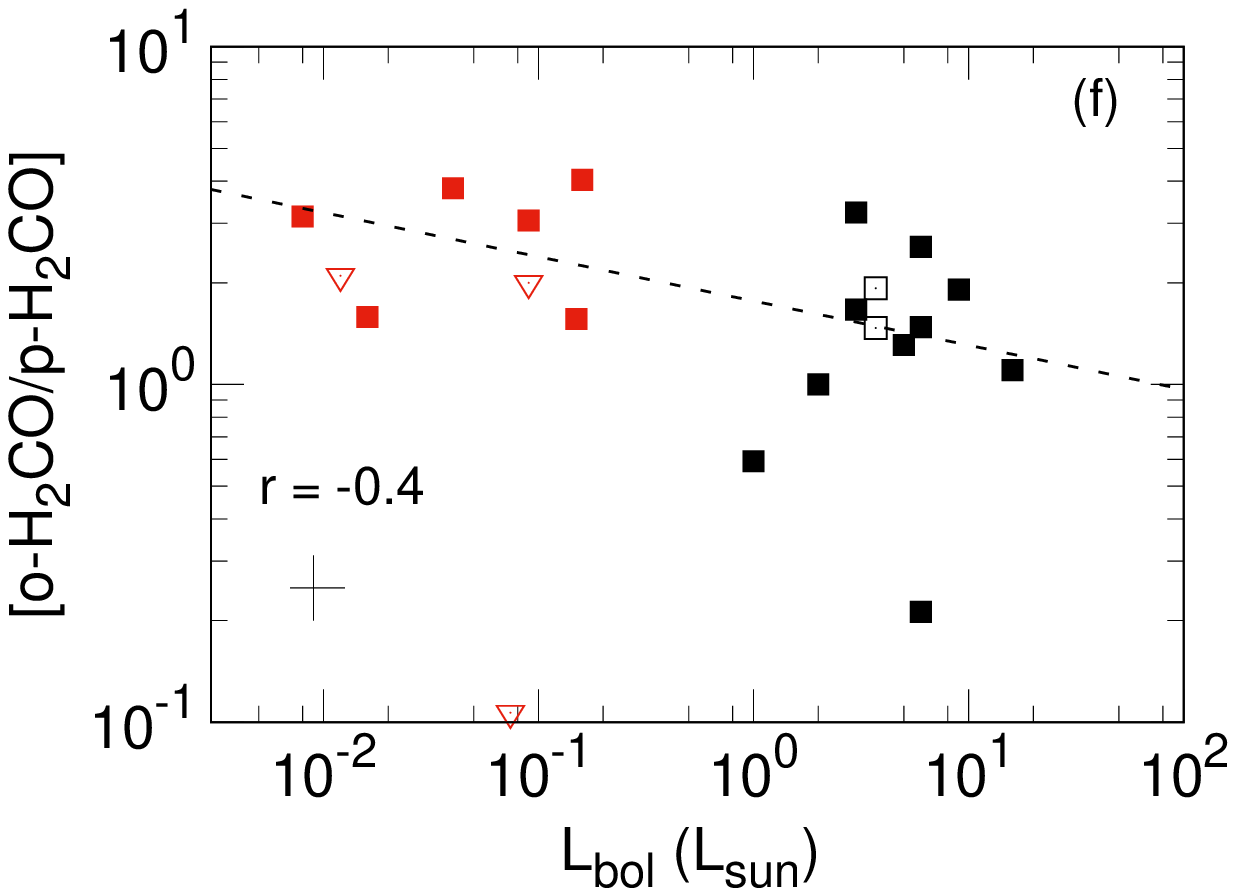}       
     \caption{The molecular abundances for the proto-BDs (filled red points) compared with Class 0 protostars (filled black points), and Class I protostars (open black points). Red diamonds denote the non-detections, which have not been included in the correlations. Typical uncertainties are estimated to be $\sim$40\% on the abundances and $\sim$30\% on L$_{\textrm{bol}}$, and are plotted in the bottom, right corner of each plot.}
  \label{lbol}       
\end{figure*}

Figure~\ref{lbol} compares the molecular abundances for the proto-BDs with low-mass Class 0/I protostars from the work of Joergensen et al. (2002; 2004). The comparison covers over four orders of magnitude range in the bolometric luminosities from $\sim$40--0.01 L$_{\sun}$, and aims to investigate any change in the trends in the molecular abundances with L$_{\textrm{bol}}$ across the sub-stellar boundary (L$_{bol} \sim$ 0.09 L$_{\sun}$). The beam-averaged measurements from Joergensen et al. (2002; 2004) have also been derived from single-dish observations.

The C$^{17}$O abundances for the proto-BDs show a similar range of about an order of magnitude as seen for the protostars (Fig.~\ref{lbol}a). A flat distribution is obtained if the proto-BD J182959 that shows the strongest CO emission ([C$^{17}$O]$\sim$2x10$^{-7}$) is excluded from the fit to all data points. While the CO abundances are nearly constant with decreasing luminosities, the large spread in the abundances could be due to various possibilities. The observed CO emission in the proto-BDs likely originates from the intermediate envelope layers ($\sim$200-400 AU) that are shielded from the central heating source and outflow/jet or external shock emission. On the other hand, the proto-BDs are known to show strong accretion activity (Riaz et al. 2015; Whelan et al. 2018), and we may be seeing the effects of UV induced photo-desorption due to an active accretion zone that could locally enhance the CO emission. Another possibility is rapid grain growth that can enhance the gas phase CO abundance regardless of the efficiency of freeze-out. Previous results on young Class II brown dwarf disks have noted stronger signs of dust processing and grain growth as compared to T Tauri disks (Riaz 2009; Riaz et al. 2012). 




Likewise, the CS abundances show no particular change between the protostars and proto-BDs, and appears to remain constant with $L_{bol}$ (Fig.~\ref{lbol}b). The constant CS abundance for the proto-BDs can be explained if the origin of emission is from the inner disk region in the proto-BDs. Since the physical scale of the densest region of the inner disk is $\leq$10 AU and is found to be nearly the same for all proto-BDs (Riaz et al. {\it in prep}), the resulting abundances are also expected to be nearly constant. While the CS abundances are similar for the proto-BDs and protostars, a notable difference is the non-detection of SO and OCS in the proto-BDs unlike the protostars. This could be related to the formation pathway of CS in the proto-BDs, as discussed in Sect.~\ref{trends}. For the case of protostars, the detection of both CS and SO could be related to later stages of chemical evolution when most of the sulphur is incorporated into these species (e.g., Aikawa et al. 2005). 



We find a tentative trend of a decline in the HCO$^{+}$, ortho-H$_{2}$CO, and para-H$_{2}$CO molecular abundances with decreasing bolometric luminosity (Fig.~\ref{lbol} cde). The mean HCO$^{+}$ abundance is 1.6x10$^{-9}$ for the proto-BDs and is $\sim$3 times lower compared to 5.6x10$^{-9}$ for the protostars. A similar difference by a factor of $\sim$3 is seen for the ortho-H$_{2}$CO species; the mean [o-H$_{2}$CO] is 0.5x10$^{-10}$ for the proto-BDs and 1.6x10$^{-9}$ for the protostars. On the other hand, the mean [p-H$_{2}$CO] is 0.01x10$^{-10}$ for the proto-BDs, $\sim$10 times lower than the mean [p-H$_{2}$CO] = 1.1x10$^{-9}$ for the protostars. The lower abundances for proto-BDs likely reflect their scaled-down physical structures compared to protostars. The physical structure as determined from SED modelling of proto-BDs indicates envelope sizes of $\sim$1000 AU. In comparison, the physical structures for Class 0/I low-mass protostars indicate envelope sizes of $\sim$5000-10,000 AU (e.g., Furlan et al. 2008; 2016). As discussed in Sect.~\ref{profiles}, the bulk of the emission in HCO$^{+}$ likely originates from the outer envelope layers, while H$_{2}$CO and CS emission have an origin from the inner envelope regions. The smaller physical scales of the envelopes for the proto-BDs could result in lower beam-averaged molecular abundances than a low-mass protostellar system.


Figure~\ref{lbol}f compares the H$_{2}$CO ortho-para ratio for the proto-BDs with protostars. The mean ratio for the proto-BDs is 2.8, a factor of $\sim$2 higher than the mean ratio of 1.6 for the protostars. This indicates a predominantly gas-phase formation of H$_{2}$CO in the inner, dense regions close to the accretion zone in the proto-BDs, as compared to H$_{2}$CO formation on dust grains in the outer cold envelope for the protostars. The proto-BDs also show a flat distribution in the ortho-to-para ratio, with a similar range in values for L$_{bol} \sim$ 0.1-0.01 L$_{\sun}$. These are tentative trends a confirmation of which requires high-resolution interferometric observations.

\subsection{Trends vs evolutionary stage}
\label{evolution}


The molecular abundances for the proto-BDs show a wide spread, which suggests source to source variations in the chemical composition. This could be due to different depletion plus time evolution of the molecular species, resulting in some of the proto-BDs being more chemically evolved than the others, even if they are in the same physical stage. The large spread could also be due to the variations in physical structure and possible differences in the emitting regions of the species. The spread seen in the CS could be due to the effects of shocks in jets/outflows that can enhance the CS abundance (e.g., Bachiller et al. 1997). 

The strong dependence of CO depletion on the age and density of a core makes it a useful probe of the core history. The significant CO depletion ($\sim$90\%; Sect.~\ref{depletion}) in J182854 and J182844 suggests that these objects are pre-substellar cores or in the early Class 0 stage. Yet, J182854 does not show any CS emission. This is in contrast to results from several chemical models on time-dependent chemistry occurring in collapsing cores at different phases of dynamical and chemical evolution that find some species (e.g., CS, H$_{2}$CO, HCN) to peak in abundance before others (e.g., HCO$^{+}$, SO, NH$_{3}$) (e.g., Morata et al. 2003; Aikawa et al. 2005). On the other hand, line modelling indicates that CS traces the densest material towards the inner disk in these objects. The non-detection of CS could also be due to the distribution of mass in the envelope and disk as approximated from SED modelling. For instance, J182959 has the most massive disk ($\sim$13 M$_{\textrm{Jup}}$) in the sample and shows the strongest emission in these high density tracers, compared to J182844 for which most of the mass resides in the envelope. 

Among the proto-BDs, J163136 is the only object that does not show emission in CS and H$_{2}$CO, suggesting that it is either a more chemically evolved object so that the abundances are too low and below the detection limit, or a large fraction of H$_{2}$CO and CS could still be frozen. In comparison, J182952 is the only object among the evolved Stage I-T/II sources that shows HCO$^{+}$ and H$_{2}$CO emission, indicating the presence of a gas-rich circumstellar disk. The detection of H$_{2}$CO also indicates that there is a large reservoir of cold chemistry in these early stage proto-BDs that appears to survive in the gas phase as they evolve into Stage II. The least abundant C$^{17}$O isotopologue is detected in all of the proto-BDs as well as the more evolved Stage I-T/II objects (Fig.~A1), and can be an important tracer of the quiescent gas at all evolutionary stages. 



\section{Conclusions}

We present results from an investigation of CO, H$_{2}$CO, HCO$^{+}$, and CS chemistry in 7 proto-BDs identified in the Serpens and Ophiuchus regions. The hydrogen column density for the proto-BDs derived from the CO gas emission is $\sim$2-15 lower than that derived from the dust continuum emission, indicating CO depletion from the gas phase. All proto-BDs show emission in HCO$^{+}$, and 85\% of the objects also show emission in the ortho- and para-H$_{2}$CO and the CS lines. The H$_{2}$CO, HCO$^{+}$, and CS molecular abundances for the proto-BDs show a weak correlation with CO. A comparison of the molecular abundances for the proto-BDs with Class 0/I protostars shows a nearly constant CS and CO abundance for L$_{\textrm{bol}} \sim$0.01-40 L$_{\sun}$. There is a tentative trend of a decline in the HCO$^{+}$, ortho- and para-H$_{2}$CO molecular abundances with decreasing L$_{\textrm{bol}}$, which possibly reflects the scaled-down physical structures for the proto-BDs compared to protostars. The mean H$_{2}$CO ortho-para ratio derived for the proto-BDs is $\sim$3 and indicates a predominantly gas-phase formation for H$_{2}$CO in the proto-BDs. Among the more evolved Stage I-T/II brown dwarfs in the sample, one object shows HCO$^{+}$ and H$_{2}$CO emission, indicating the presence of a gas-rich circumstellar disk. None of the other Stage I-T/II objects show emission in any of the high-density tracers; however, C$^{17}$O is detected in all targets including the Stage I-T/II objects.

\section*{Acknowledgements}

B.R. acknowledges funding from the Deutsche Forschungsgemeinschaft (DFG) - Projekt number 402837297. This work is based on observations carried out under project number 139-15 and 120-17 with the IRAM 30m telescope. IRAM is supported by INSU/CNRS (France), MPG (Germany) and IGN (Spain). B.R. would like to thank the IRAM staff for help provided during the observations. 







\onecolumn

\appendix

\section{CO and isotopologues}
\label{12co-13co-c18o}

\begin{figure*}
  \centering              
     \includegraphics[width=1.8in]{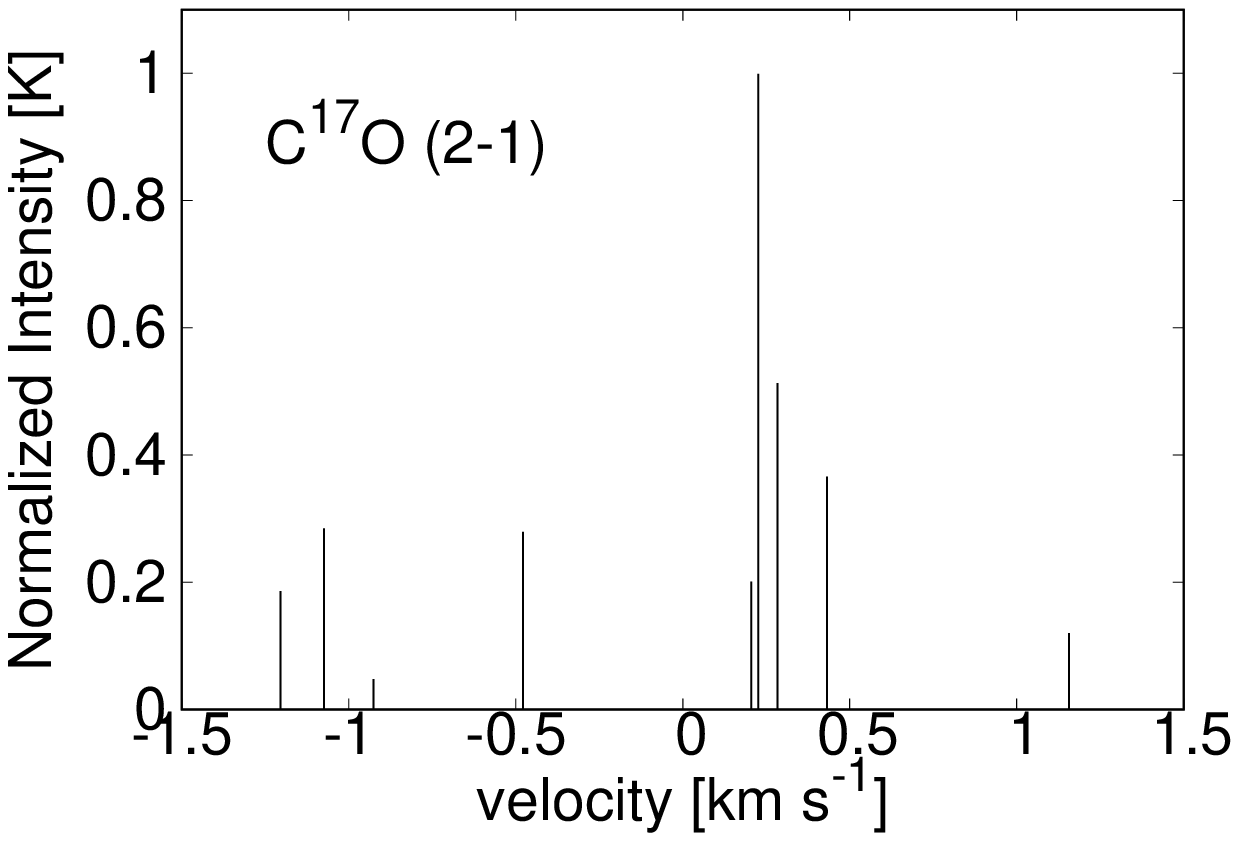}     \\
     \includegraphics[width=1.8in]{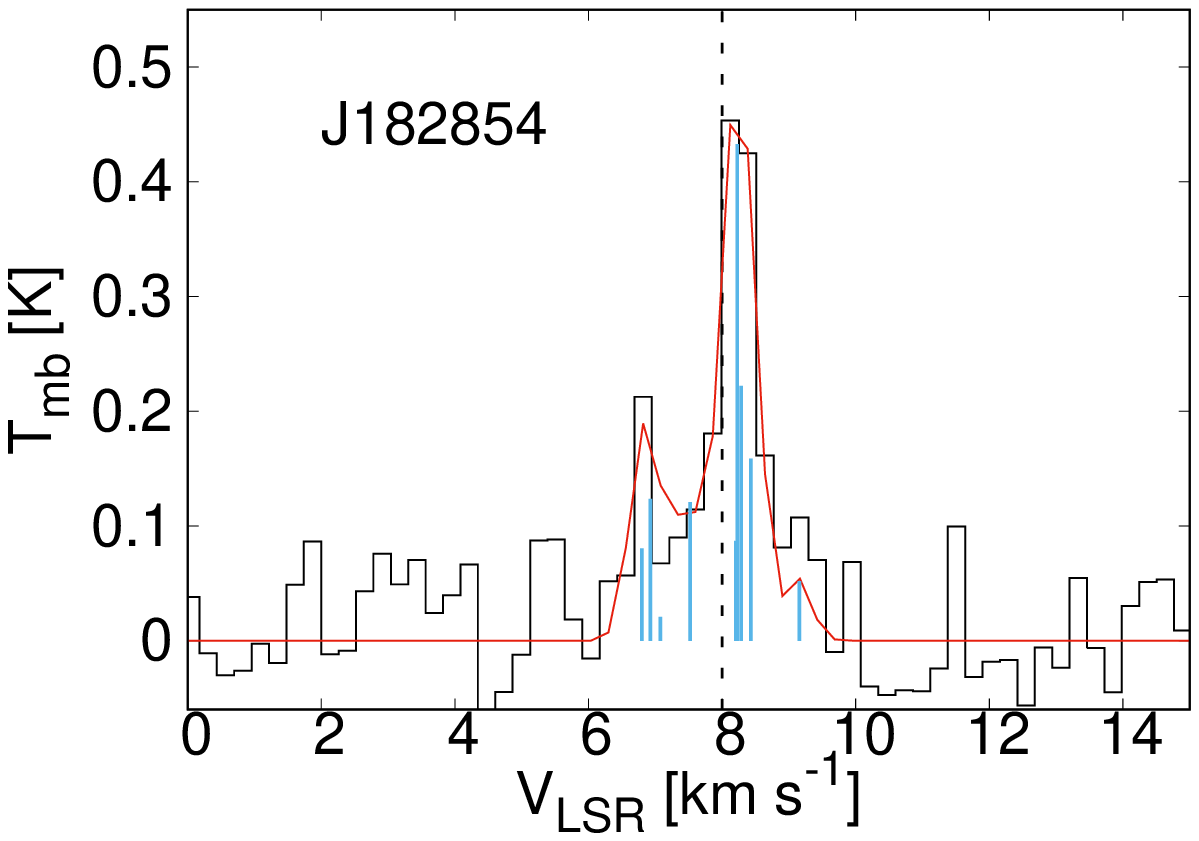}
     \includegraphics[width=1.8in]{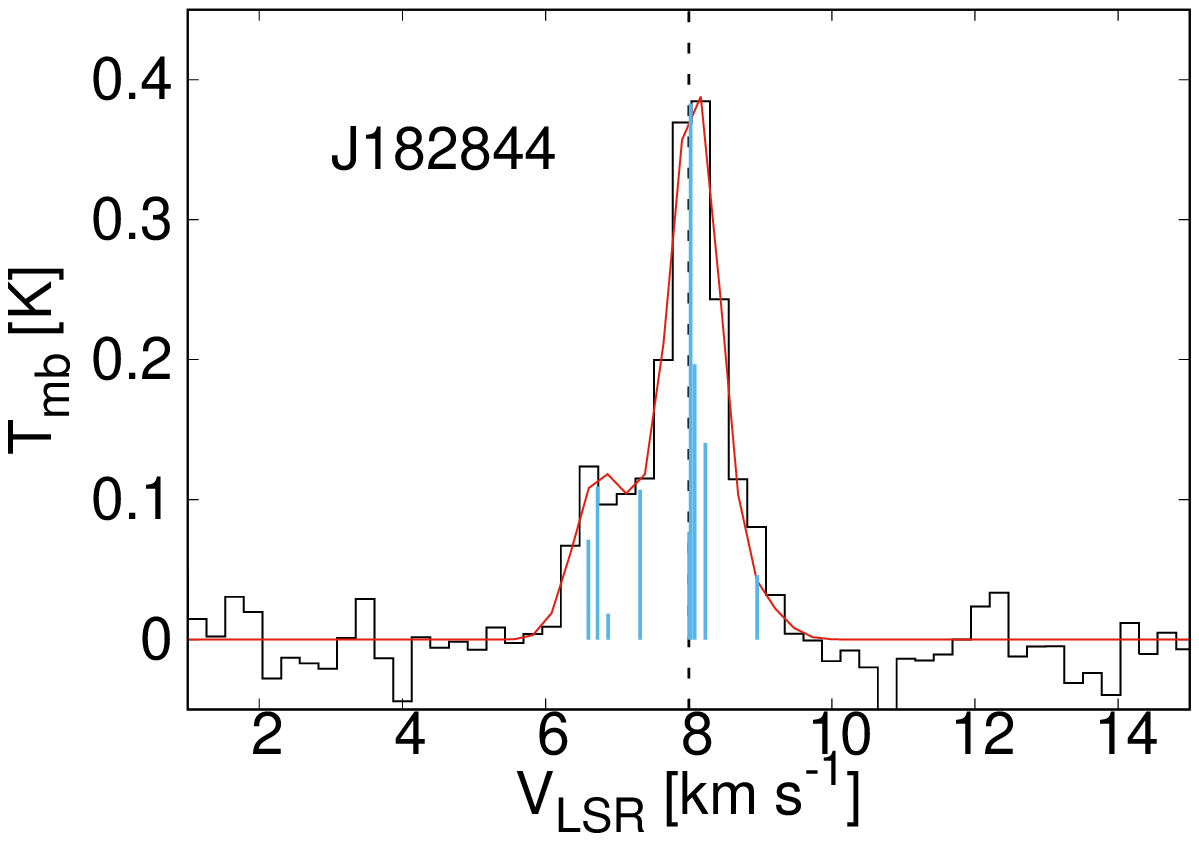}
     \includegraphics[width=1.8in]{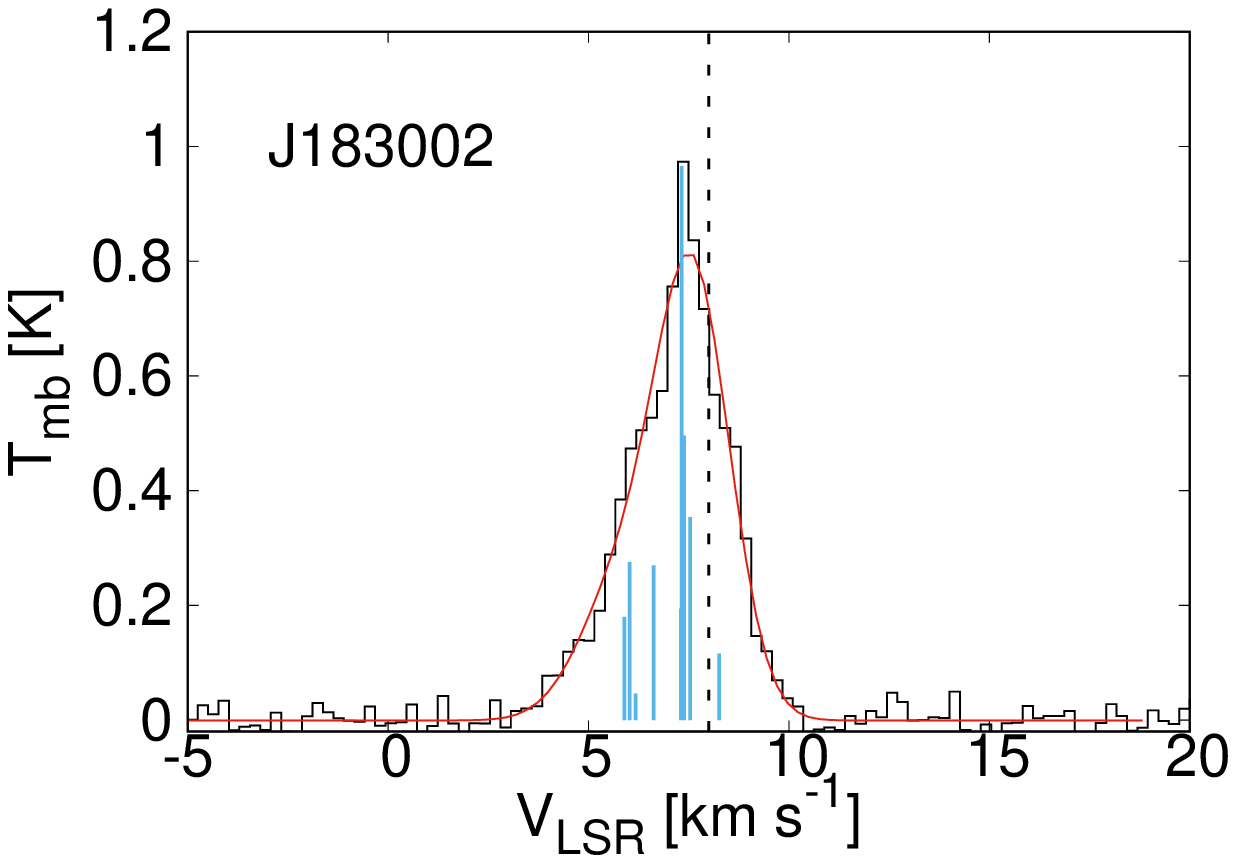}     
     \includegraphics[width=1.8in]{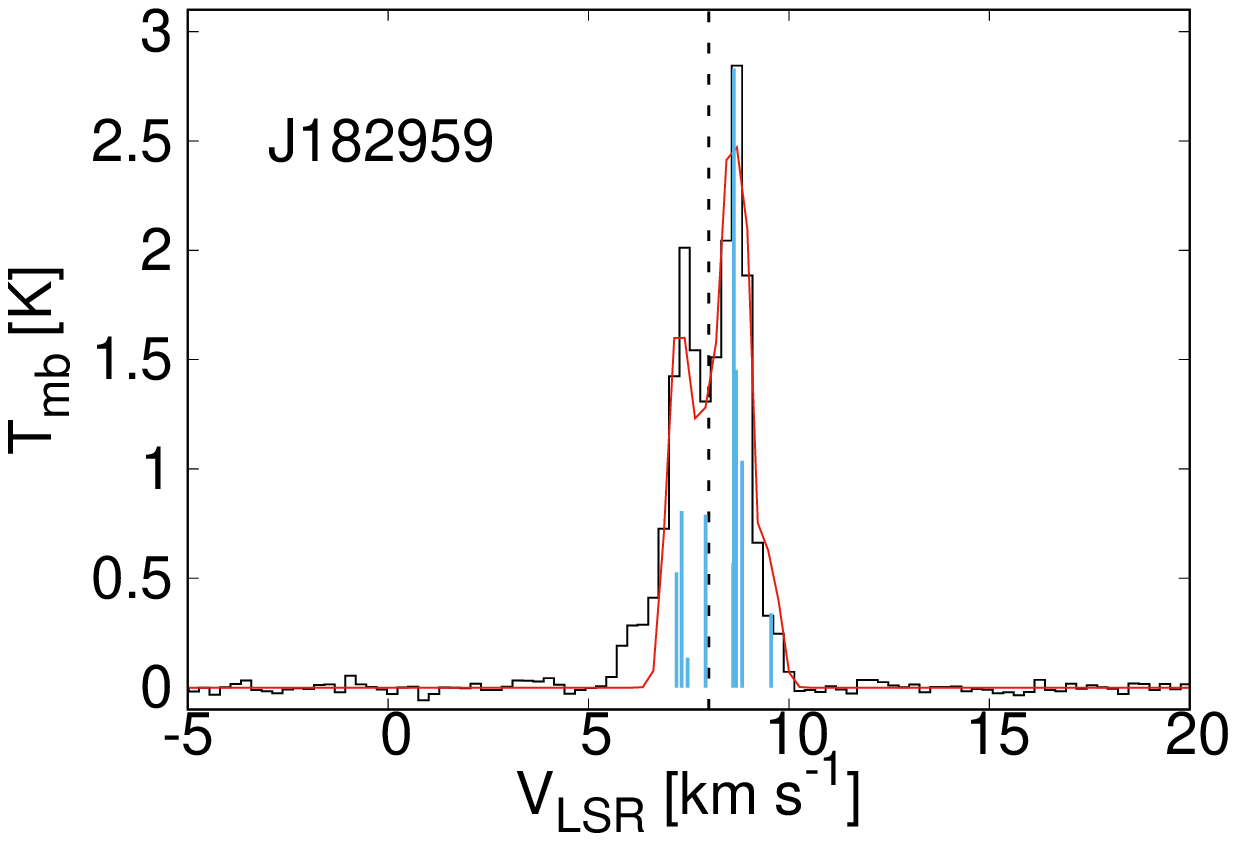}
     \includegraphics[width=1.8in]{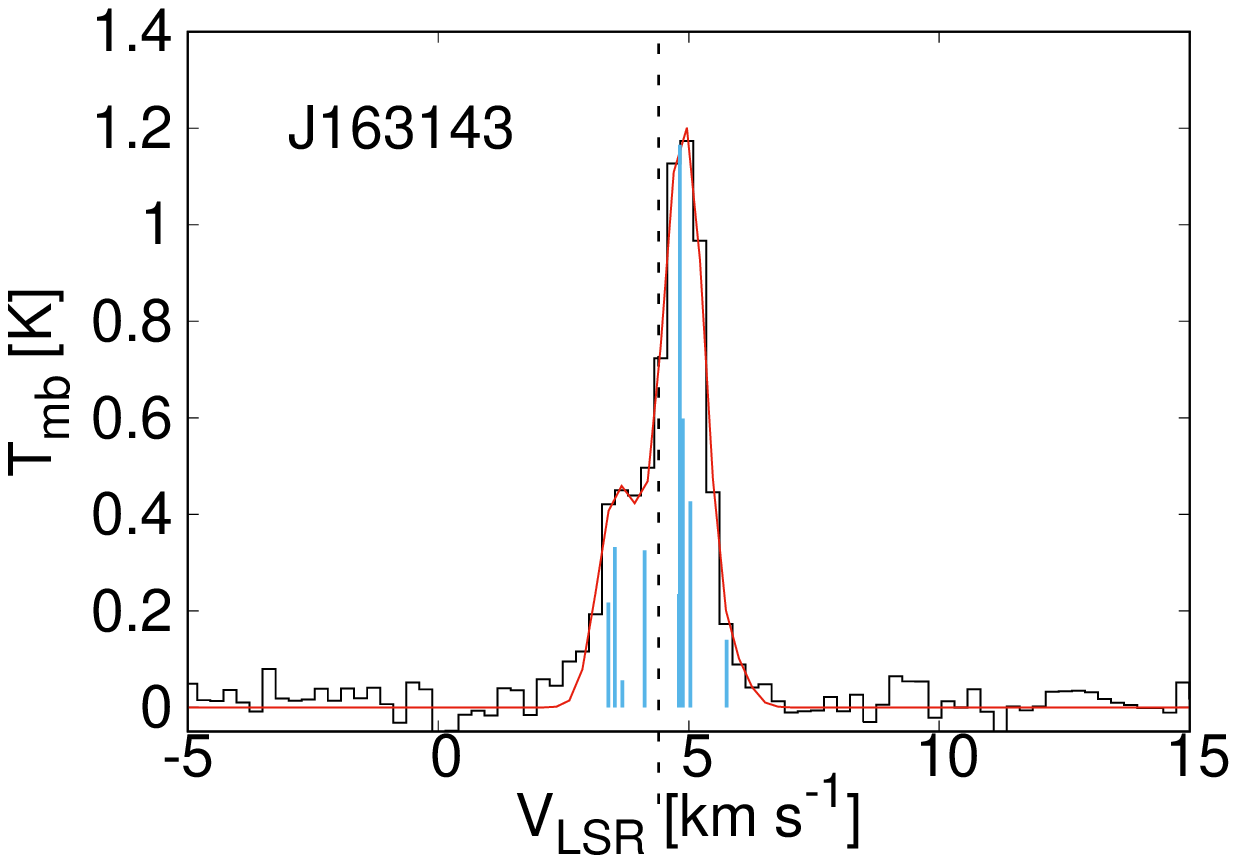}   
     \includegraphics[width=1.8in]{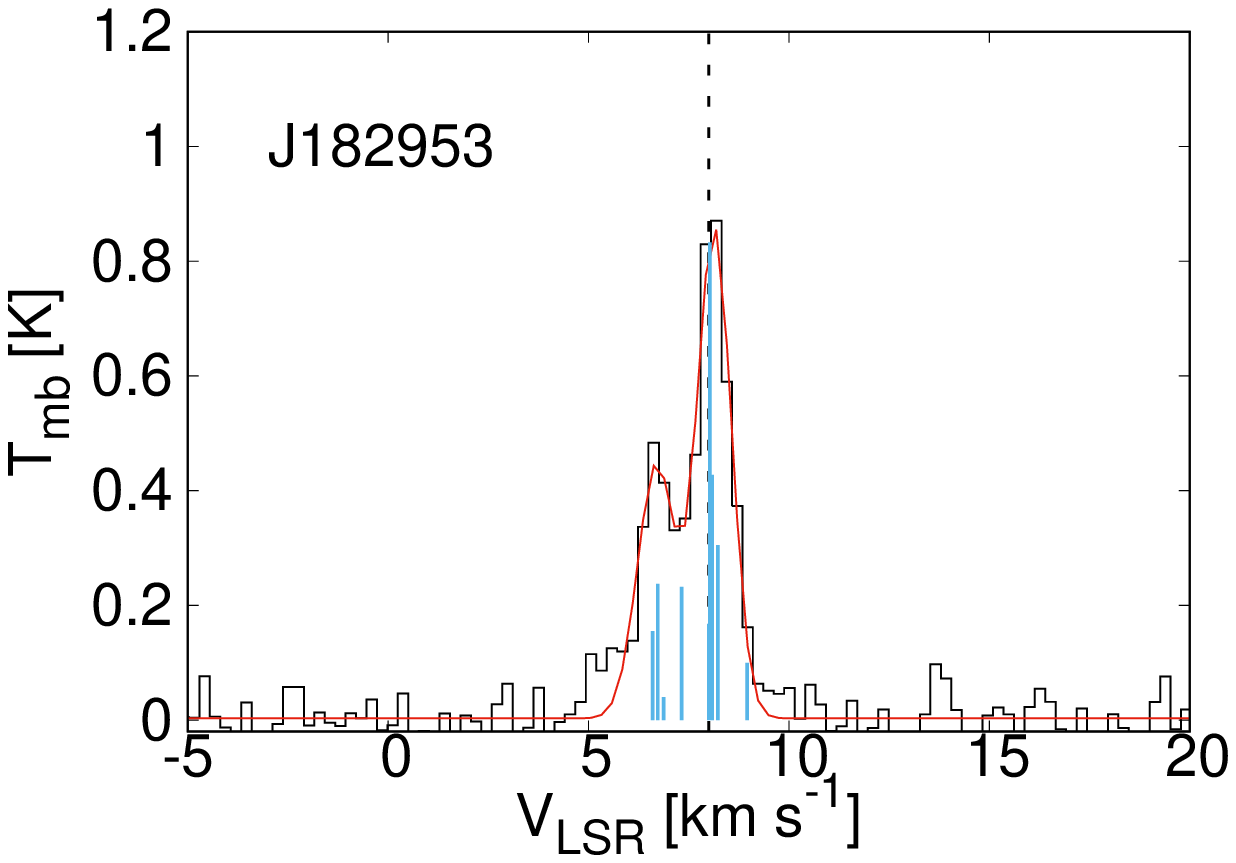}
     \includegraphics[width=1.8in]{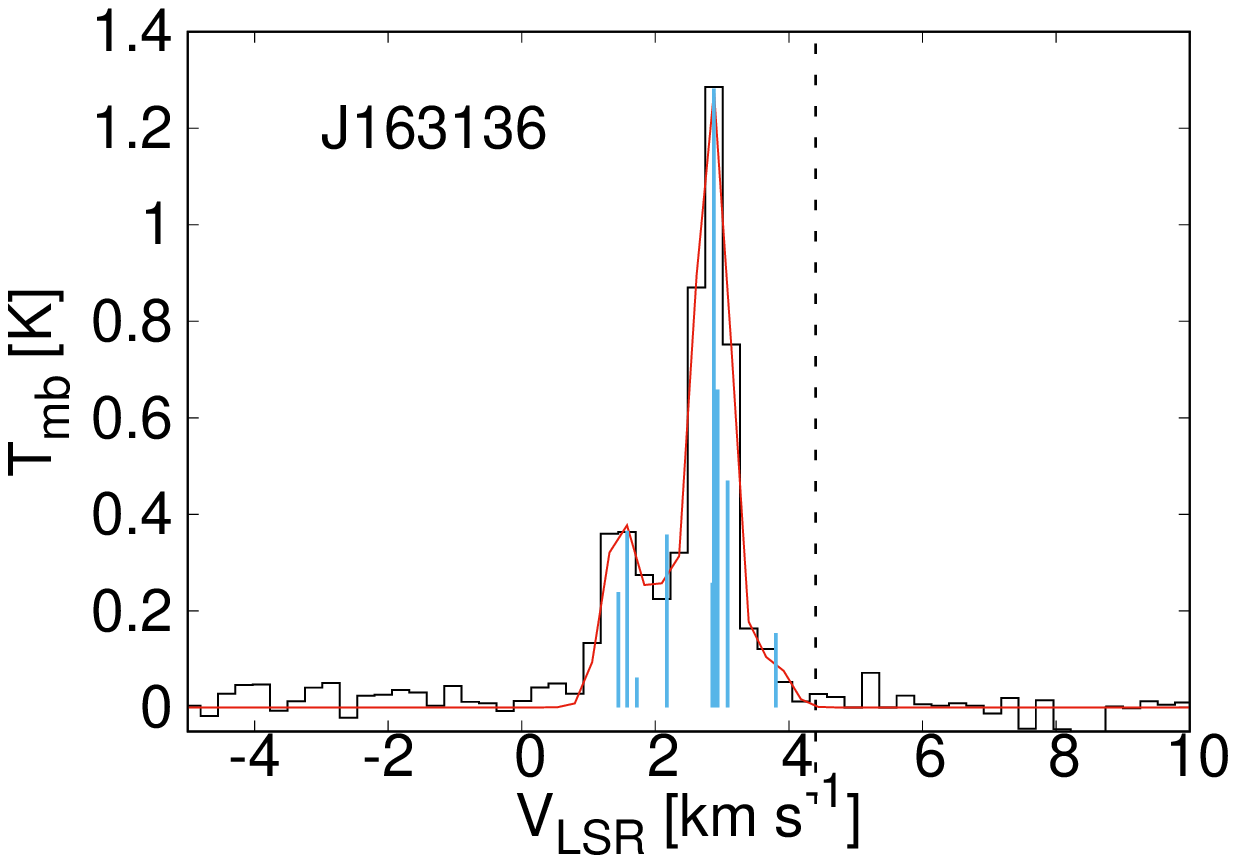}        
     \includegraphics[width=1.8in]{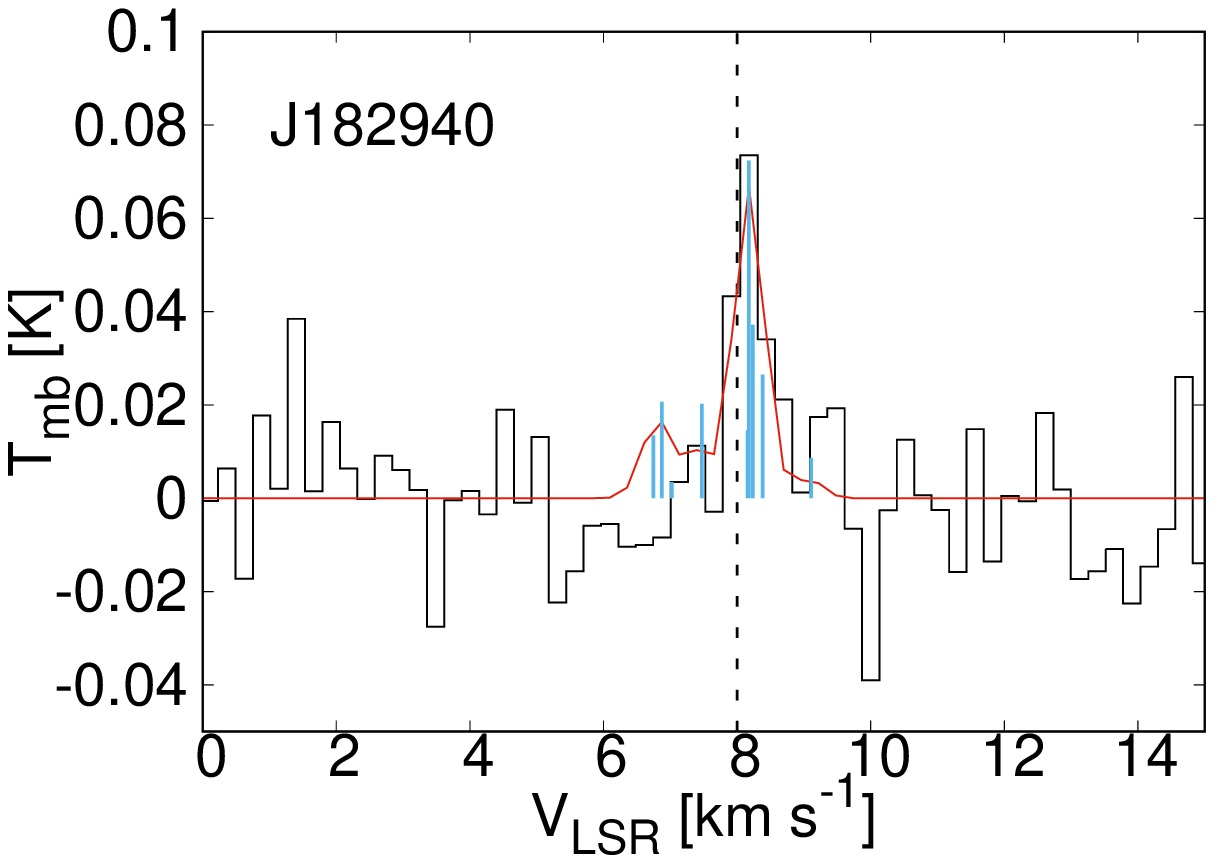}     
     \includegraphics[width=1.8in]{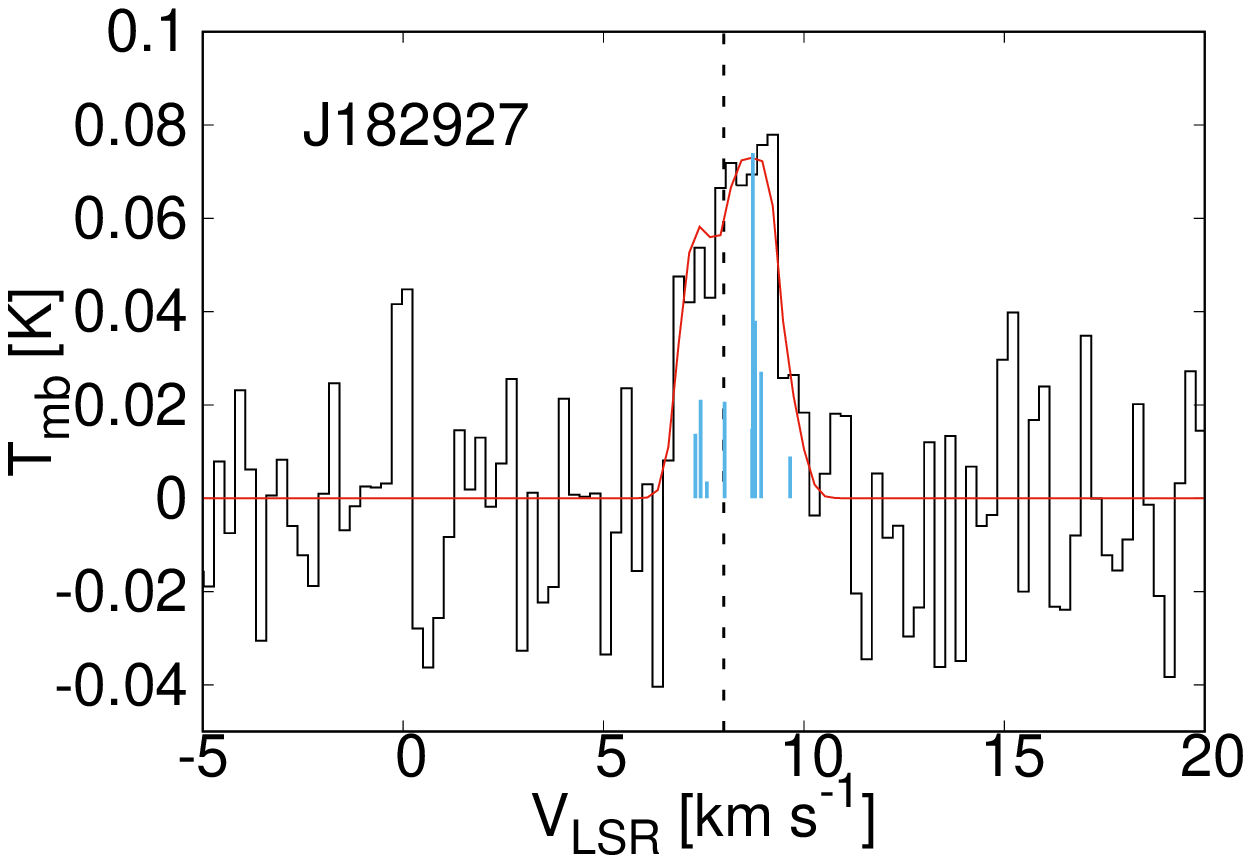}               
     \includegraphics[width=1.8in]{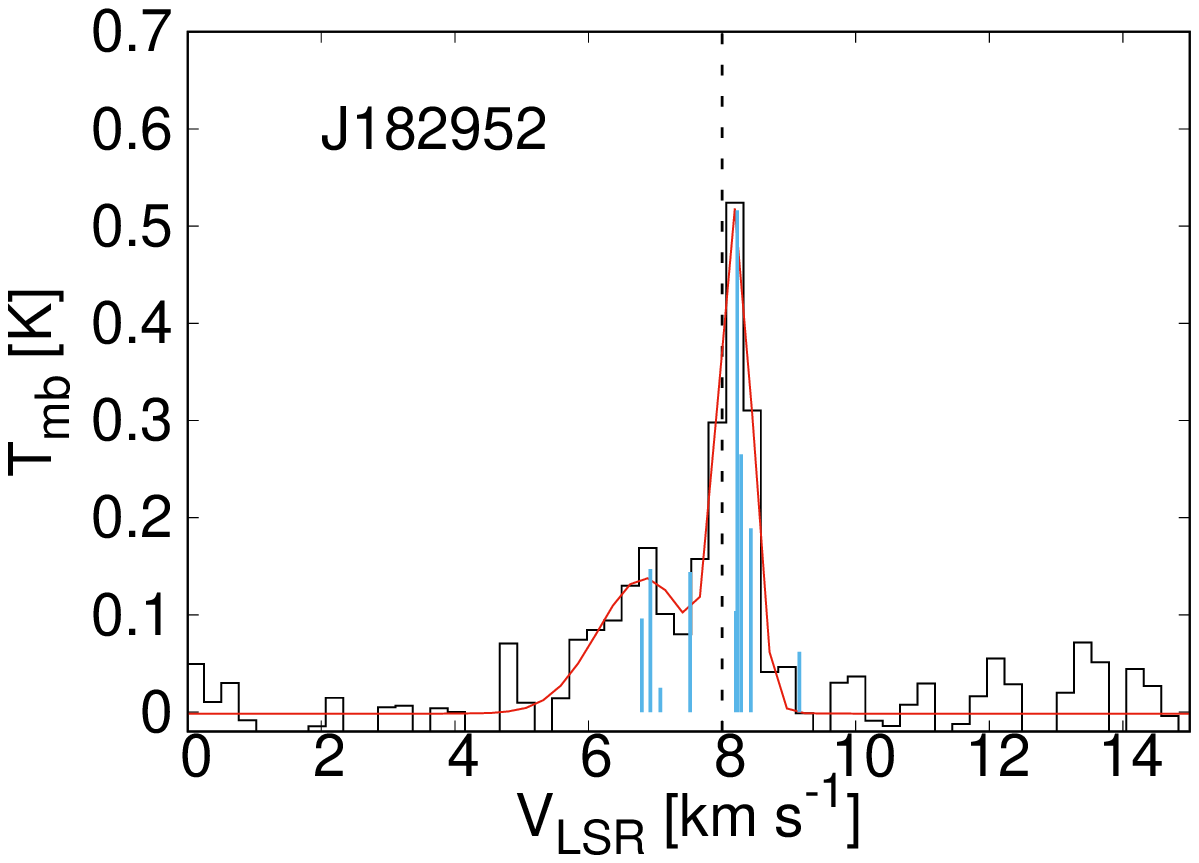}      
     \caption{The observed C$^{17}$O (2-1) spectra (black) with the Gaussian fits (red). Top panel shows the C$^{17}$O hyperfine structure expected under optically thin LTE conditions with the main hyperfine lines labelled. Blue lines in the observed spectra mark the hyperfine components shown in the top panel, and have been scaled to match the peak intensity in the observed spectrum. Black dashed line marks the cloud systemic velocity of $\sim$8 km/s and $\sim$4.4 km/s in Serpens and Ophiuchus, respectively. }
     \label{c17o-figs}
  \end{figure*}

\begin{figure*}
  \centering              
     \includegraphics[width=1.8in]{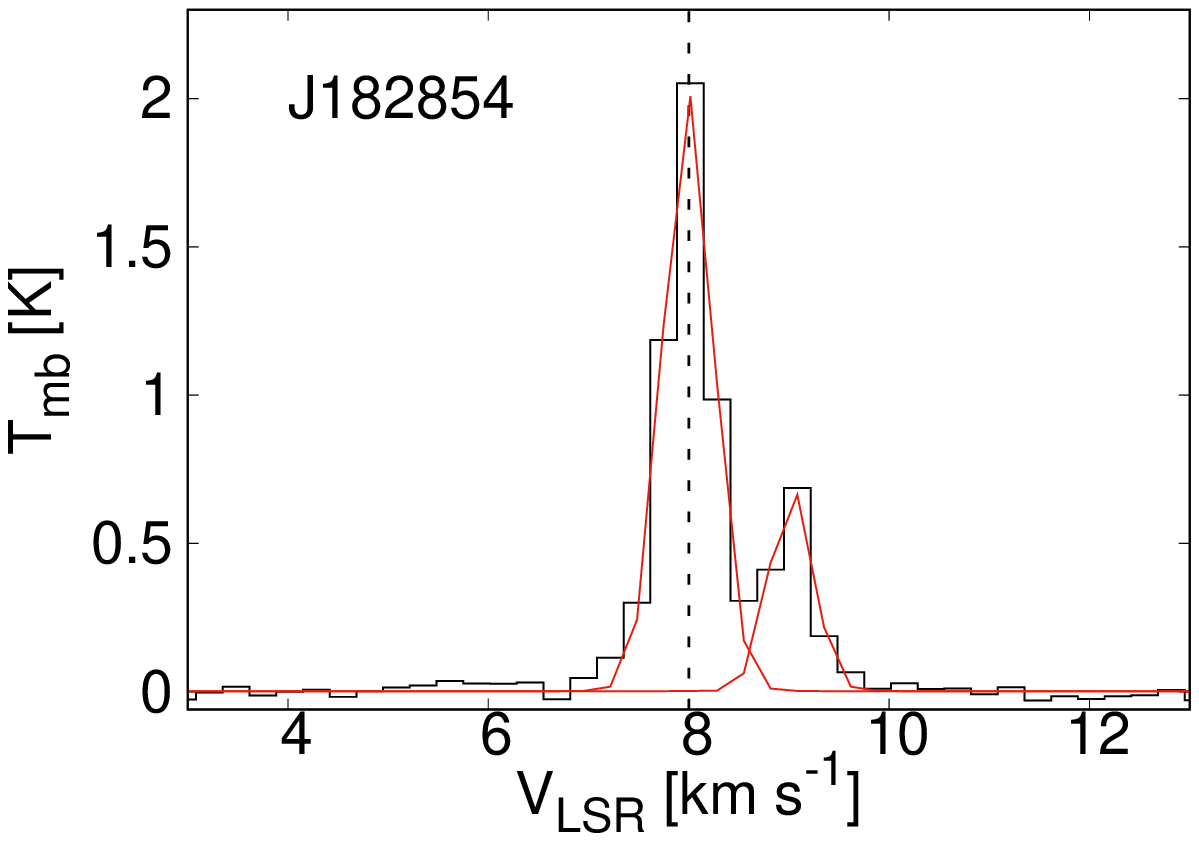}
     \includegraphics[width=1.8in]{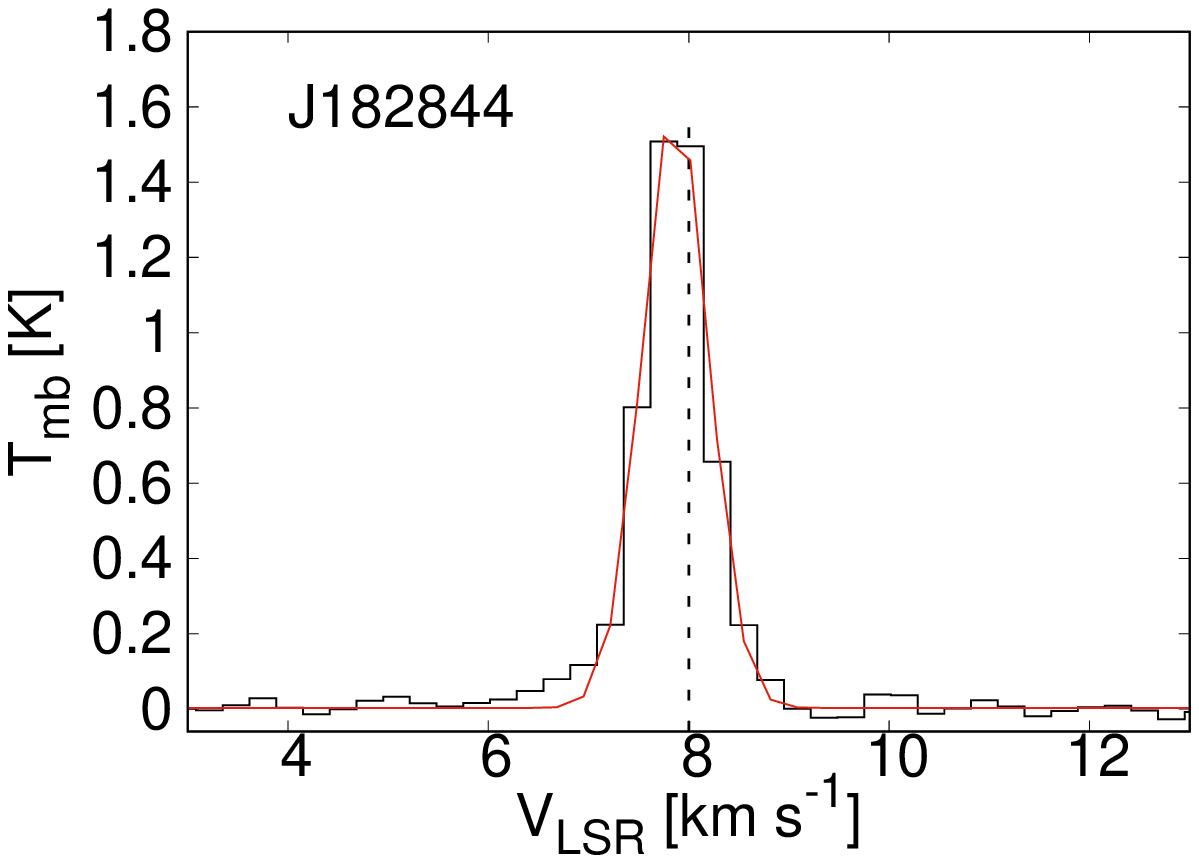}
     \includegraphics[width=1.8in]{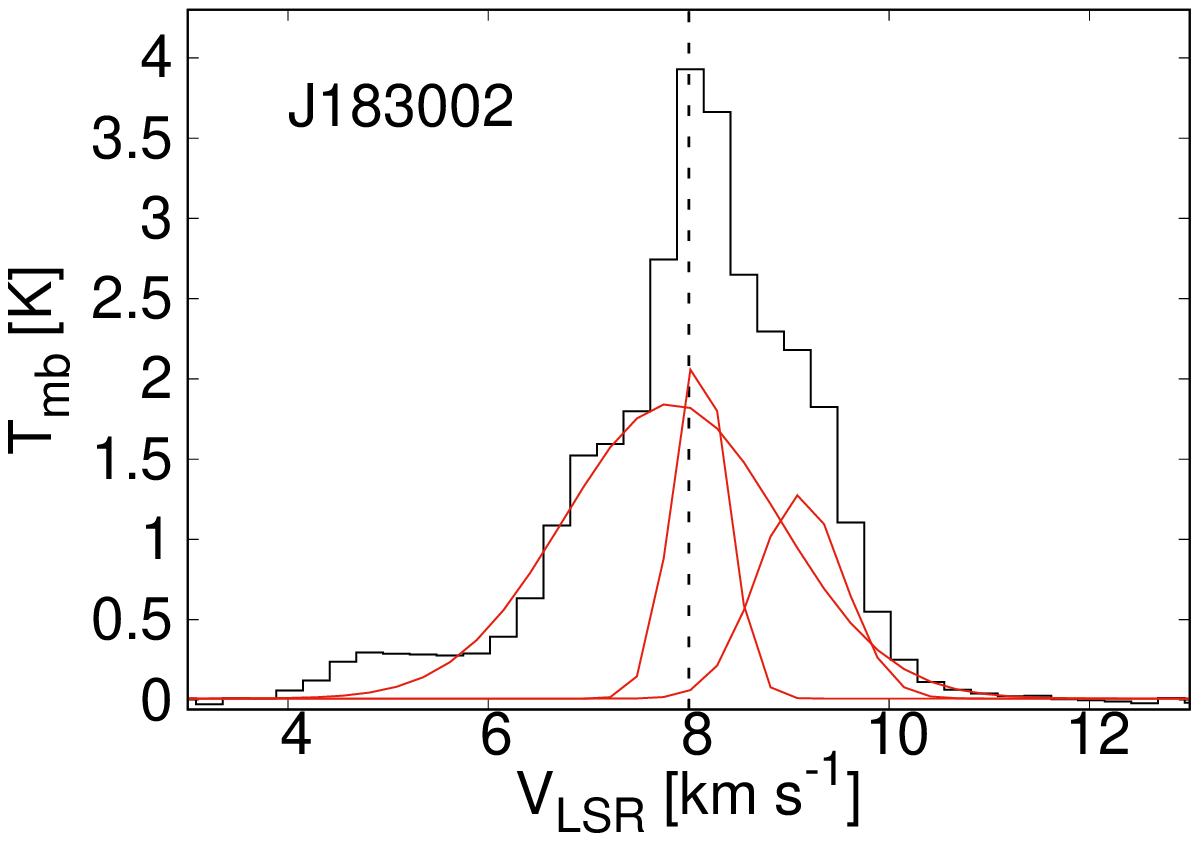}     
     \includegraphics[width=1.8in]{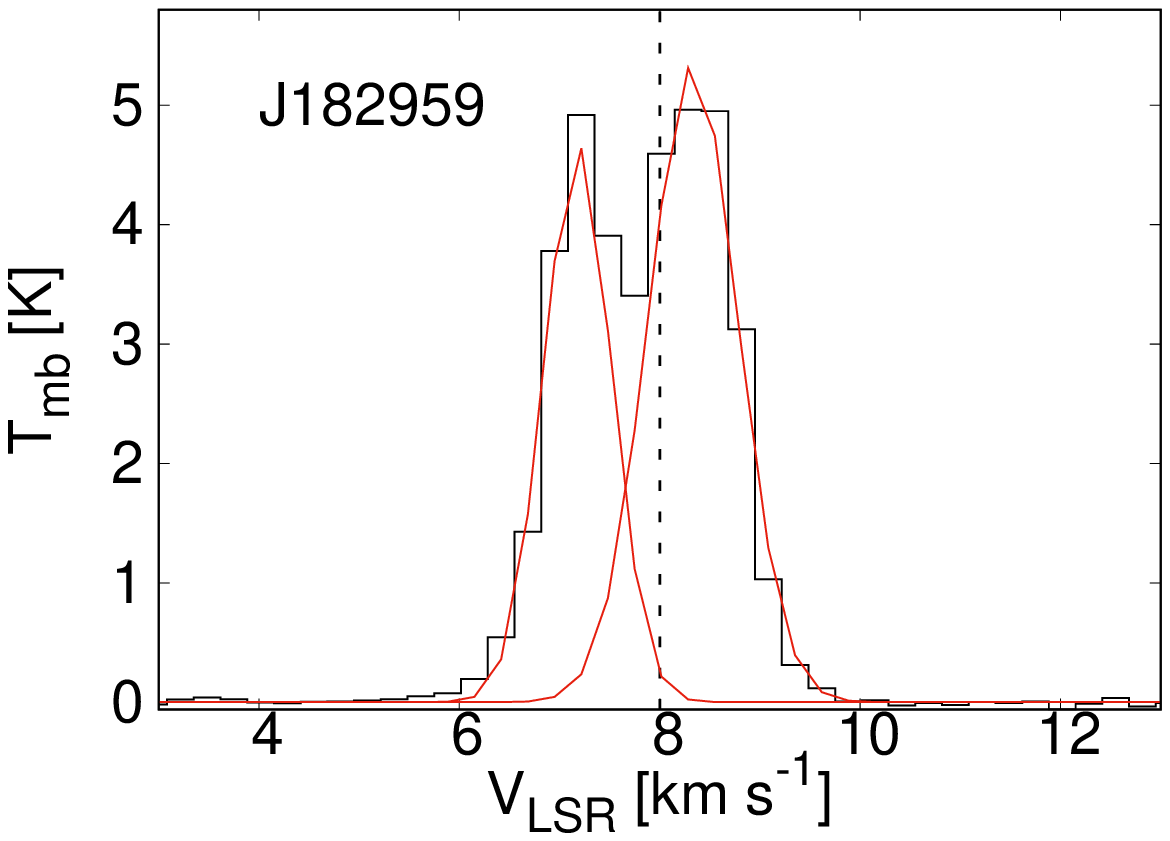}
     \includegraphics[width=1.8in]{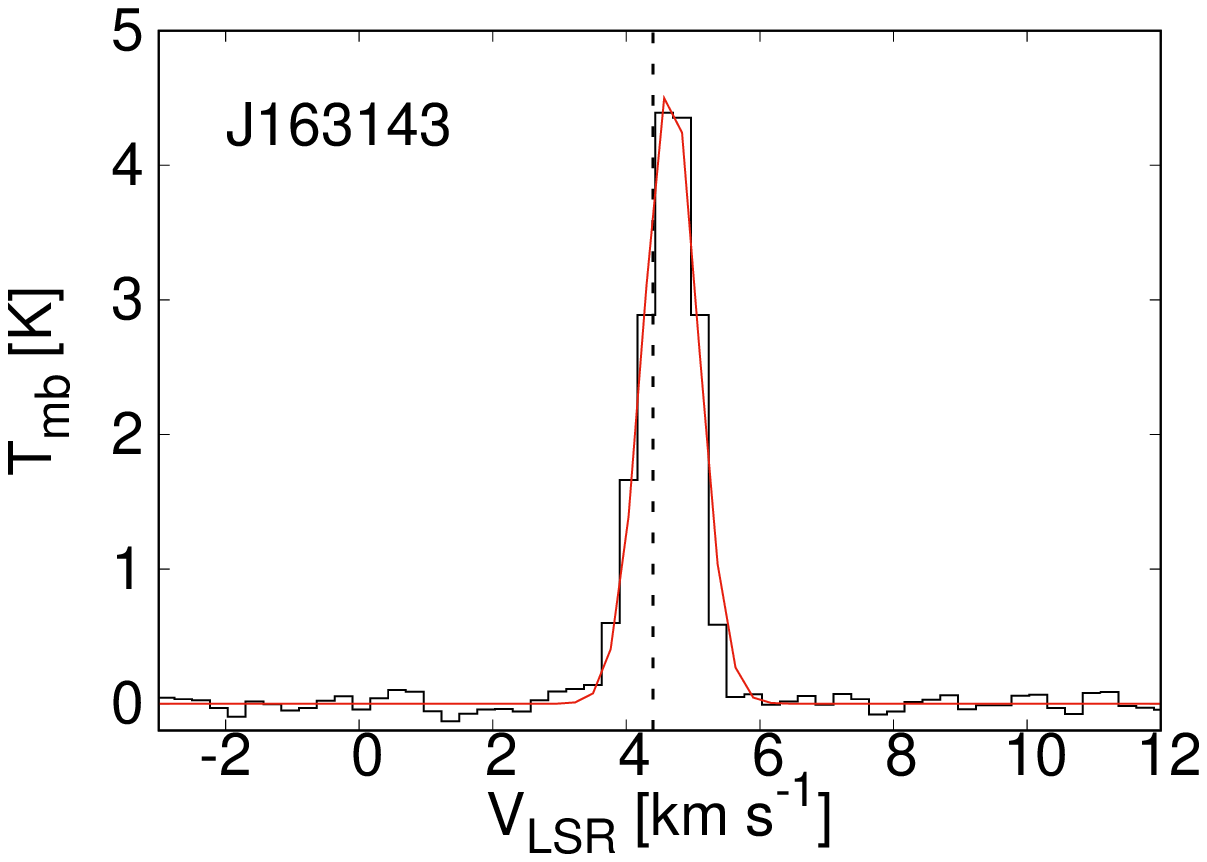}   
     \includegraphics[width=1.8in]{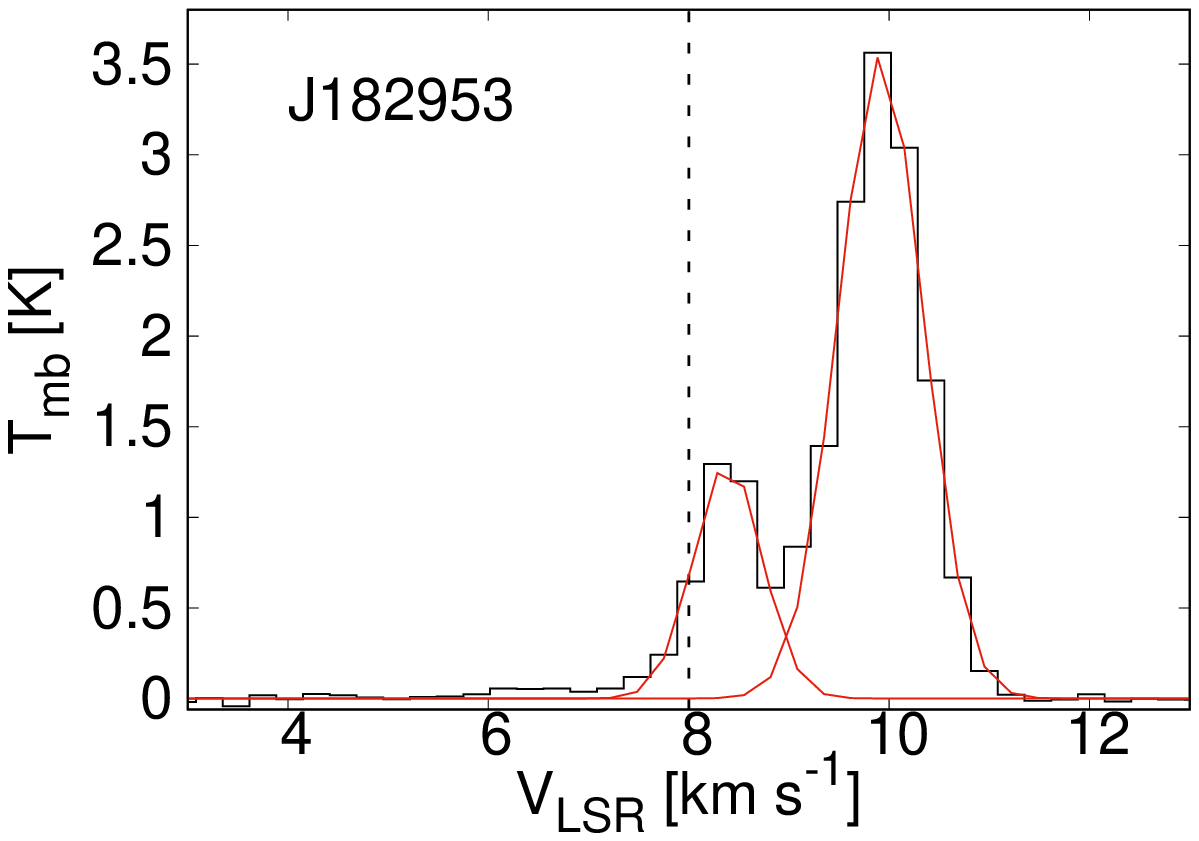}
     \includegraphics[width=1.8in]{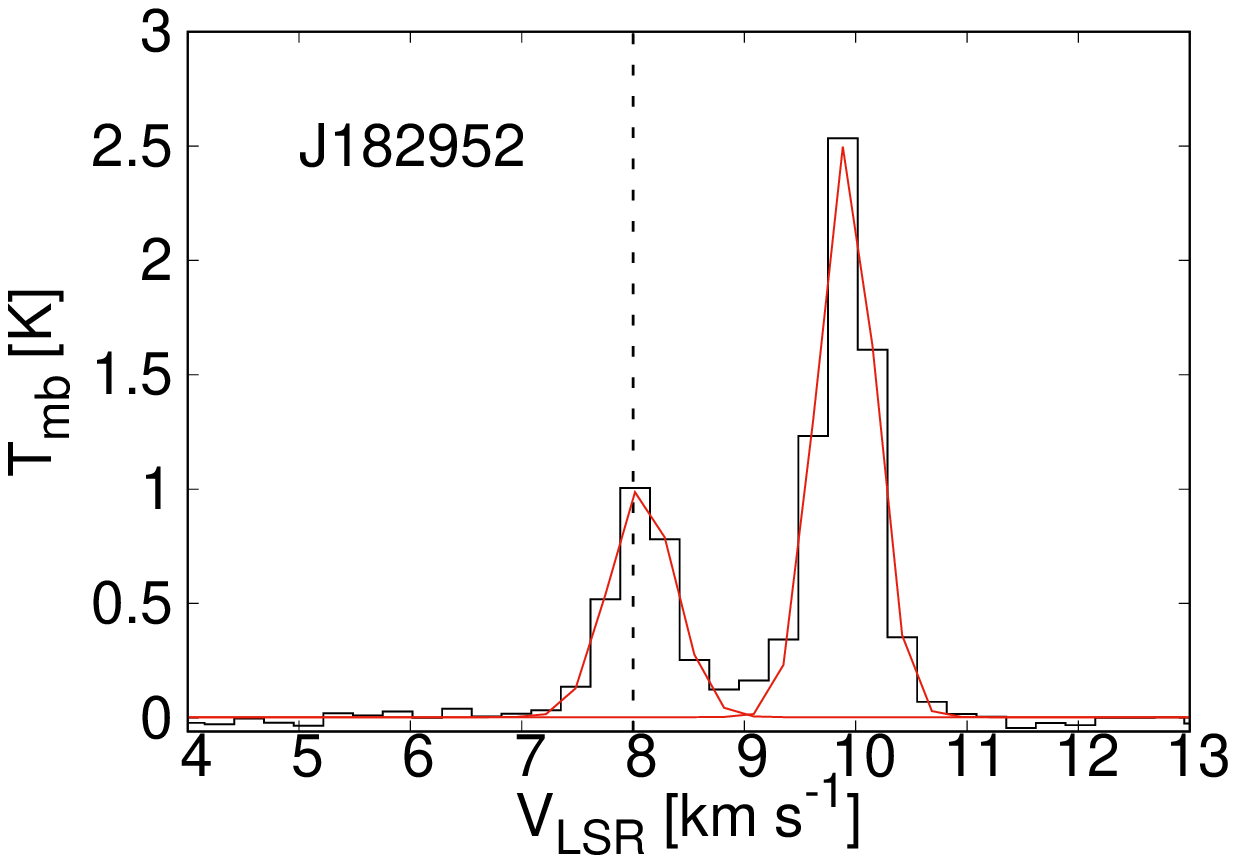}      
     \caption{The observed C$^{18}$O (2-1) spectra (black) with the Gaussian fits (red). Black dashed line marks the cloud systemic velocity of $\sim$8 km/s and $\sim$4.4 km/s in Serpens and Ophiuchus, respectively.  }     
     \label{c18o-figs}
  \end{figure*}

In the order of decreasing relative abundance, $^{12}$CO is the most abundant (99.76\%) isotope, followed by the $^{13}$CO (1.3\%), C$^{18}$O (0.18\%), and C$^{17}$O (0.05\%) isotopomers. C$^{17}$O (2-1) shows a hyperfine structure composed of 9 components that are spectrally unresolved (Fig.~\ref{c17o-figs}). A detailed discussion on the C$^{17}$O spectra is presented in Sect.~3.1.1. C$^{18}$O has an intermediate optical depth and shows more reasonable line profiles compared to $^{12}$CO and $^{13}$CO (Fig.~\ref{c18o-figs}). The proto-BDs J182844 and J163143 show single peaked profiles with a maximum temperature at $\sim$1.5-4 K, which is much weaker than the high peak intensities seen in the $^{12}$CO and $^{13}$CO lines. The remaining objects show double-peaked profiles, except the broad spectrum for J183002 that can be fit with three single Gaussians (Fig.~\ref{c18o-figs}). J182959 shows self-absorption at the line center which is likely caused by the low excitation temperature layer int he outer envelope surrounding the proto-BD core. Also notable is a blue-shifted extended wing in the J182854 and J182844 profiles, suggesting that C$^{18}$O is sufficiently abundant that it can also marginally trace the outflow in the line wings. 

\begin{figure*}
  \centering              
     \includegraphics[width=1.8in]{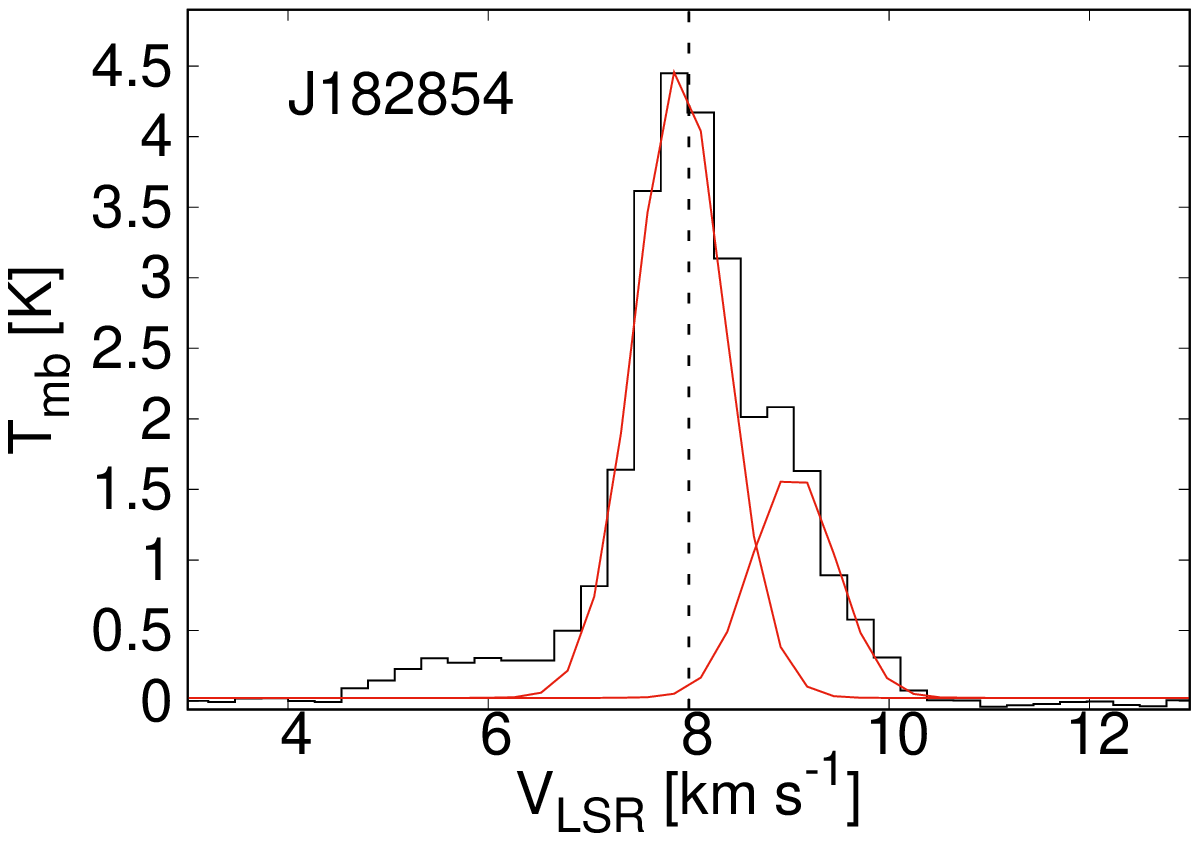}
     \includegraphics[width=1.8in]{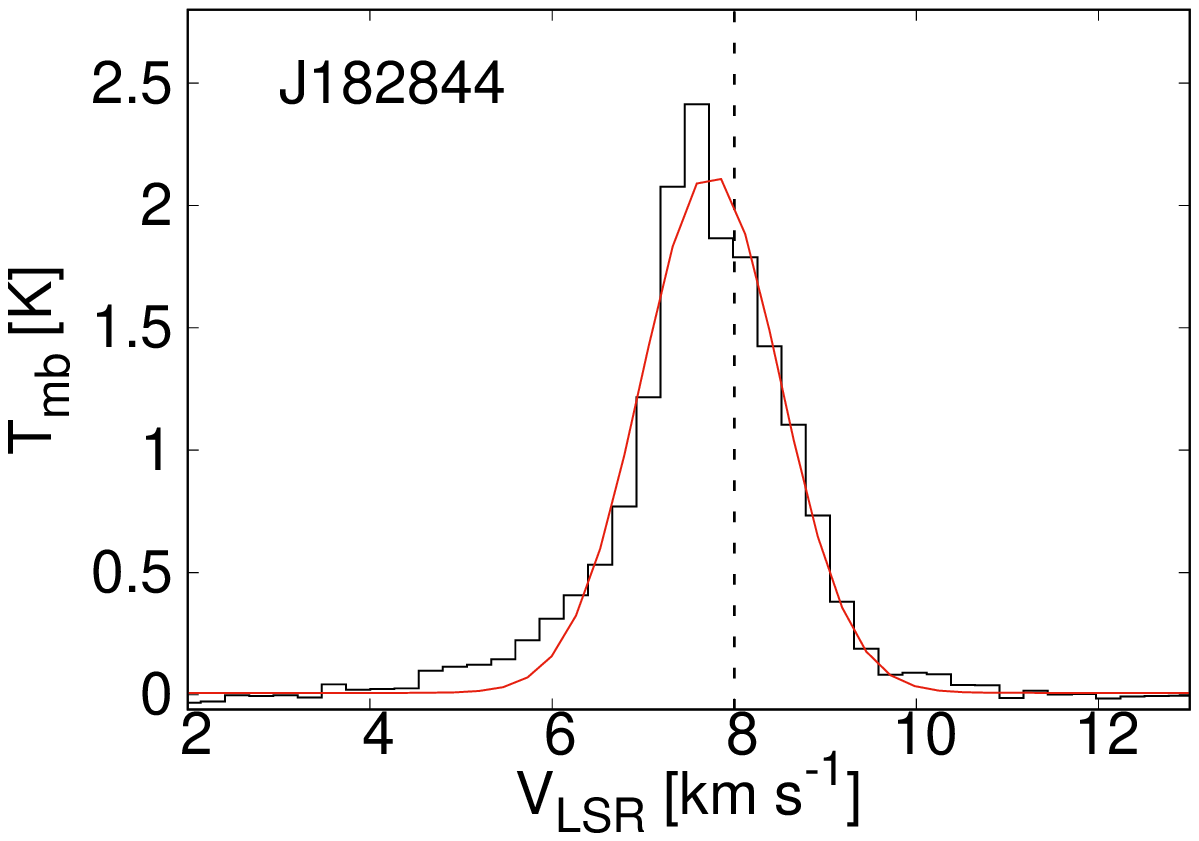}
     \includegraphics[width=1.8in]{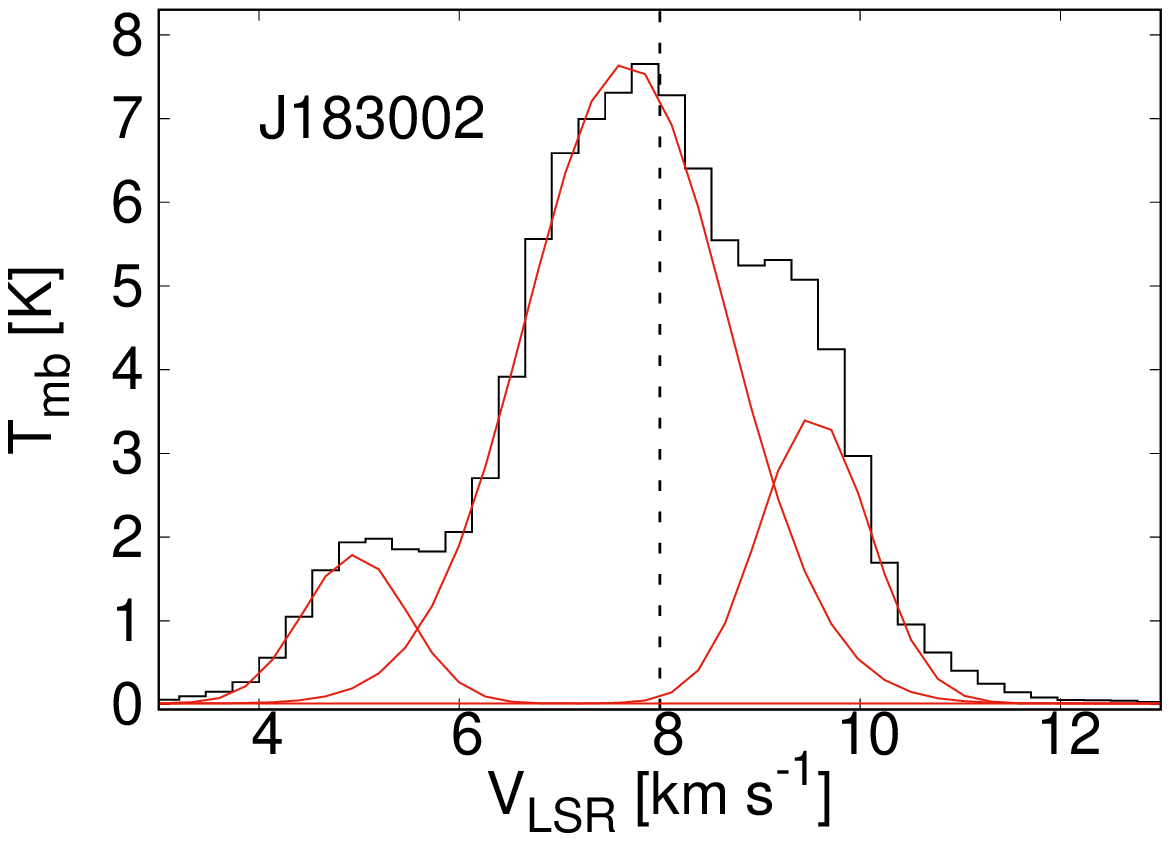}     
     \includegraphics[width=1.8in]{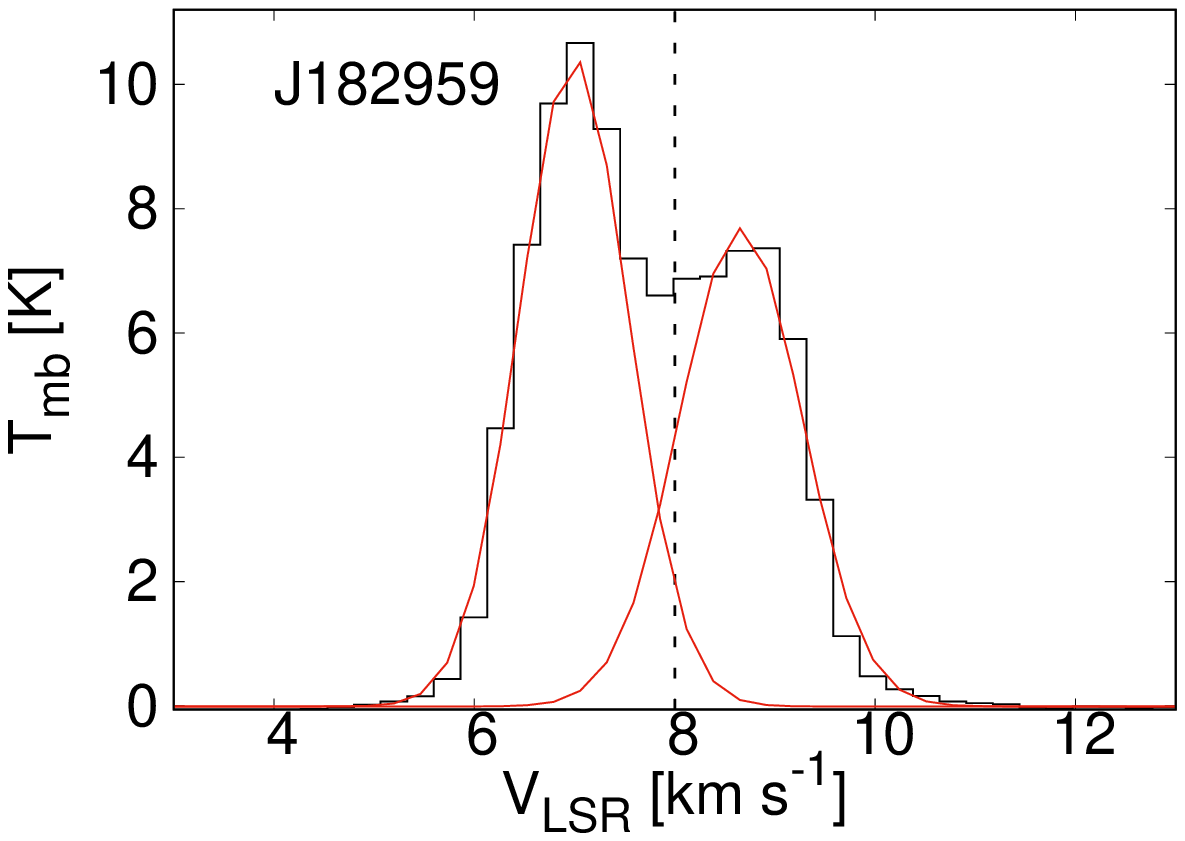}
     \includegraphics[width=1.8in]{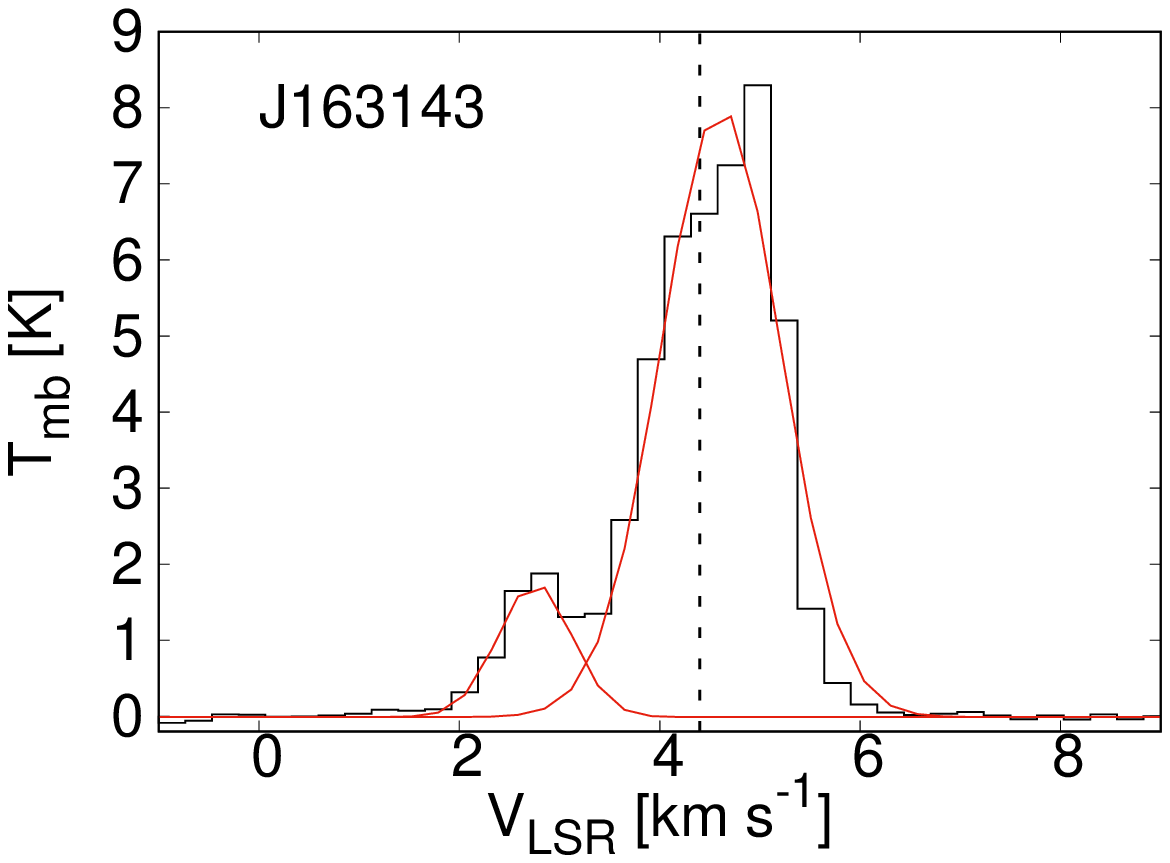}   
     \includegraphics[width=1.8in]{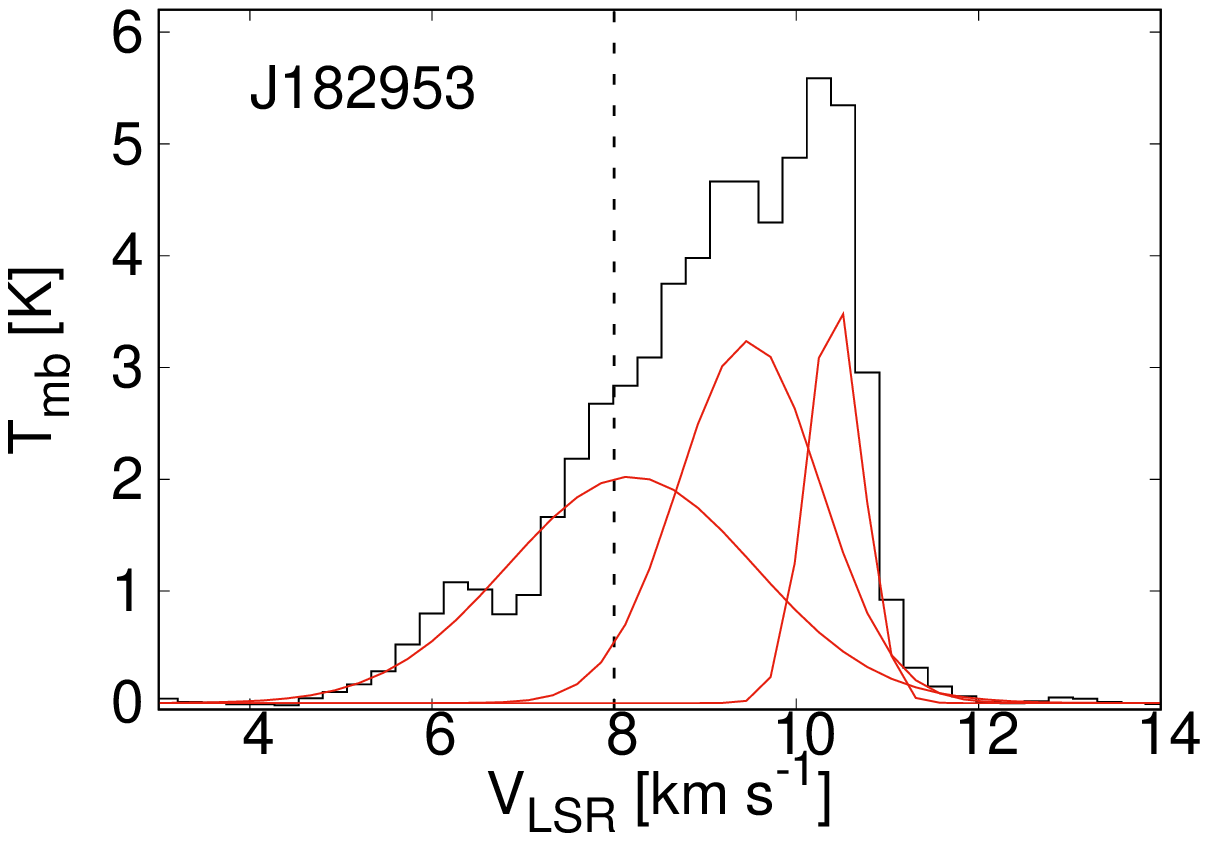}
     \includegraphics[width=1.8in]{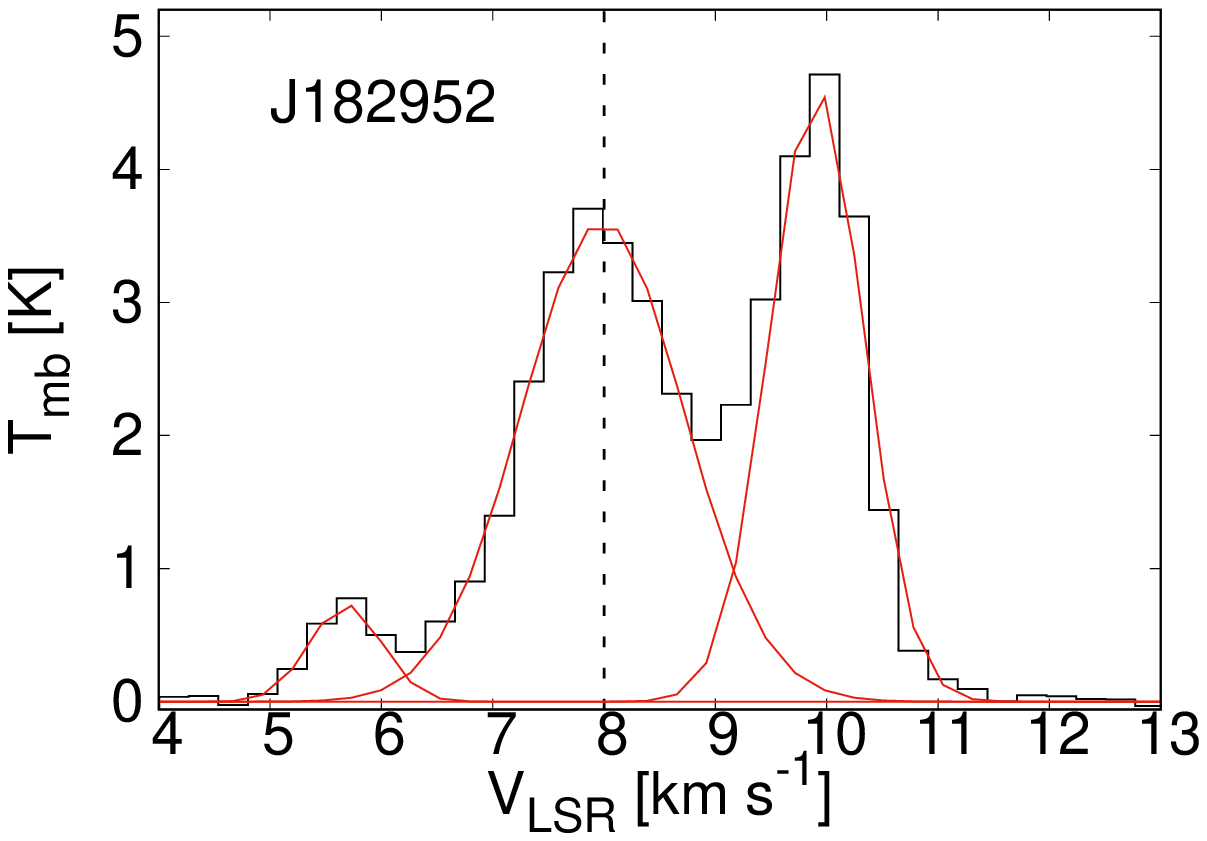}      
     \caption{The observed $^{13}$CO (2-1) spectra (black) with the Gaussian fits (red). Black dashed line marks the cloud systemic velocity of $\sim$8 km/s and $\sim$4.4 km/s in Serpens and Ophiuchus, respectively.  }
     \label{13co-figs}
  \end{figure*}

\begin{figure*}
  \centering                   
     \includegraphics[width=1.8in]{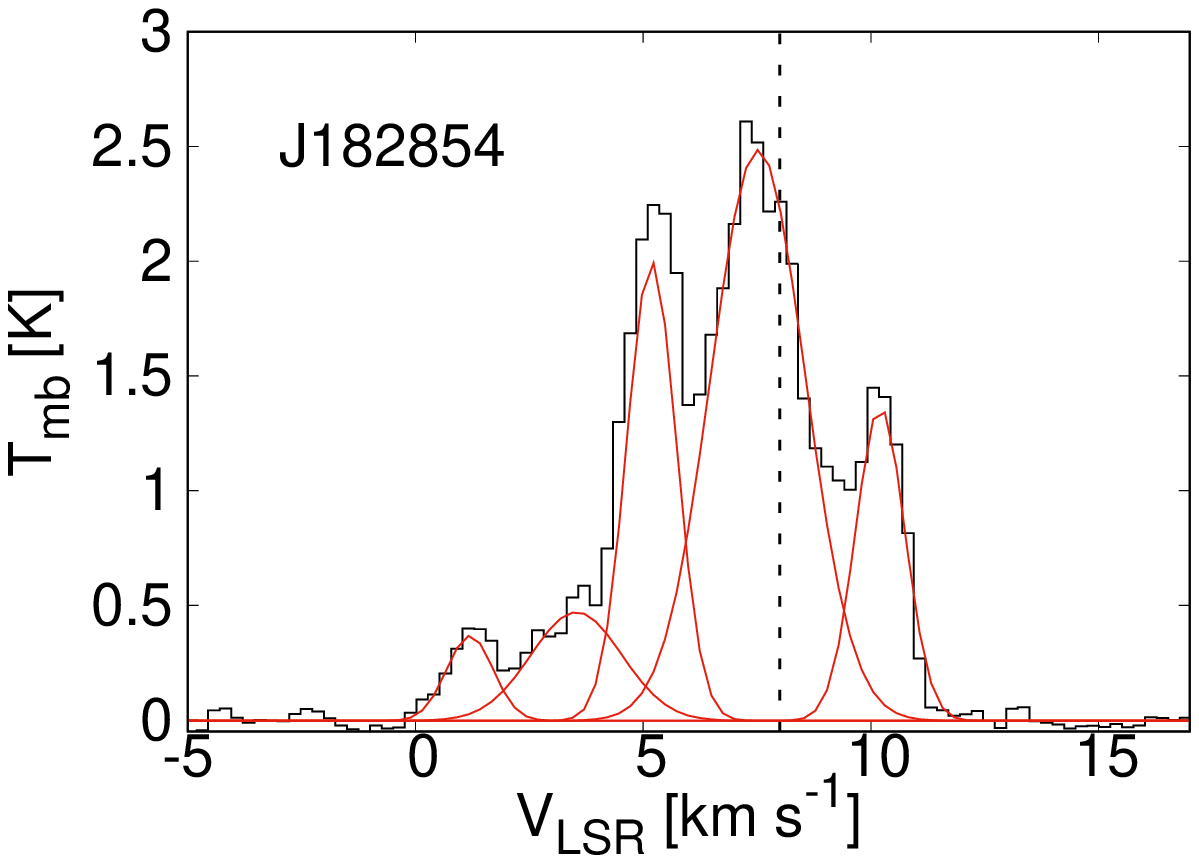}
     \includegraphics[width=1.8in]{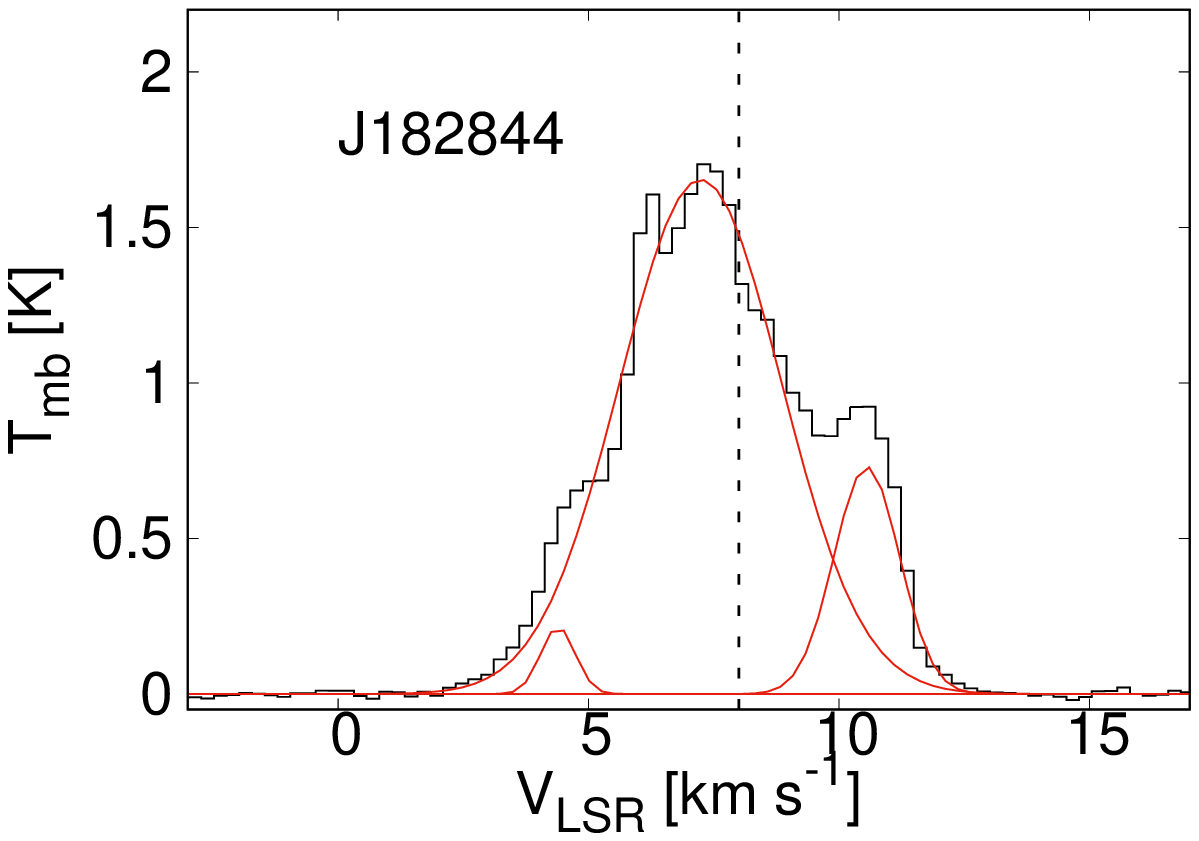}    
     \includegraphics[width=1.8in]{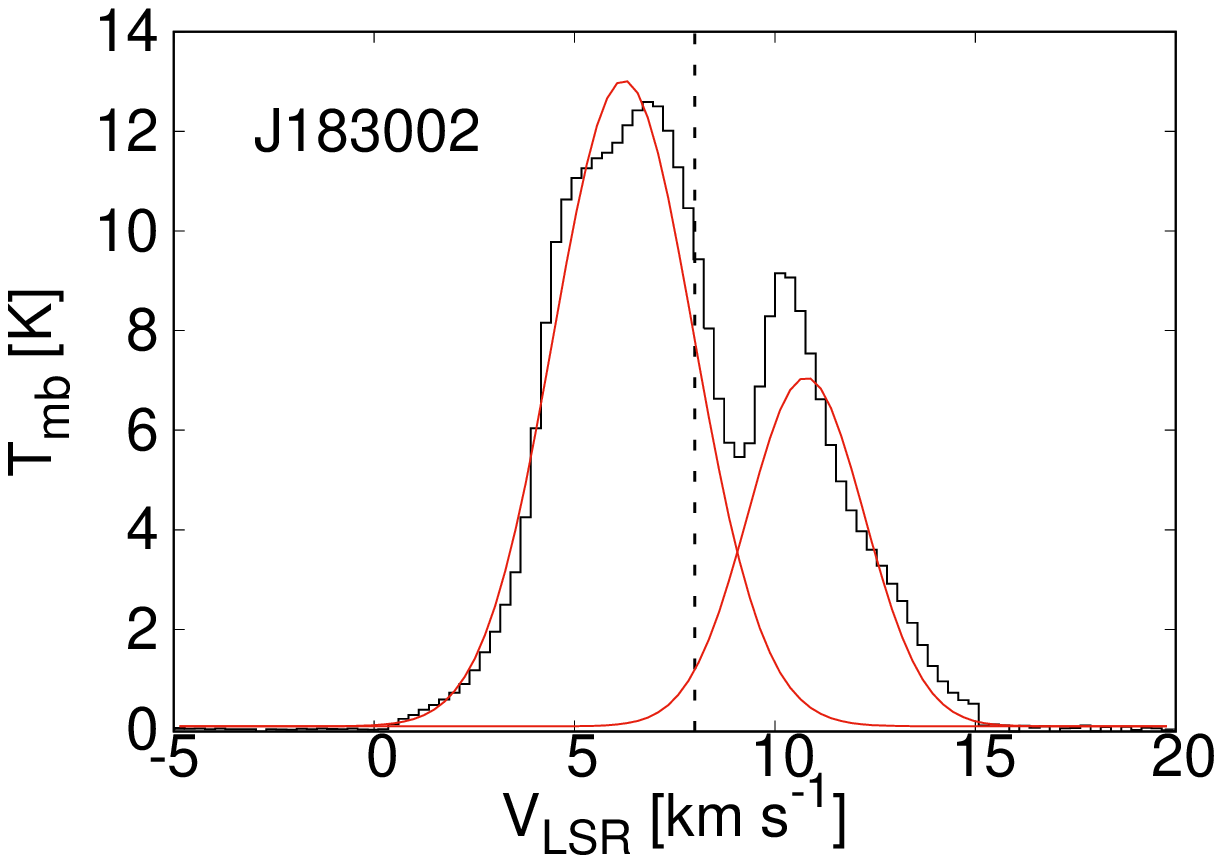}              
     \includegraphics[width=1.8in]{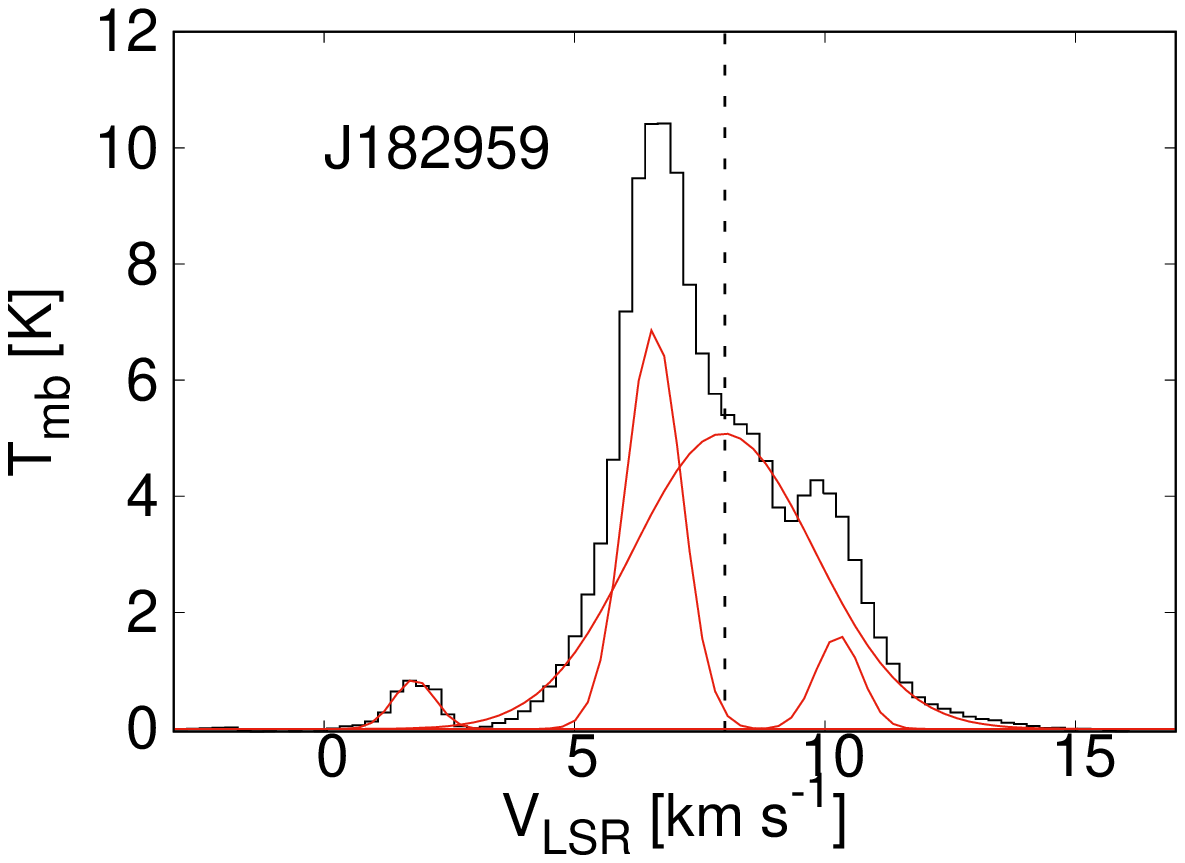}
     \includegraphics[width=1.8in]{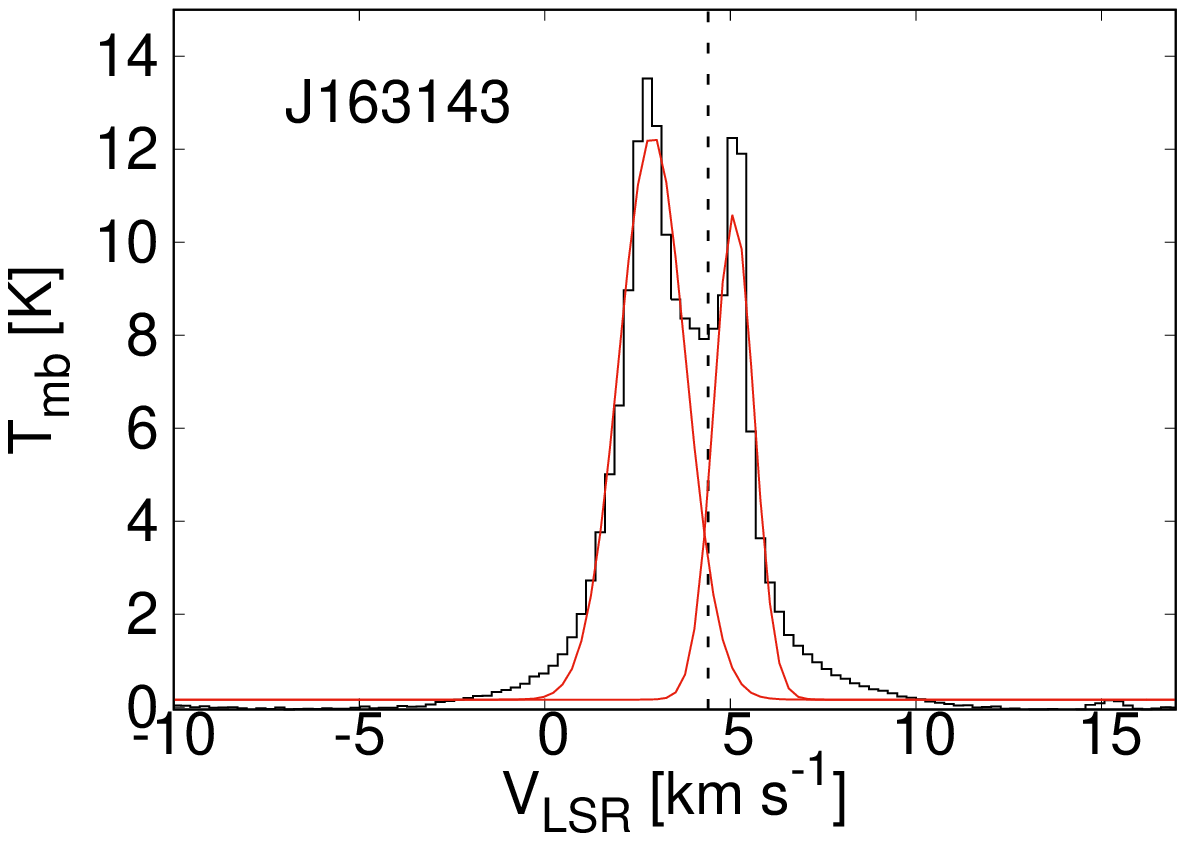}   
     \includegraphics[width=1.8in]{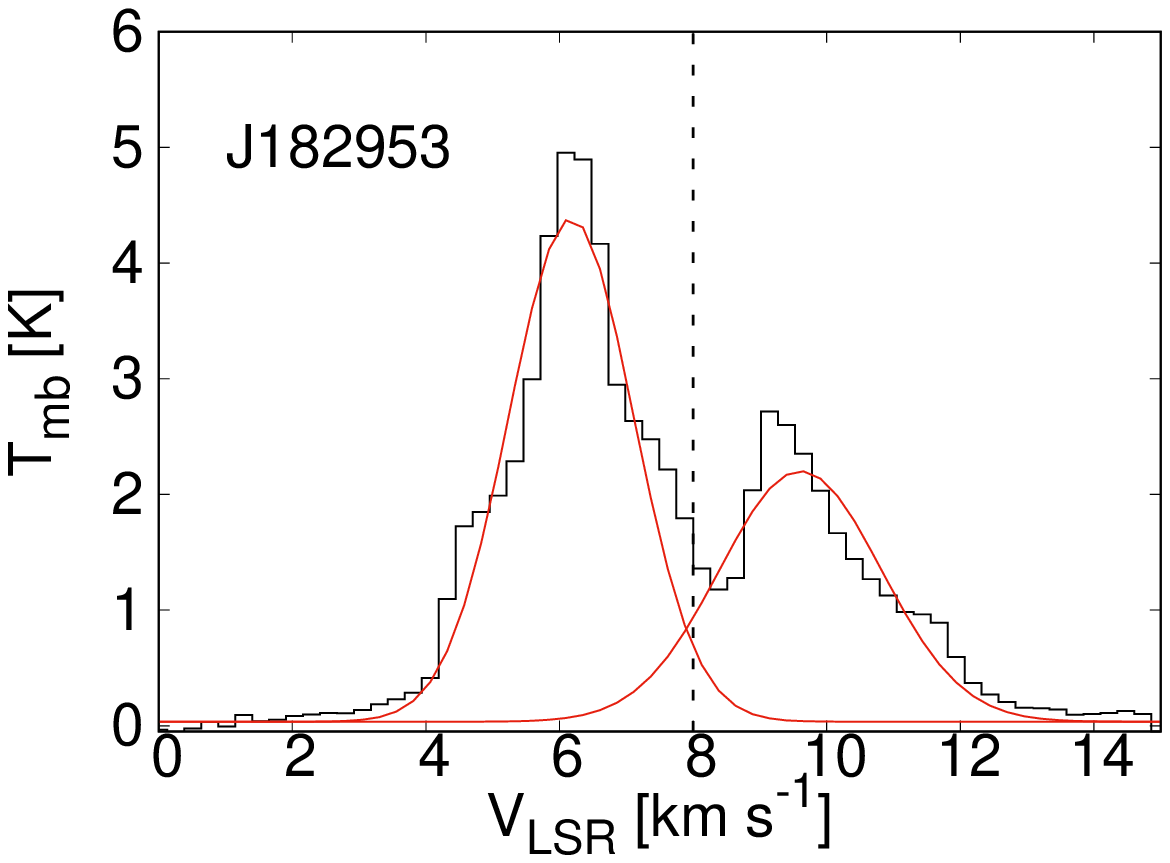}     
     \includegraphics[width=1.8in]{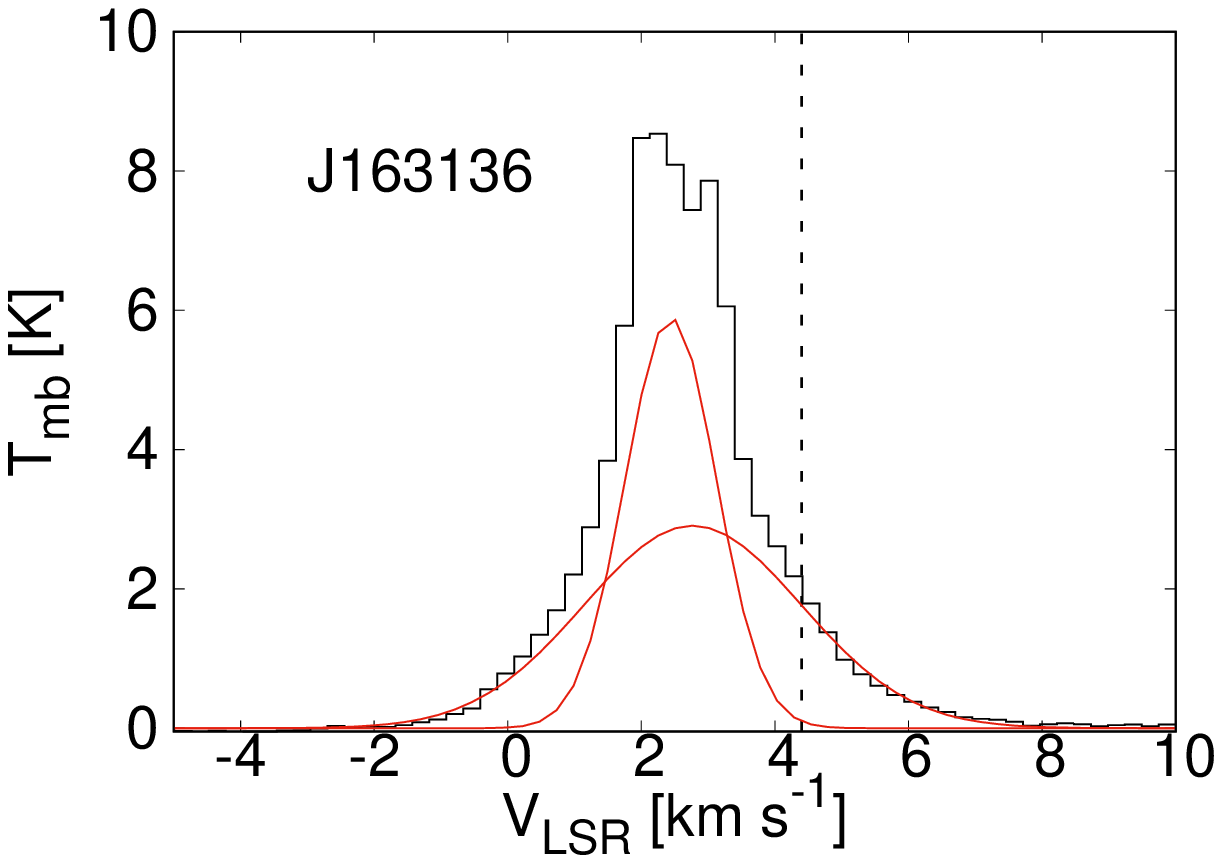}           
     \includegraphics[width=1.8in]{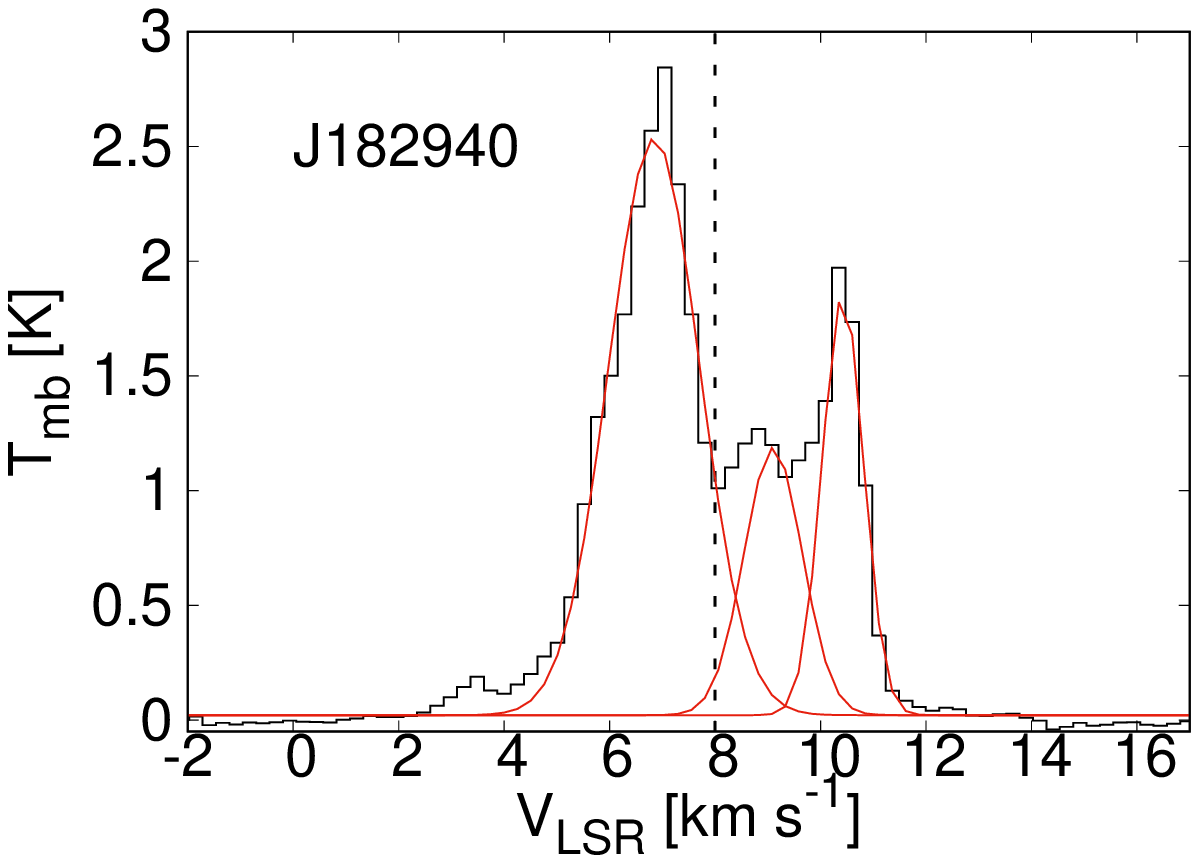}
     \includegraphics[width=1.8in]{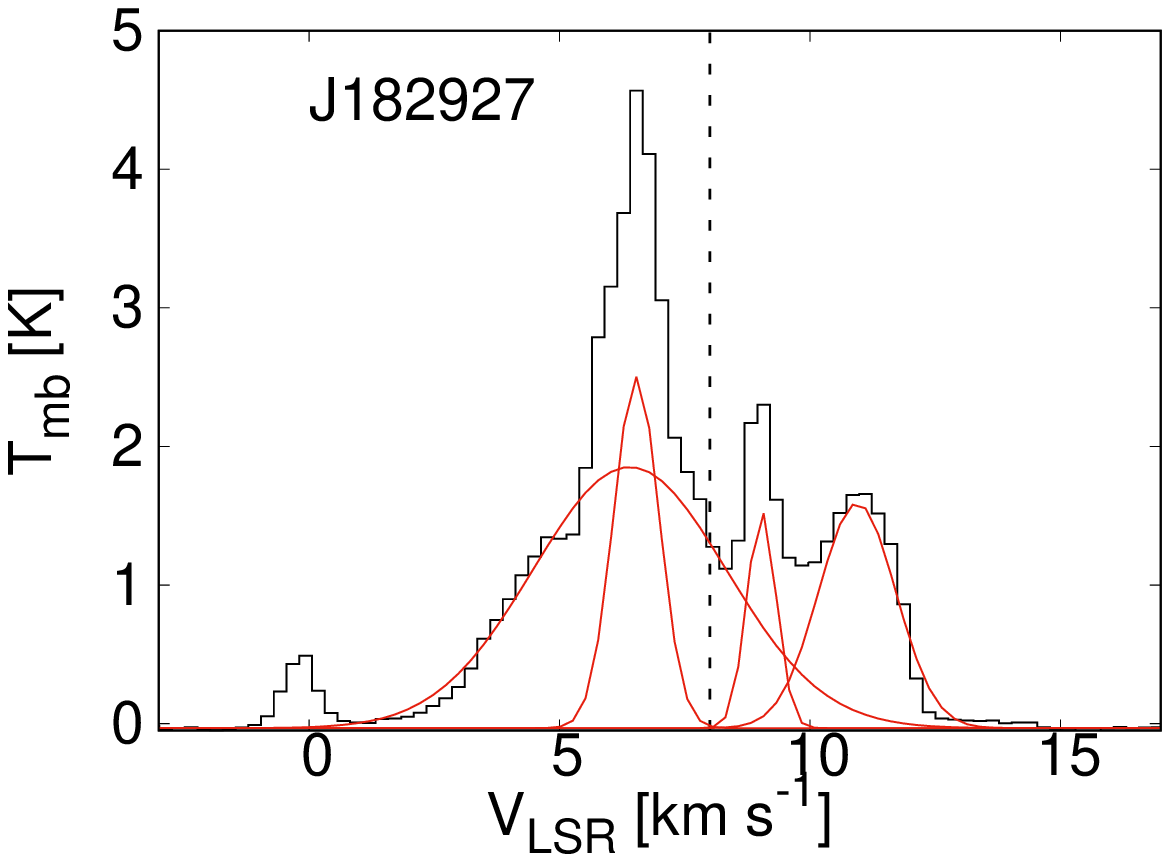}        
     \includegraphics[width=1.8in]{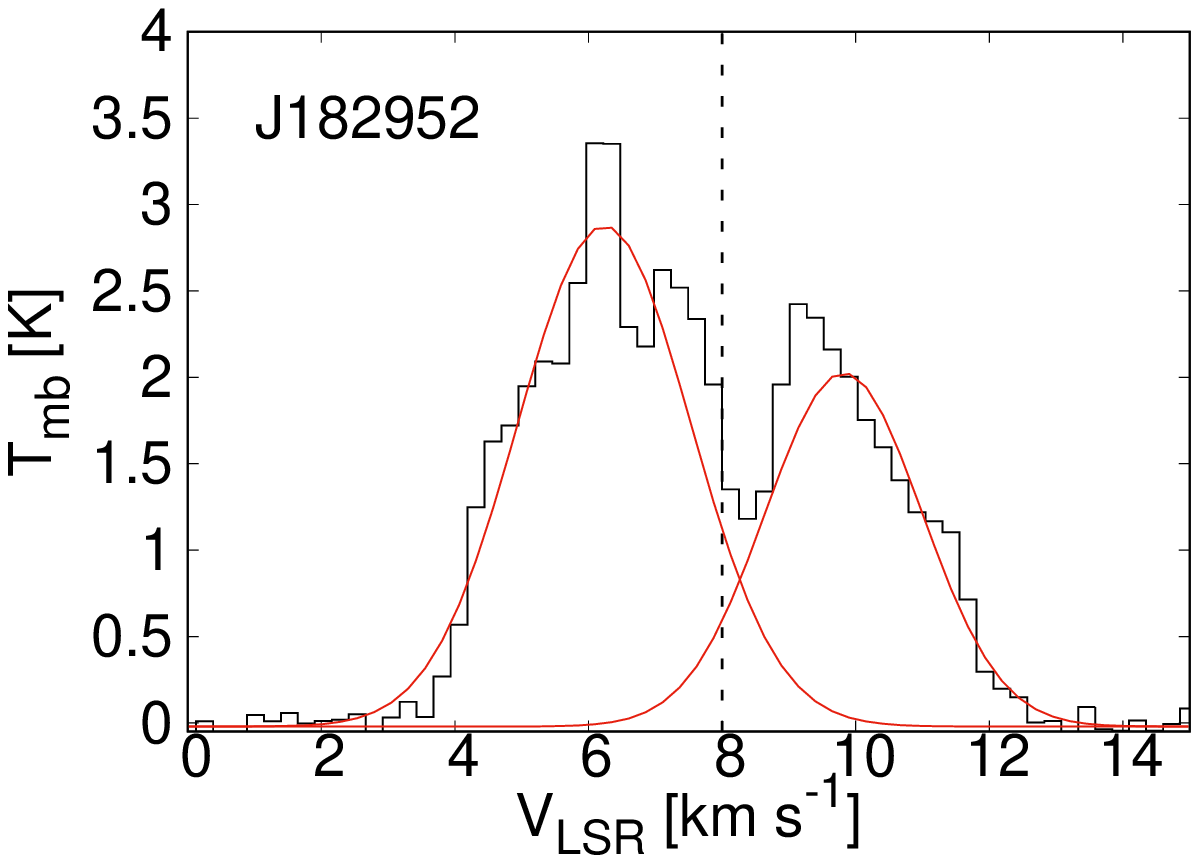}      
     \caption{The observed $^{12}$CO (2-1) spectra (black) with the Gaussian fits (red). Black dashed line marks the cloud systemic velocity of $\sim$8 km/s and $\sim$4.4 km/s in Serpens and Ophiuchus, respectively. }
     \label{12co-figs}  
  \end{figure*}

The $^{13}$CO and $^{12}$CO spectra are complex and show multiple peaks in the line profiles (Figs.~\ref{13co-figs};~\ref{12co-figs}). One exception is J182844 that shows a broad, single Gaussian shaped $^{13}$CO profile. $^{12}$CO (2-1) is a low-density tracer and is more likely to be affected by large optical depths compared to the other isotopomers. It shows very heavy self-absorption close to the cloud systemic velocity. The broad, multi-component spectra in $^{12}$CO and $^{13}$CO indicate a contribution from clouds with different velocity components as well as foreground and unrelated cloud material along the line of sight. Most sources also show extended wings in the $^{12}$CO and $^{13}$CO profiles which could be a contribution from the outflowing gas.

\section{High-density tracers}

\begin{figure*}
  \centering              
     \includegraphics[width=1.8in]{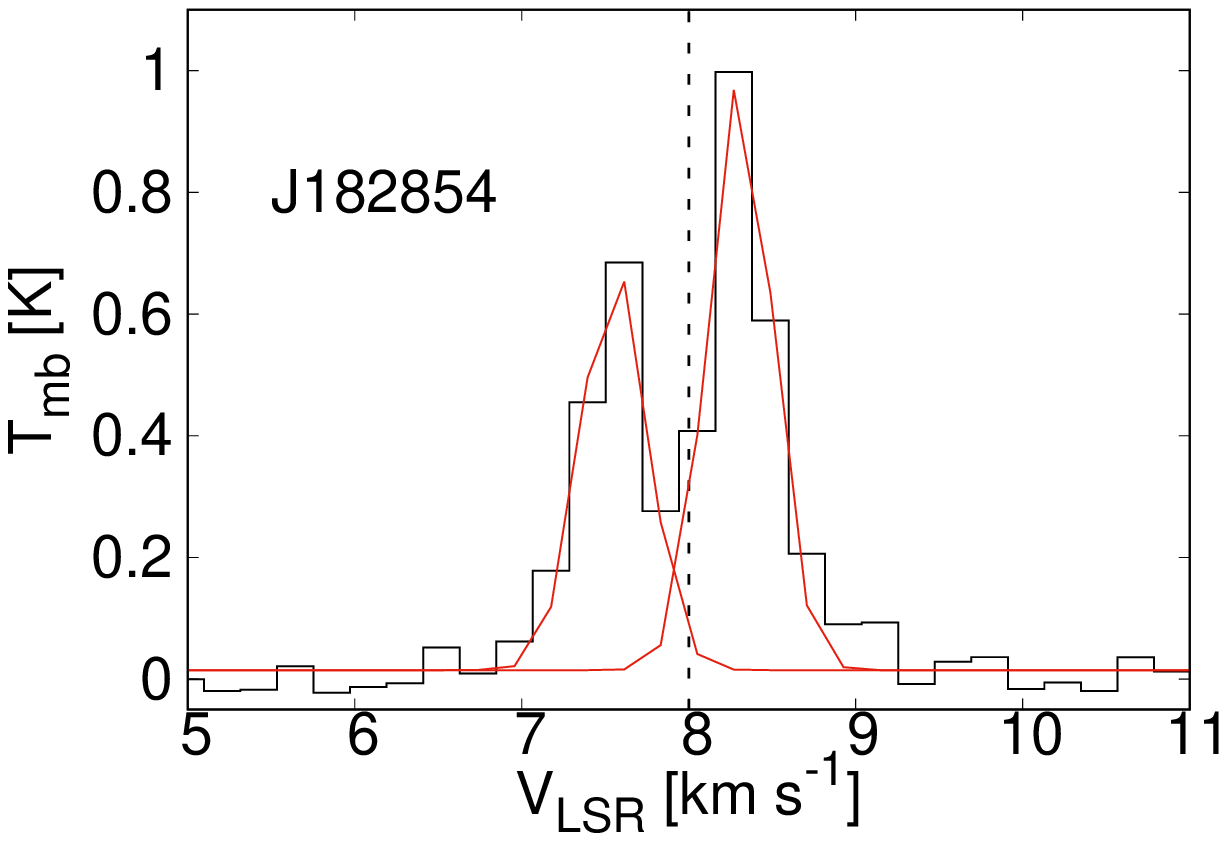}
     \includegraphics[width=1.8in]{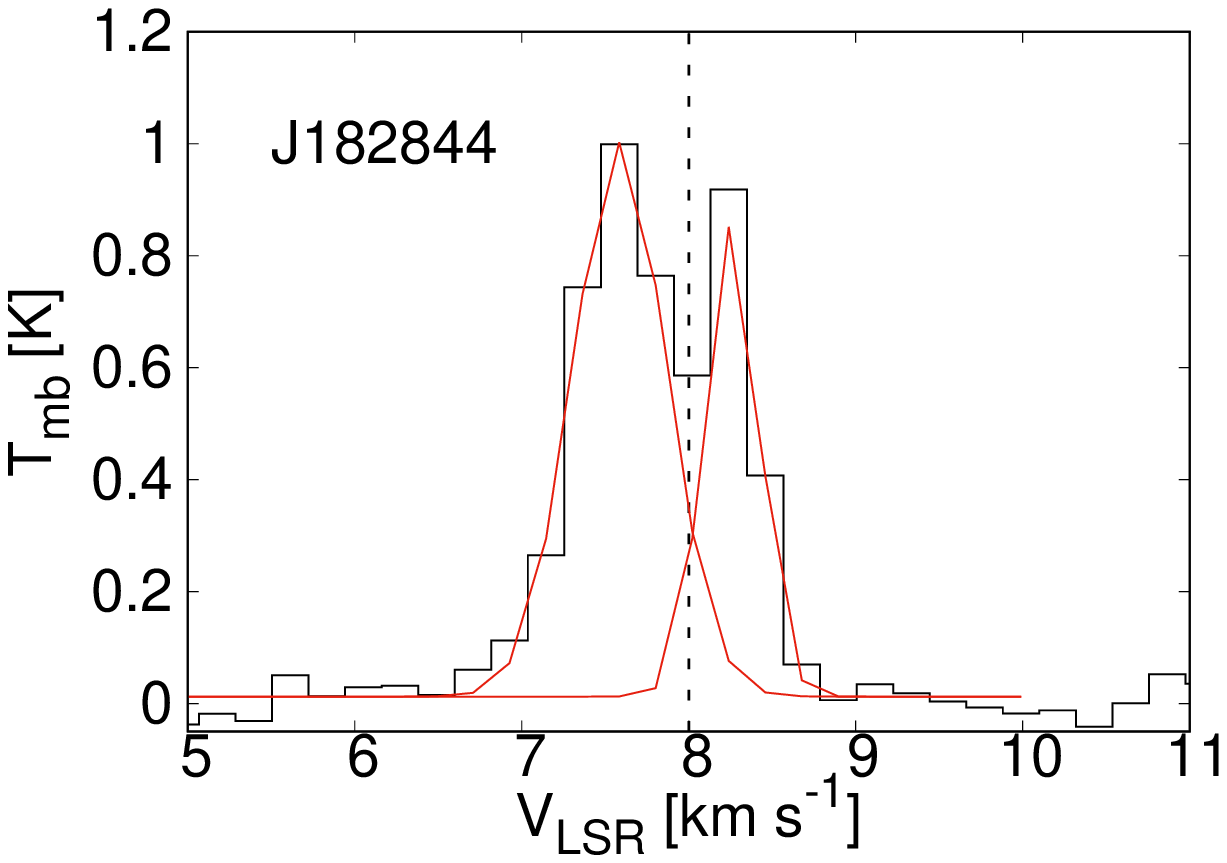}
     \includegraphics[width=1.8in]{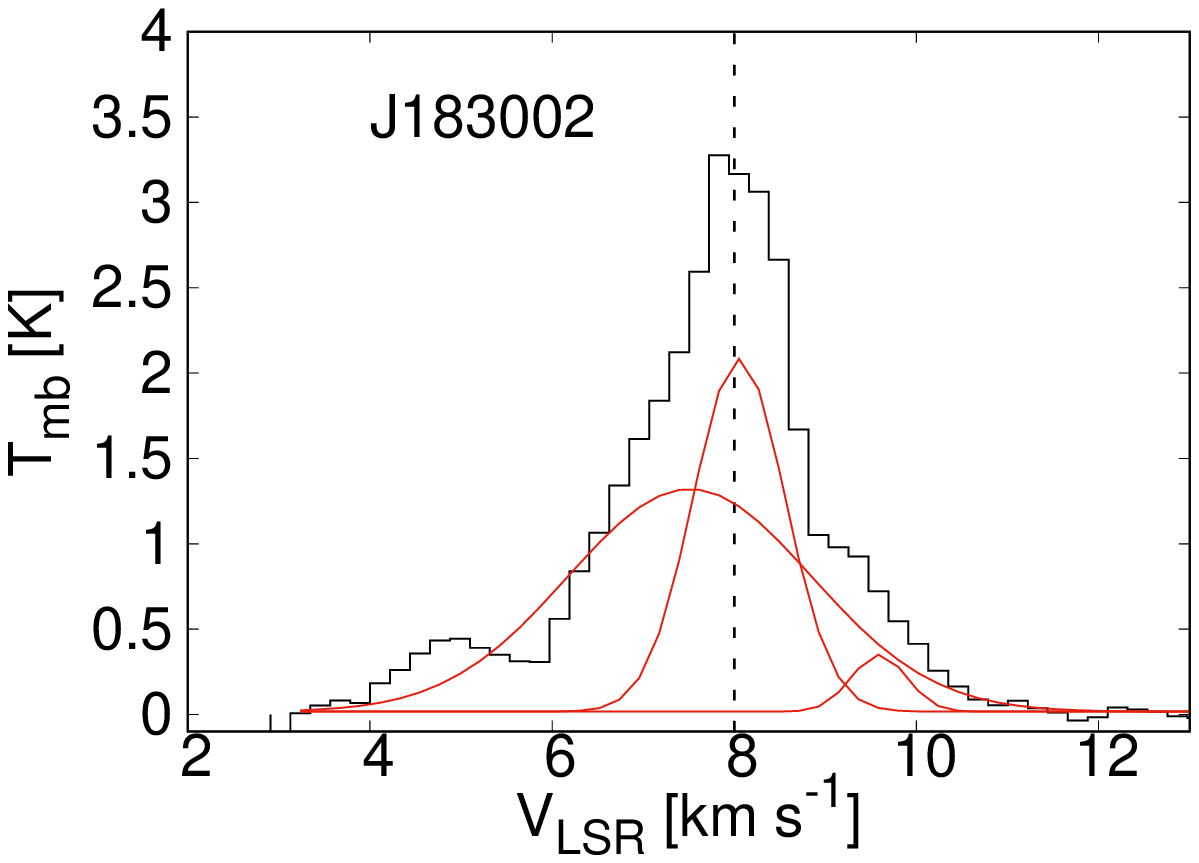}
     \includegraphics[width=1.8in]{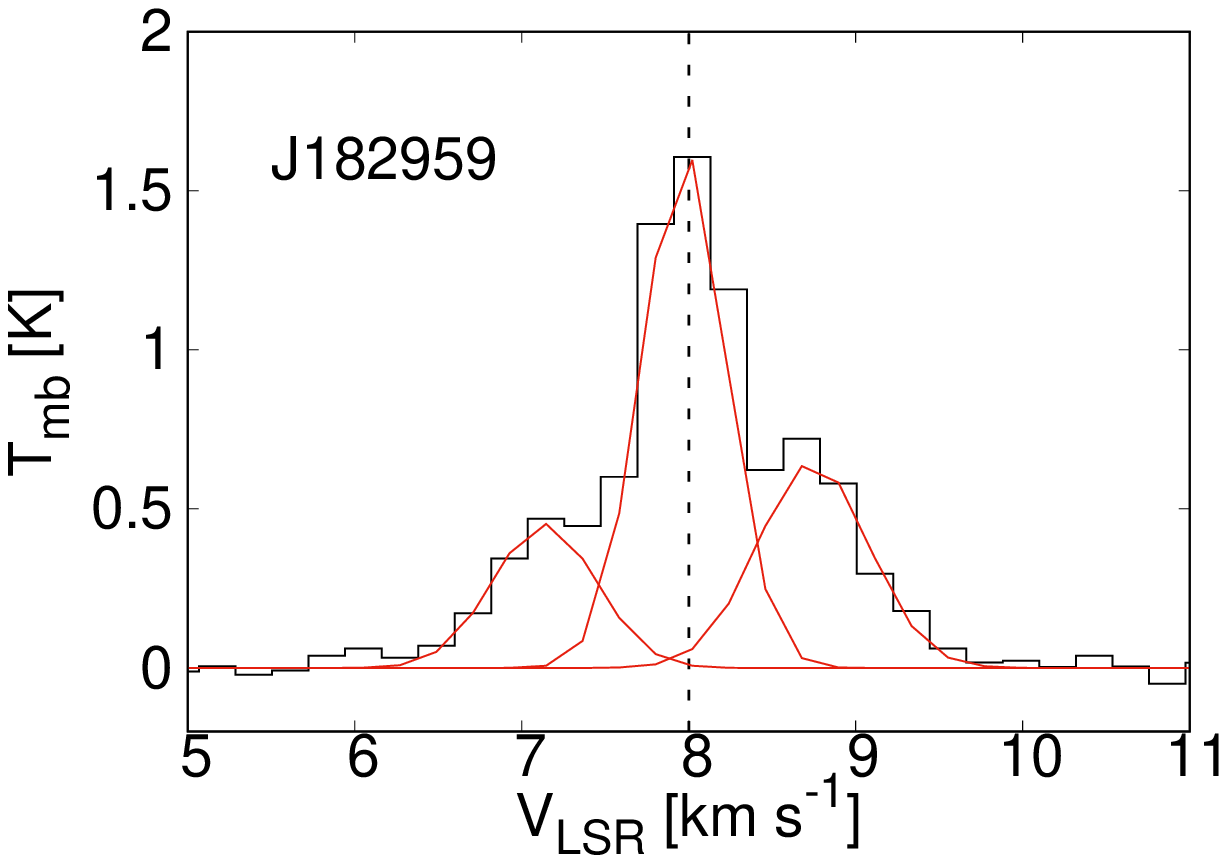}     
     \includegraphics[width=1.8in]{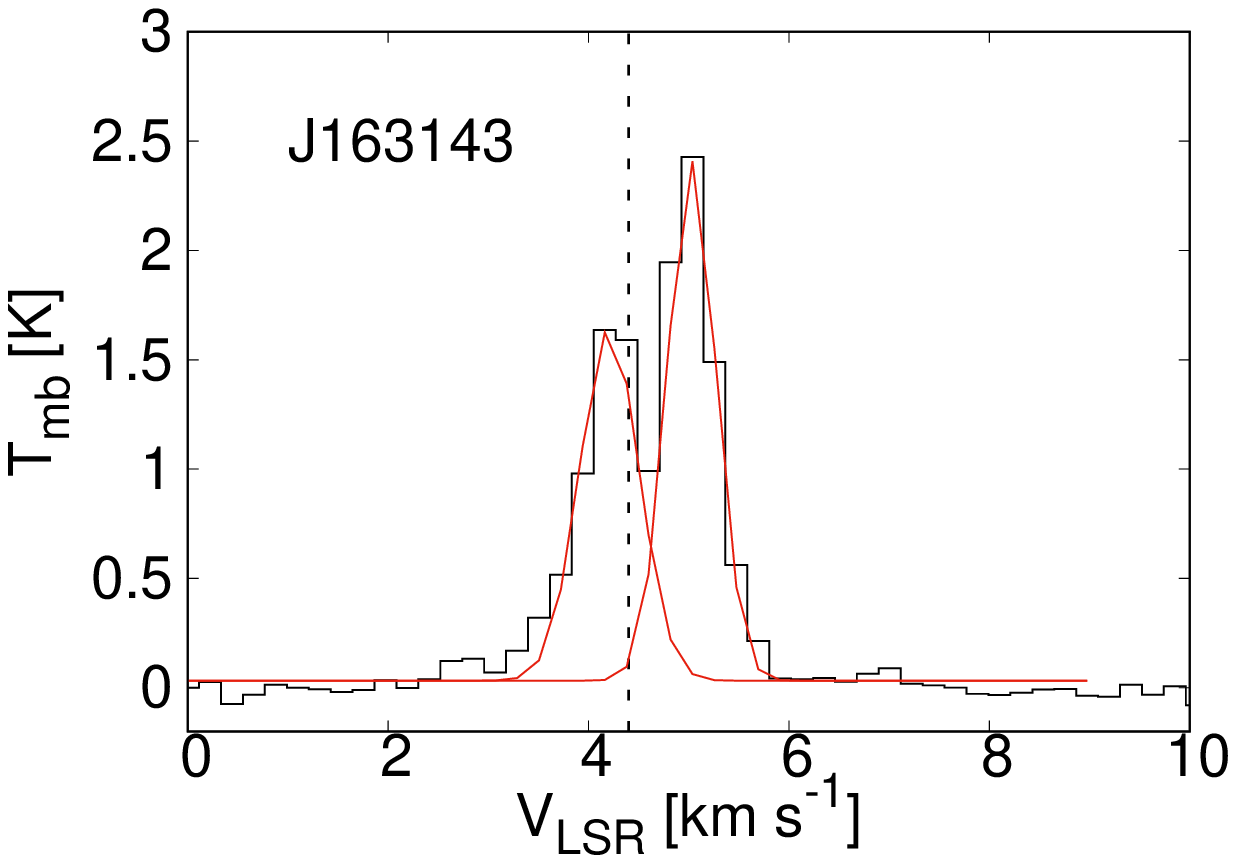}
     \includegraphics[width=1.8in]{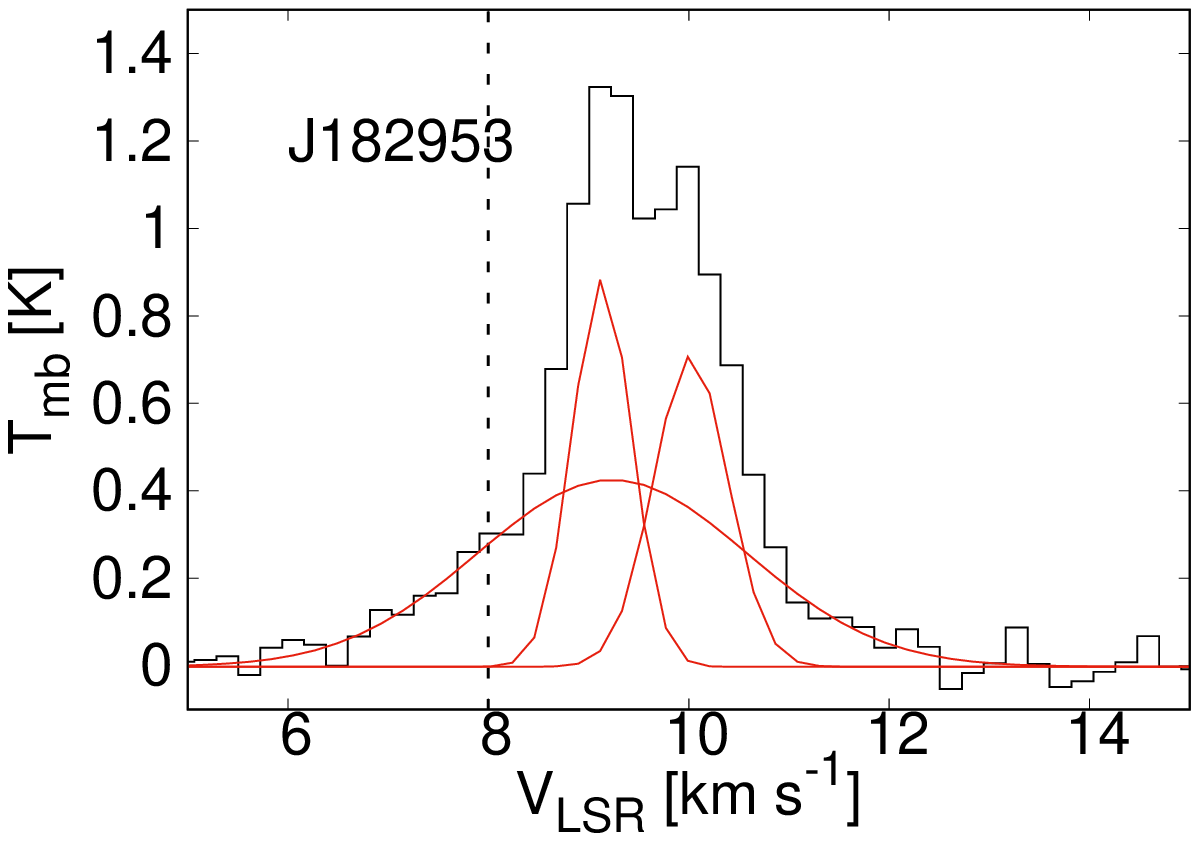}
     \includegraphics[width=1.8in]{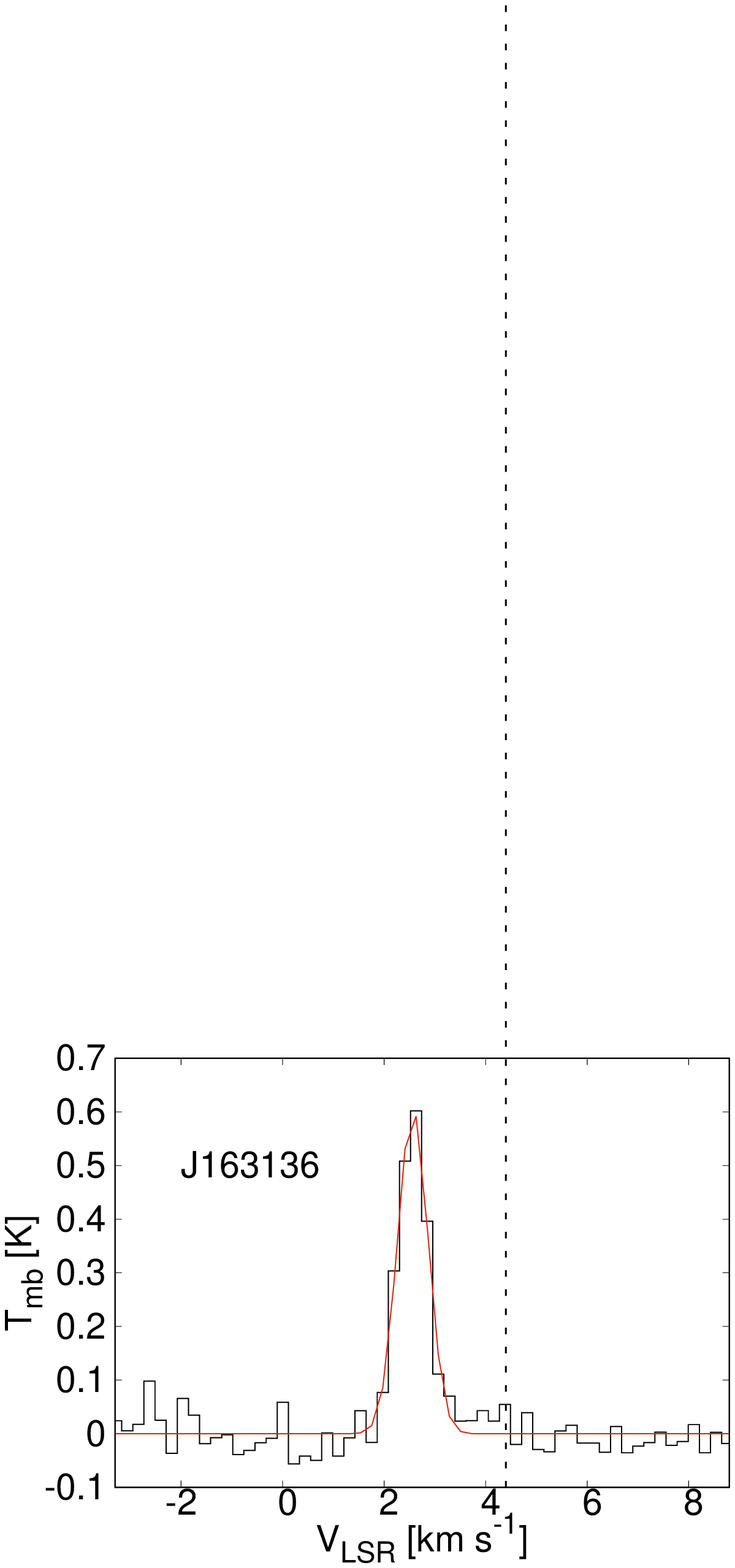}     
     \includegraphics[width=1.8in]{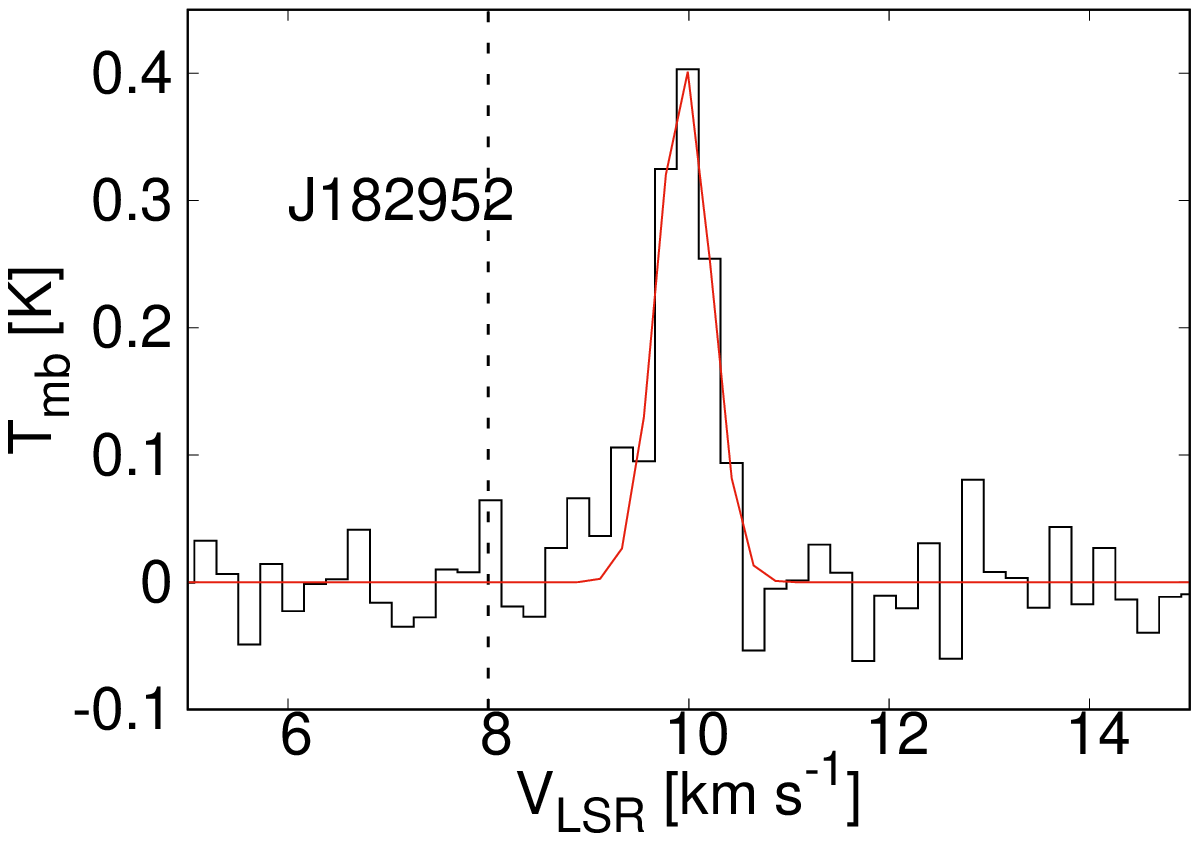}
     \caption{The observed HCO$^{+}$ (3-2) spectra (black) with the Gaussian fits (red). Black dashed line marks the cloud systemic velocity of $\sim$8 km/s and $\sim$4.4 km/s in Serpens and Ophiuchus, respectively. }
     \label{hcop-figs}     
  \end{figure*}

\begin{figure*}
  \centering              
     \includegraphics[width=1.8in]{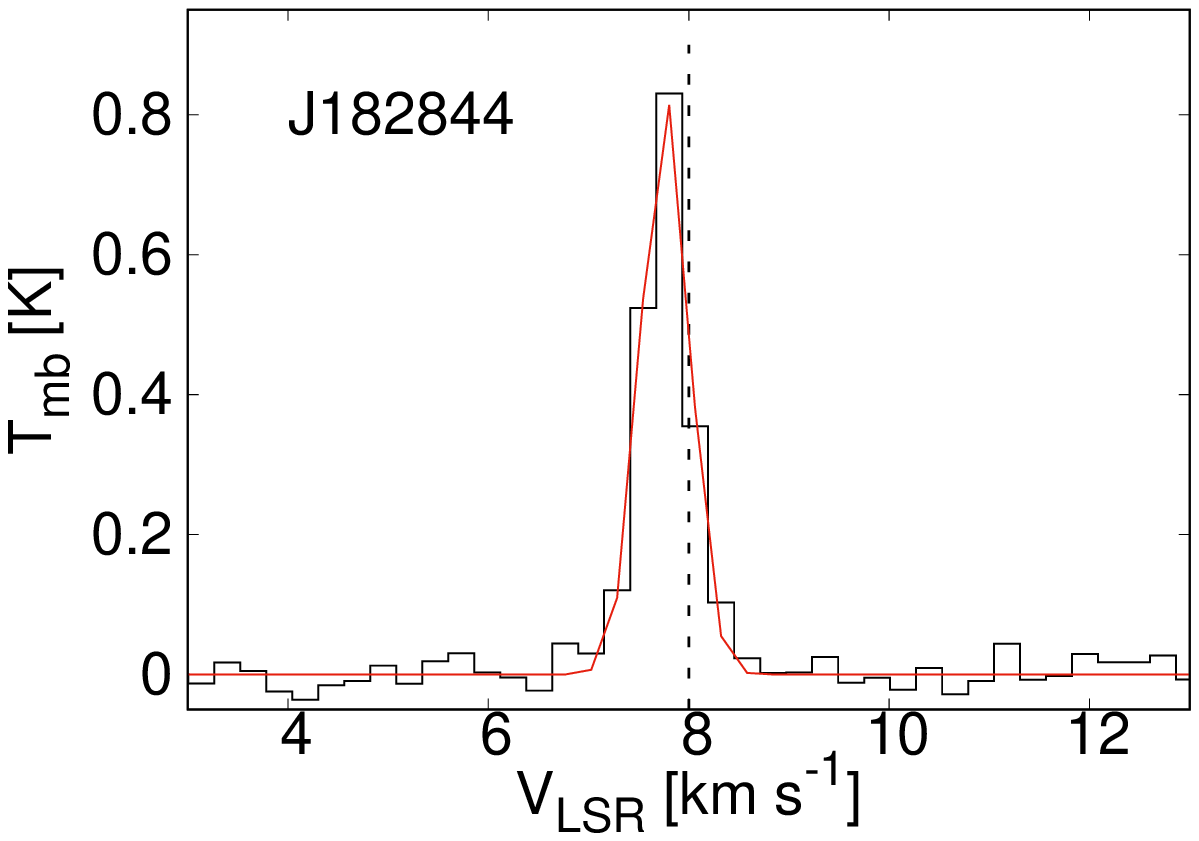}
     \includegraphics[width=1.8in]{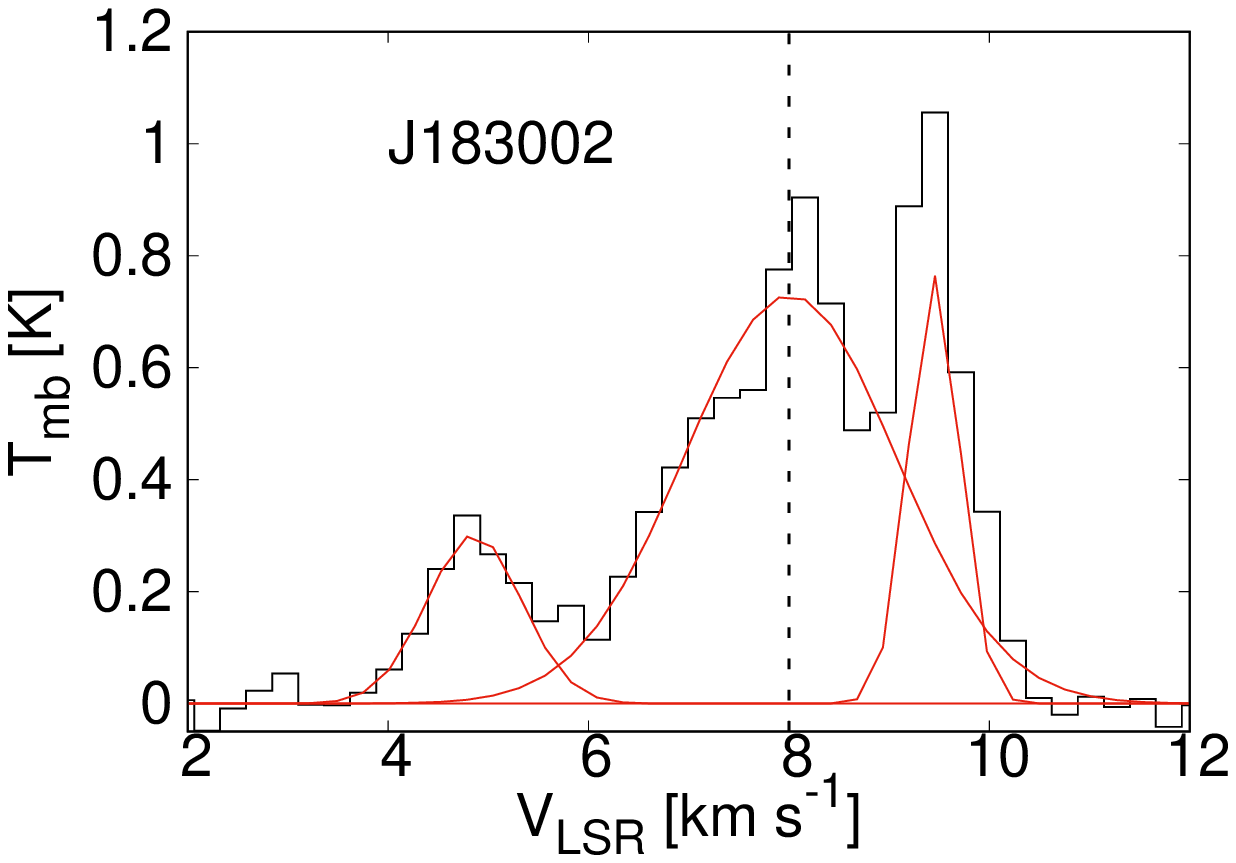}     
     \includegraphics[width=1.8in]{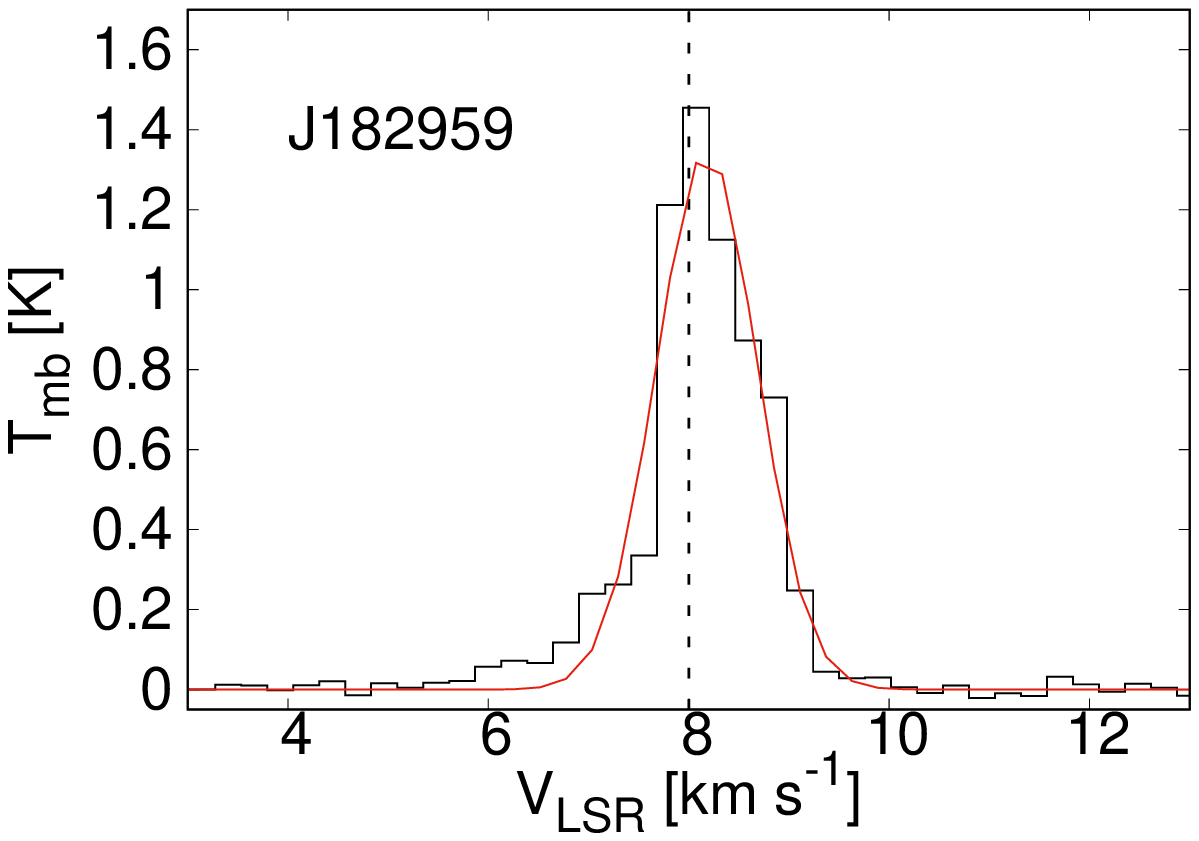}
     \includegraphics[width=1.8in]{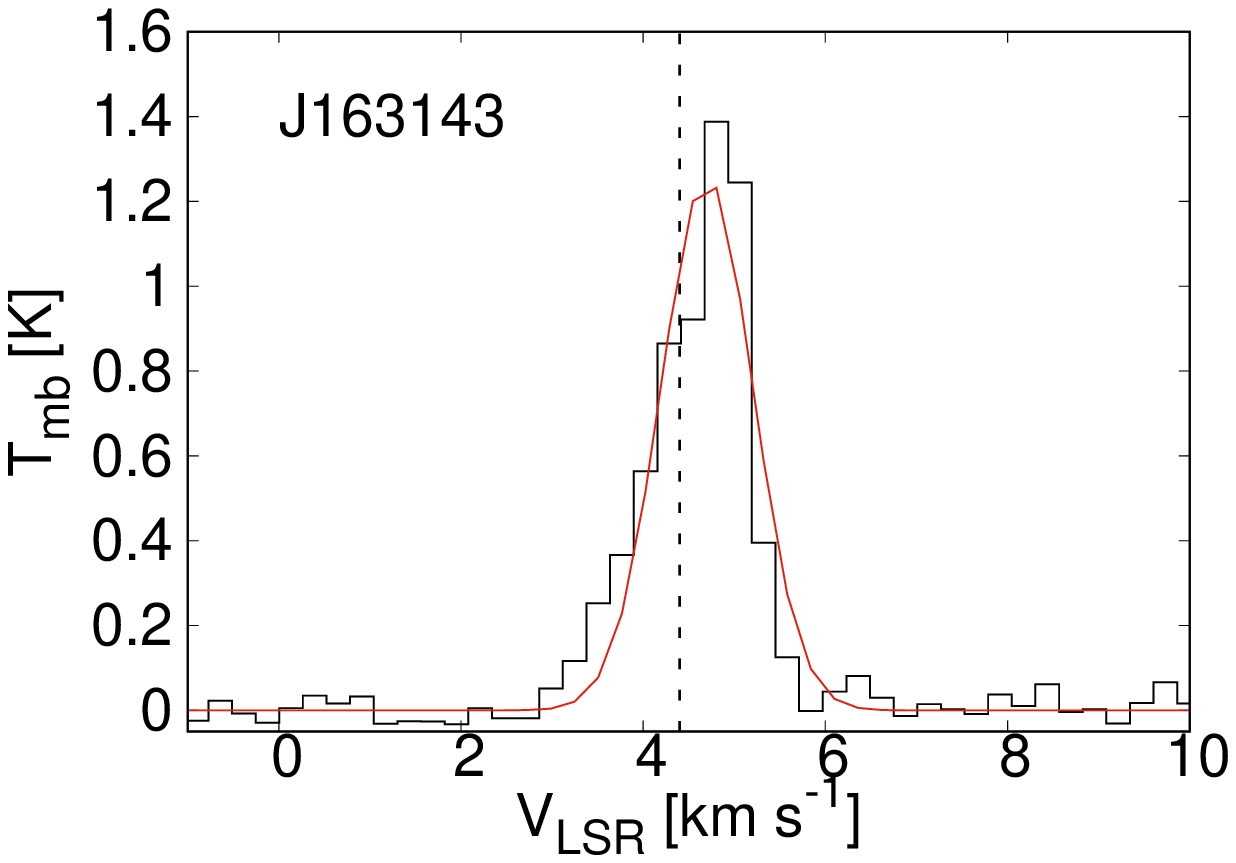}     
     \includegraphics[width=1.8in]{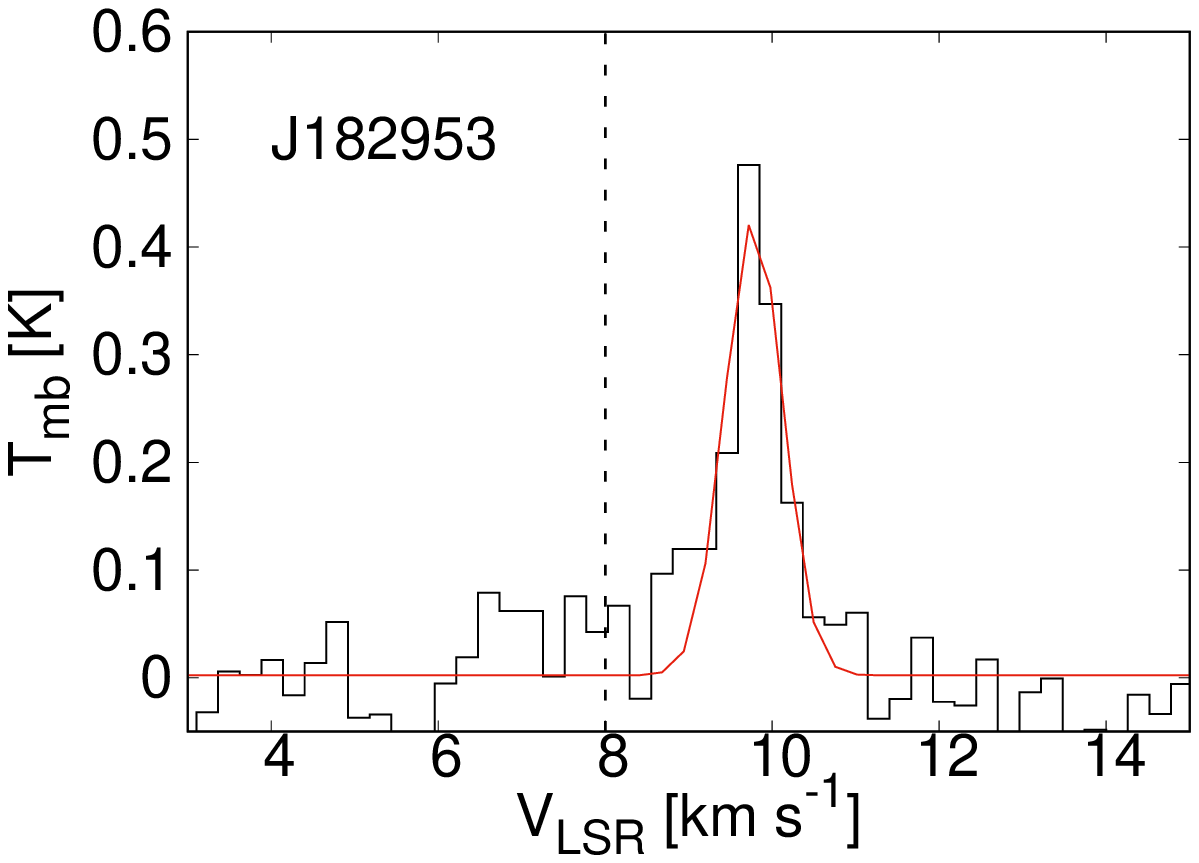}
     \includegraphics[width=1.8in]{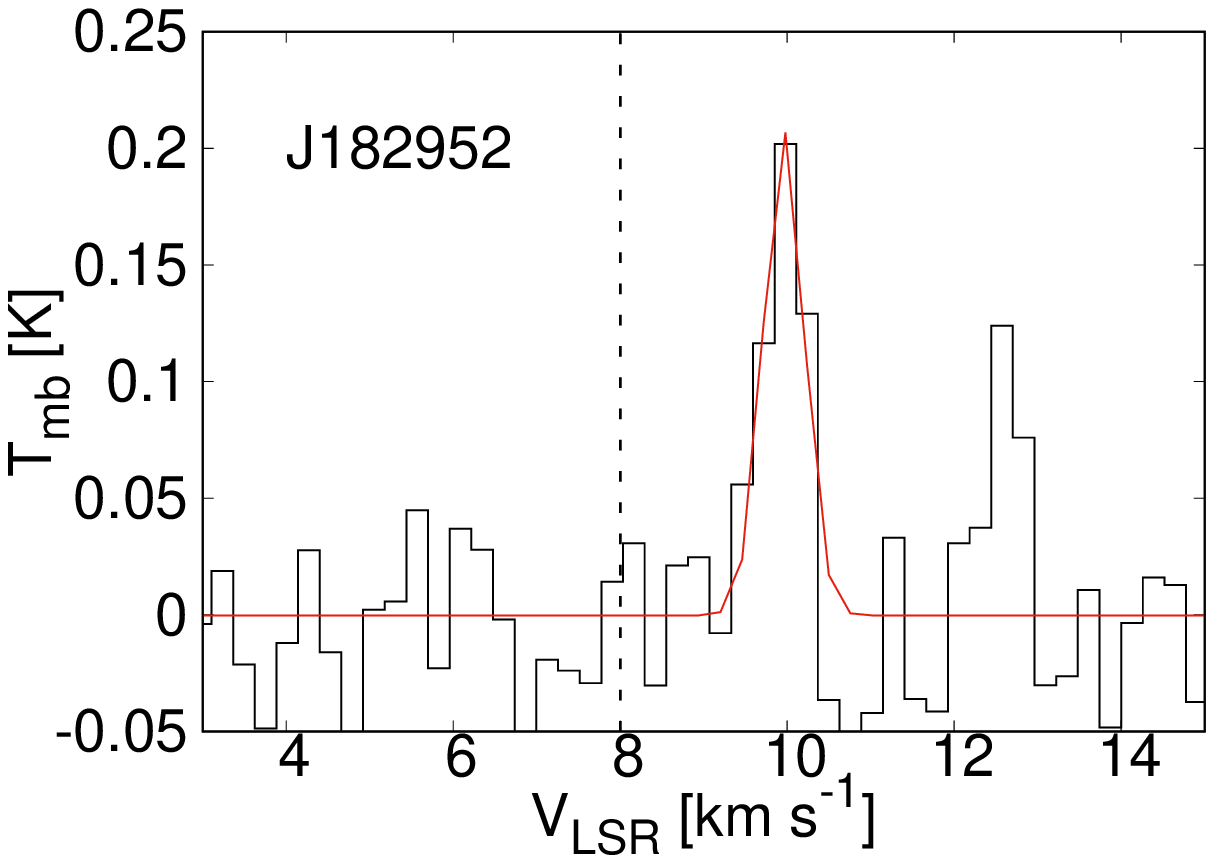}     
     \caption{The observed ortho-H$_{2}$CO (3-2) spectra (black) with the Gaussian fits (red). Black dashed line marks the cloud systemic velocity of $\sim$8 km/s and $\sim$4.4 km/s in Serpens and Ophiuchus, respectively. }
     \label{o-h2co-figs}     
  \end{figure*}

\begin{figure*}
  \centering              
     \includegraphics[width=1.8in]{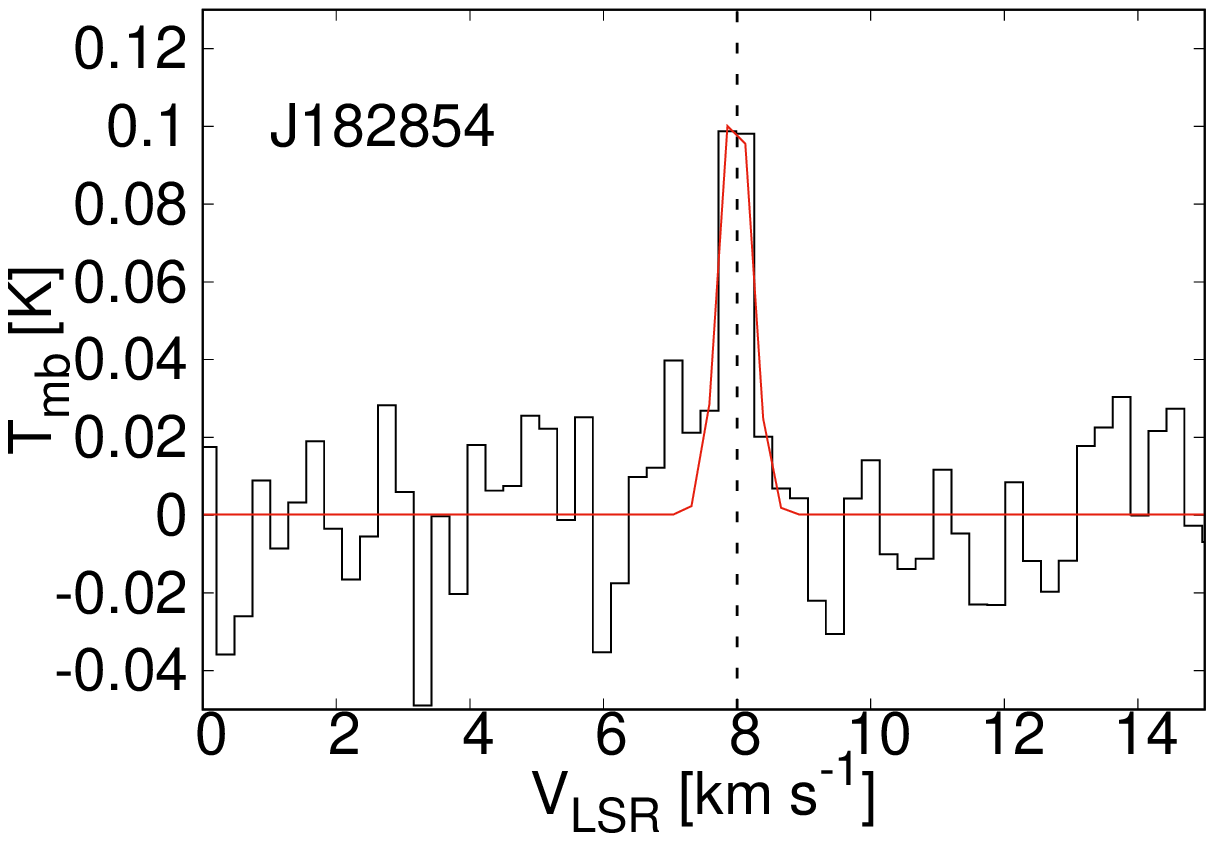}
     \includegraphics[width=1.8in]{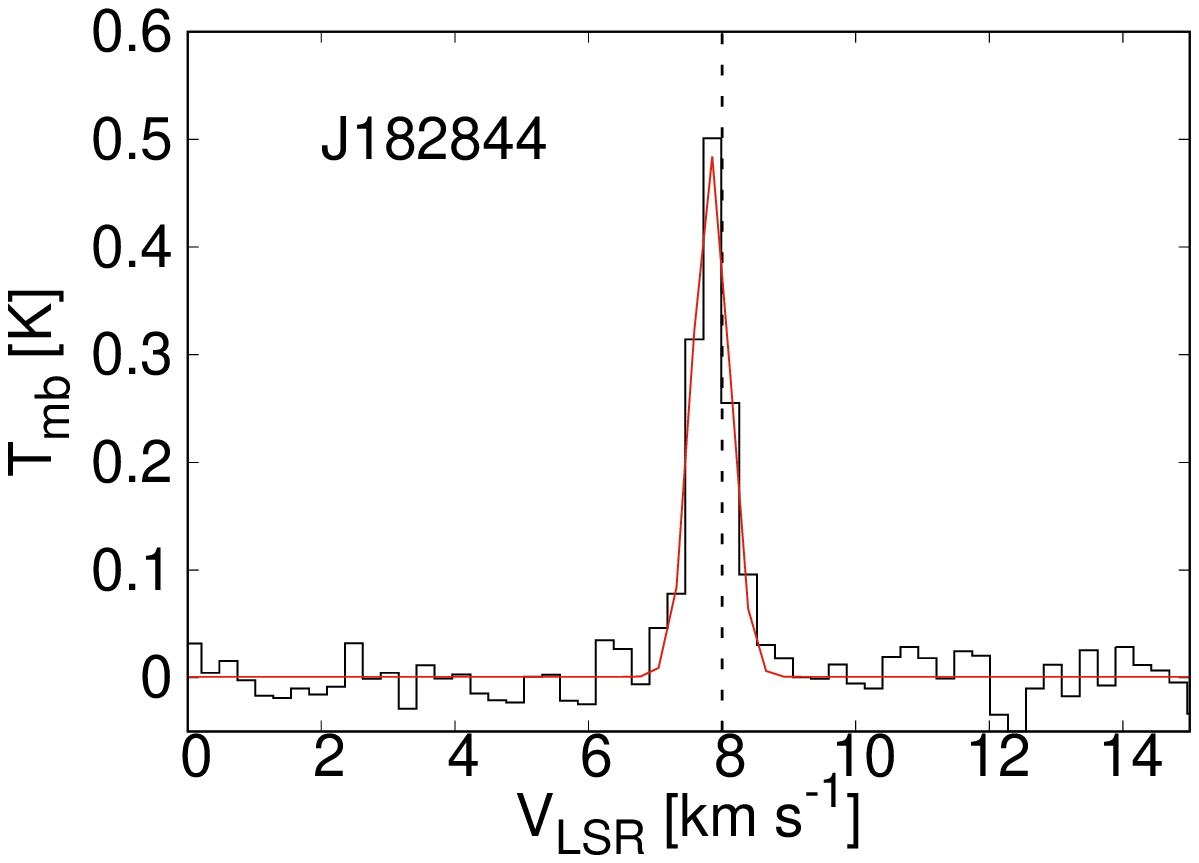}
     \includegraphics[width=1.8in]{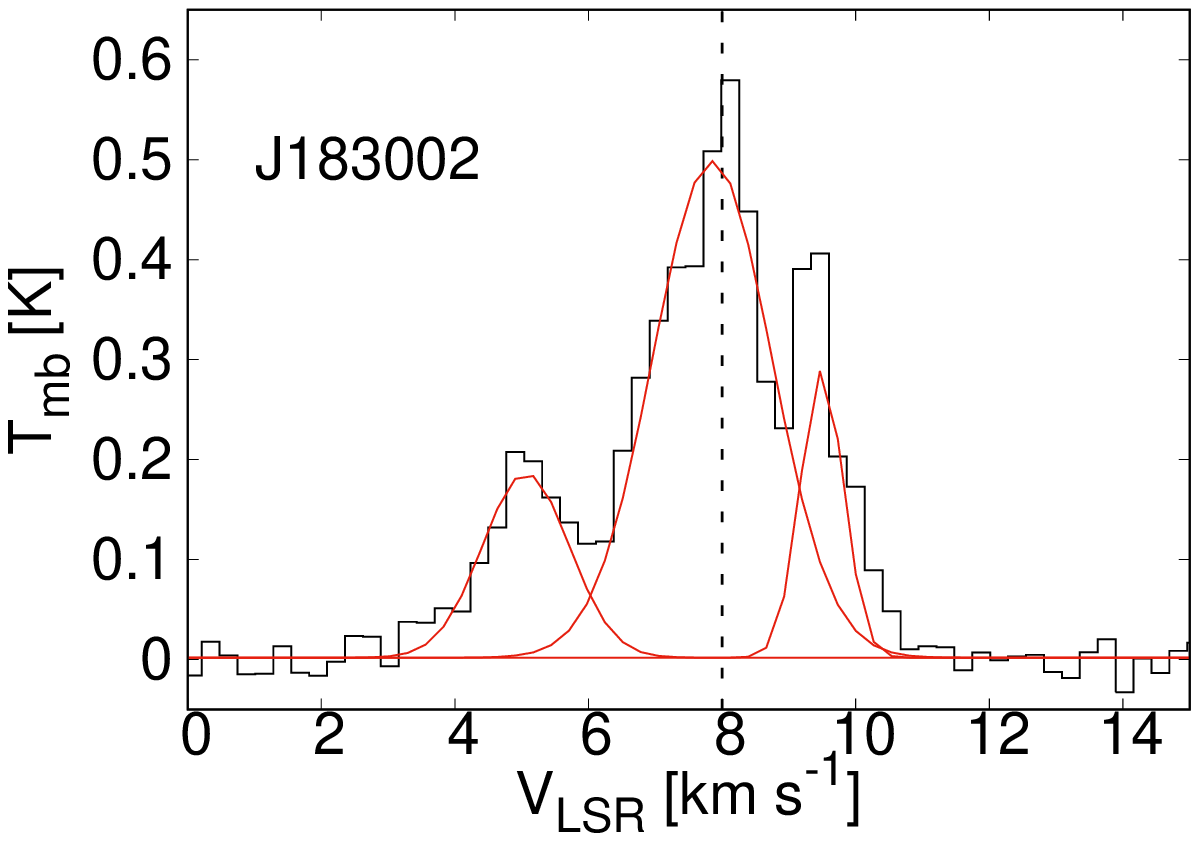}     
     \includegraphics[width=1.8in]{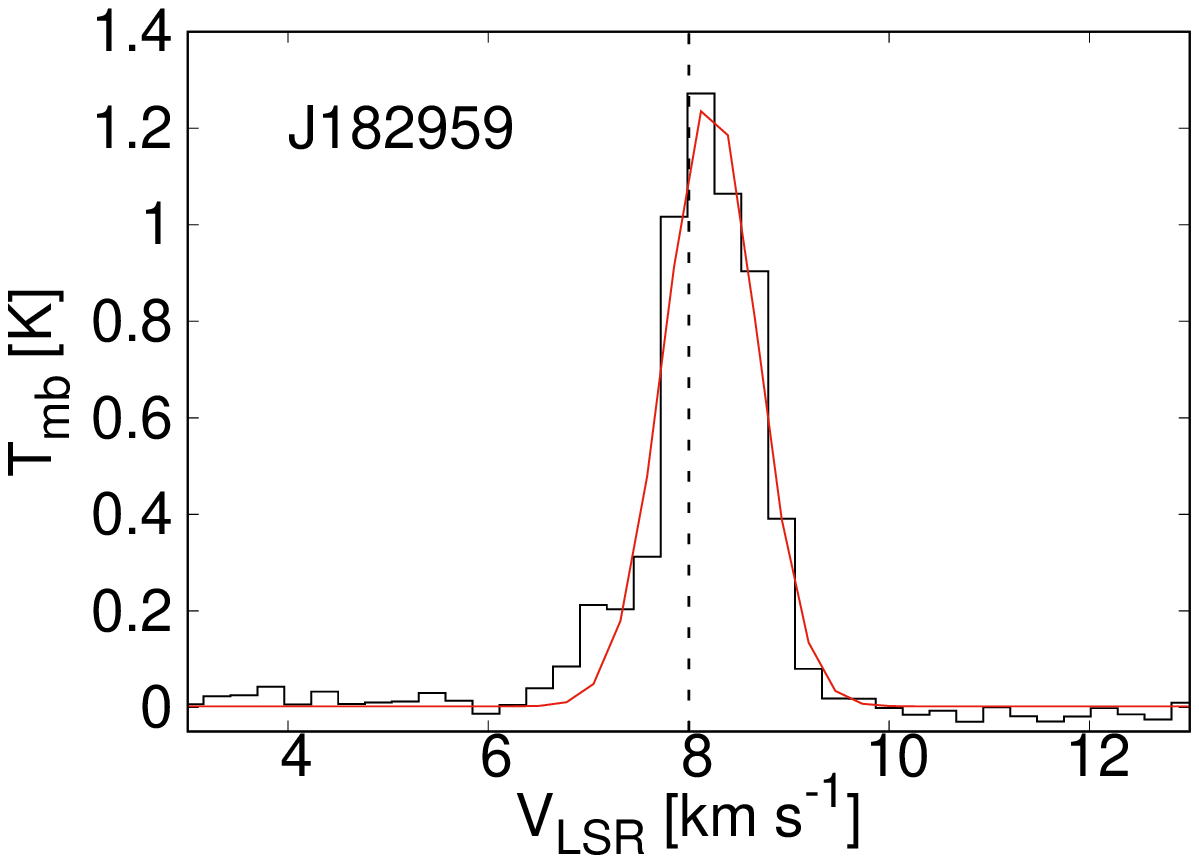}
     \includegraphics[width=1.8in]{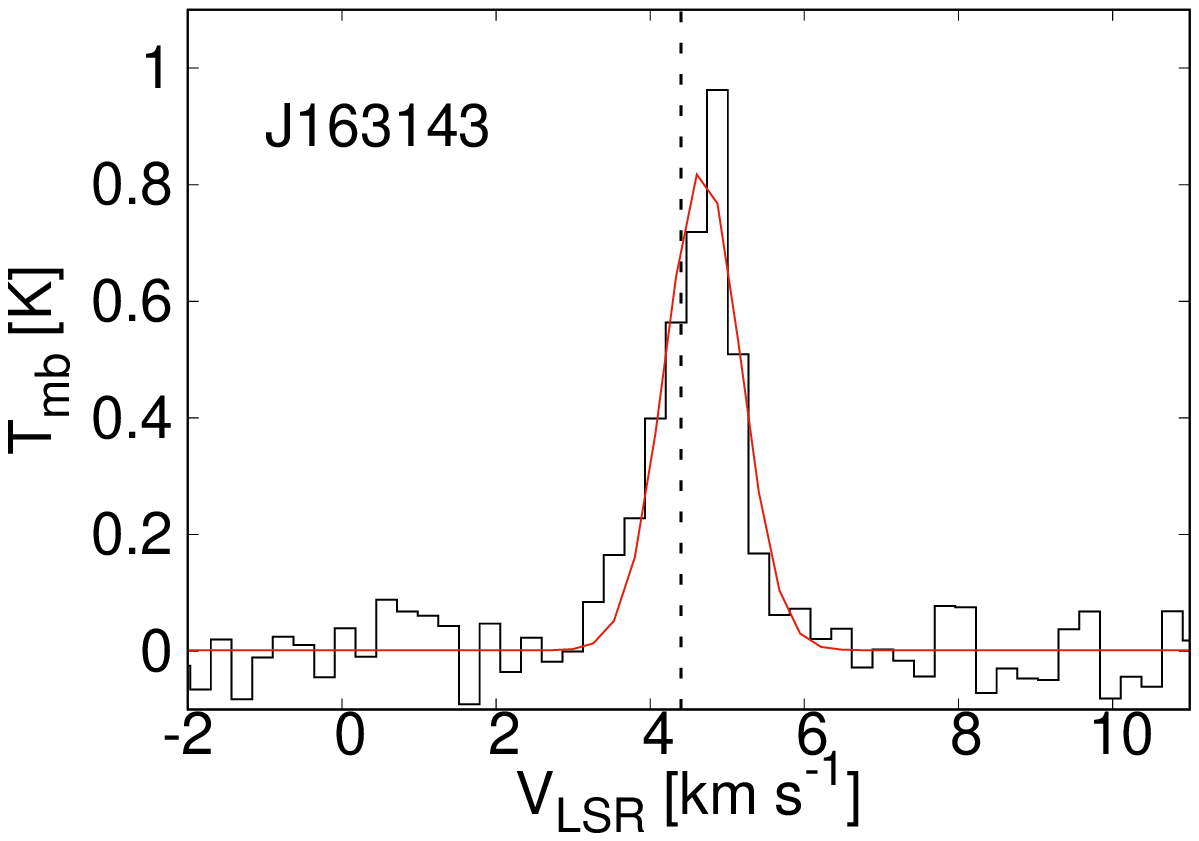}     
     \includegraphics[width=1.8in]{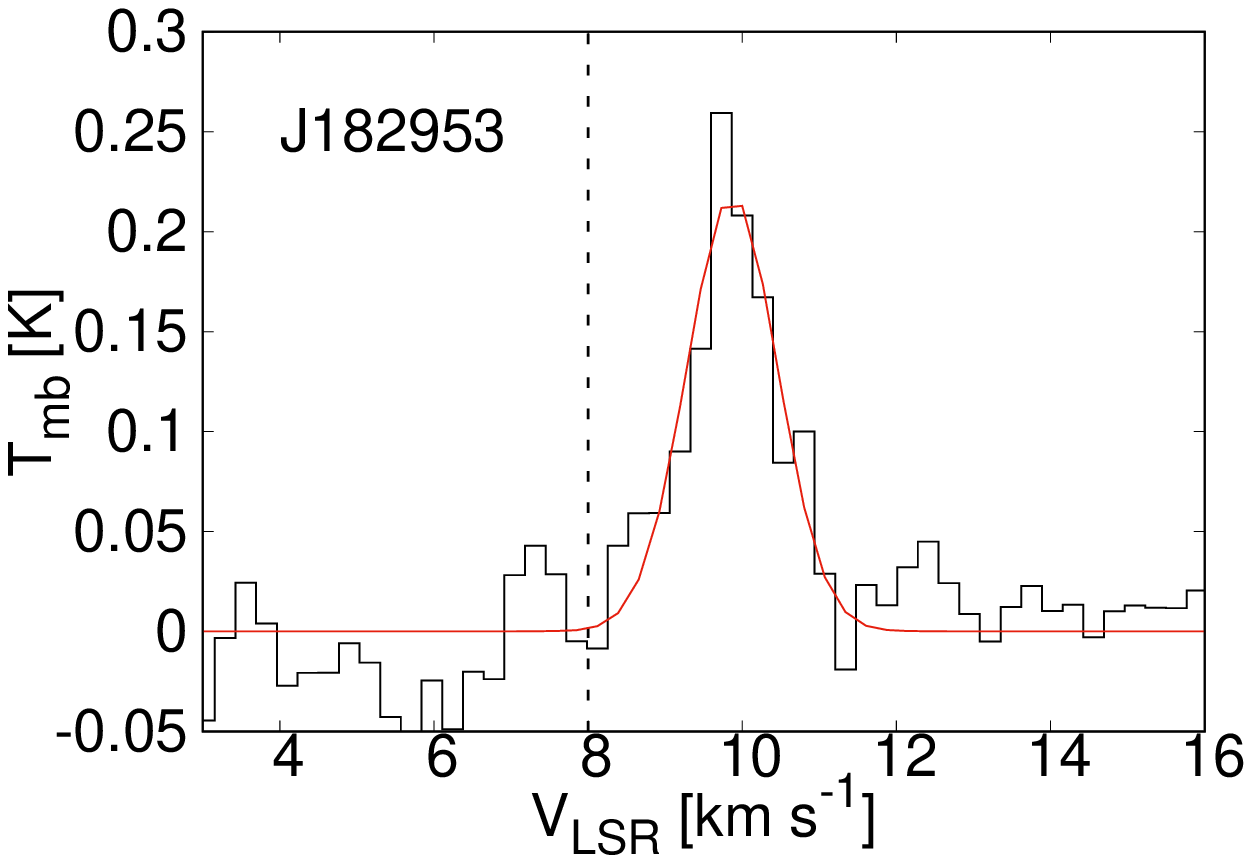}
     \includegraphics[width=1.8in]{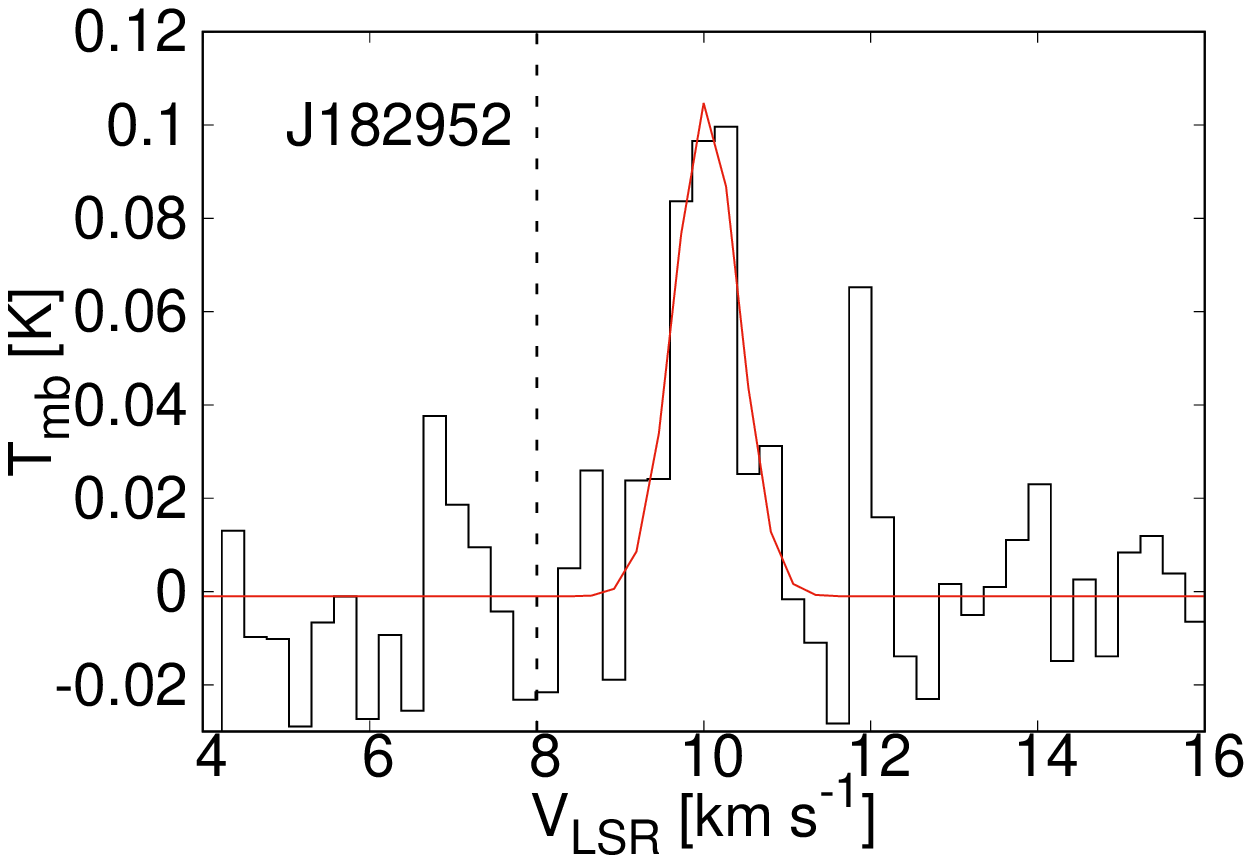}     
     \caption{The observed para-H$_{2}$CO (3-2) spectra (black) with the Gaussian fits (red). Black dashed line marks the cloud systemic velocity of $\sim$8 km/s and $\sim$4.4 km/s in Serpens and Ophiuchus, respectively.  }
     \label{p-h2co-figs}     
  \end{figure*}

\begin{figure*}
  \centering              
     \includegraphics[width=1.8in]{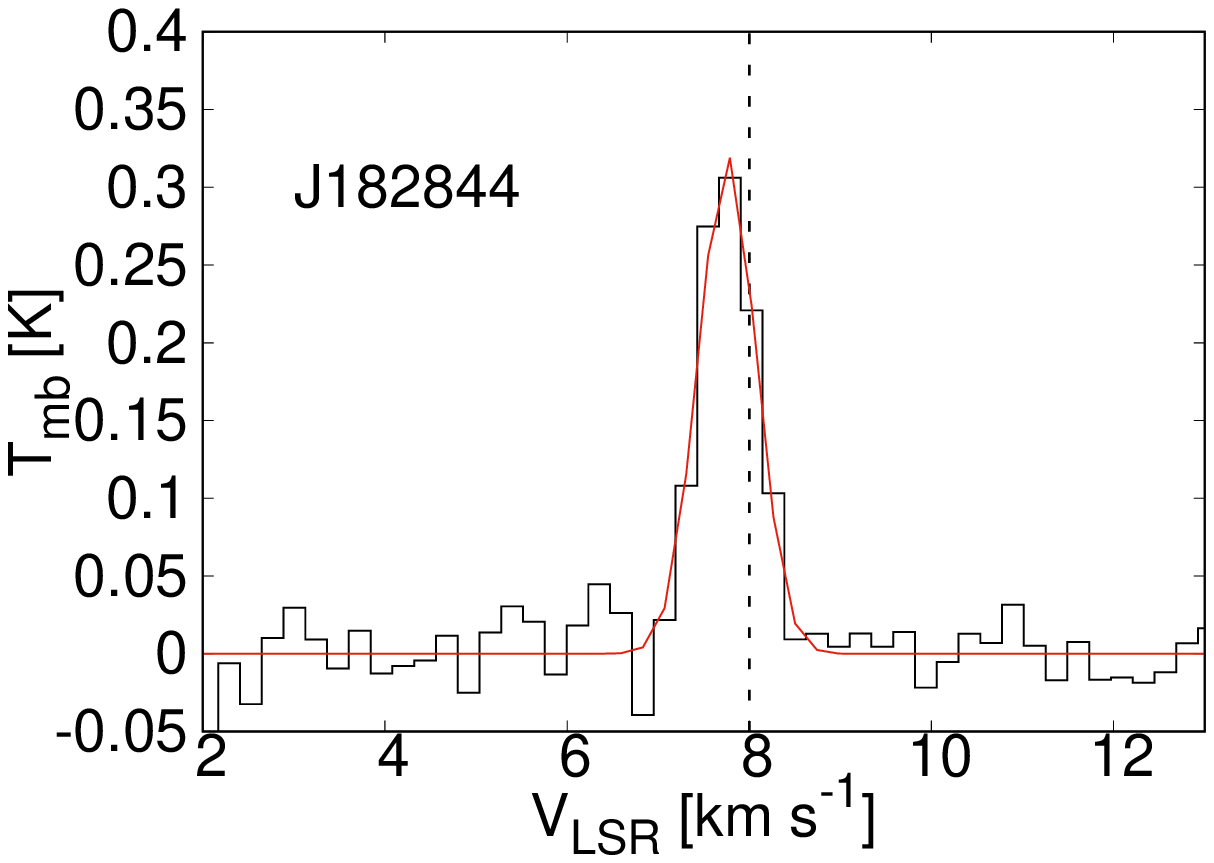}
     \includegraphics[width=1.8in]{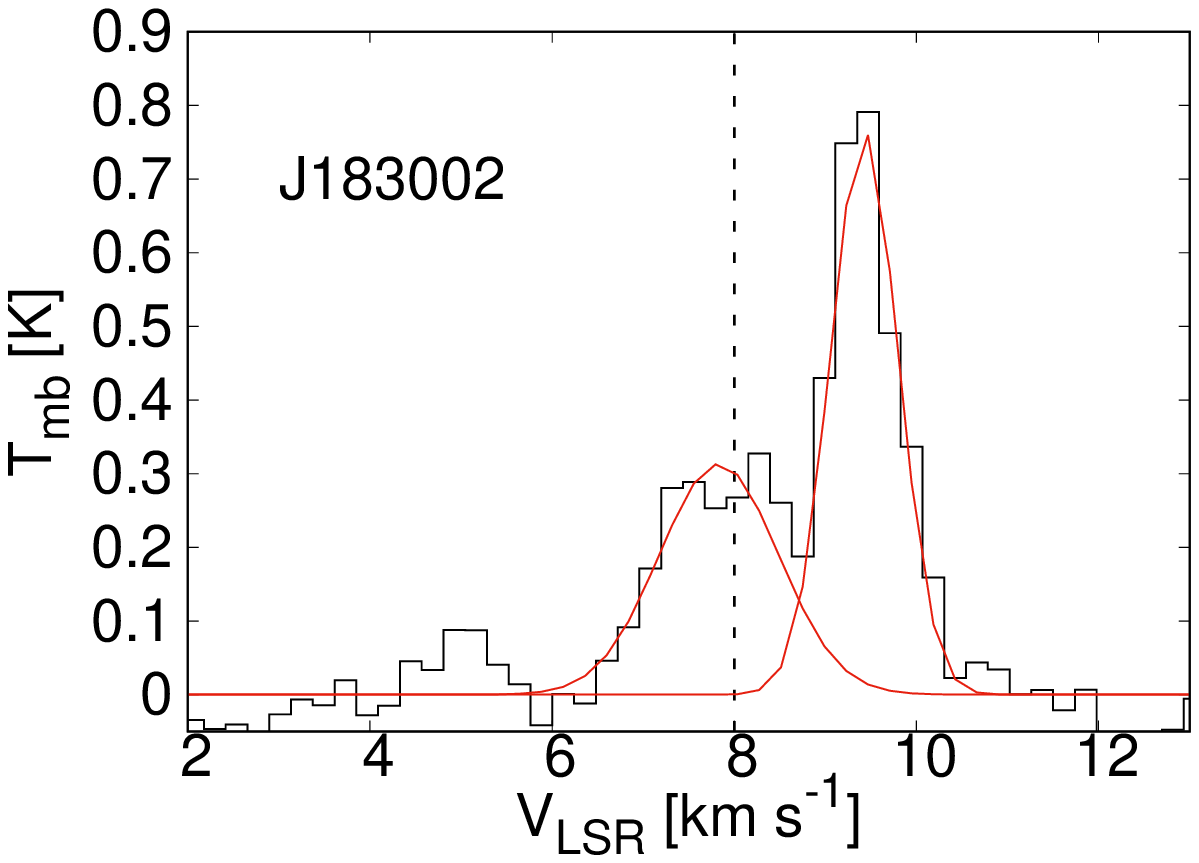}          
     \includegraphics[width=1.8in]{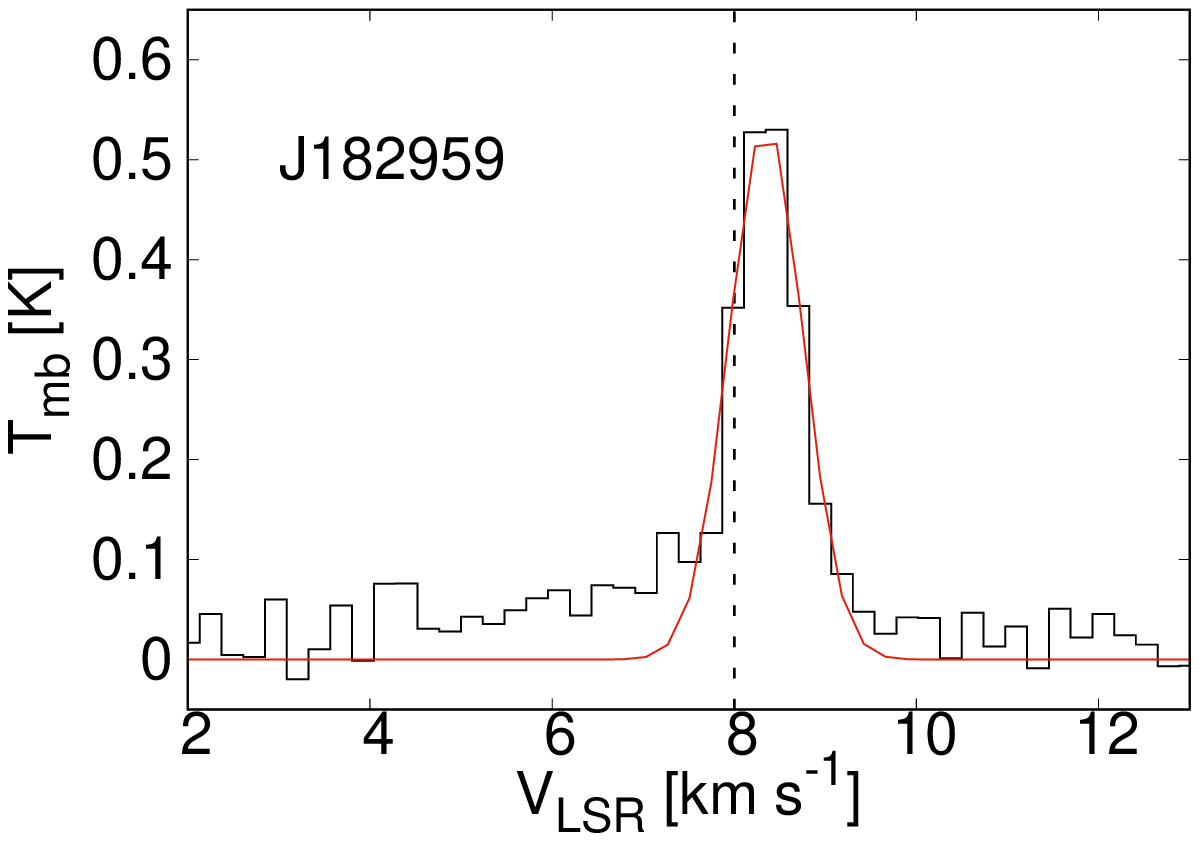}
     \includegraphics[width=1.8in]{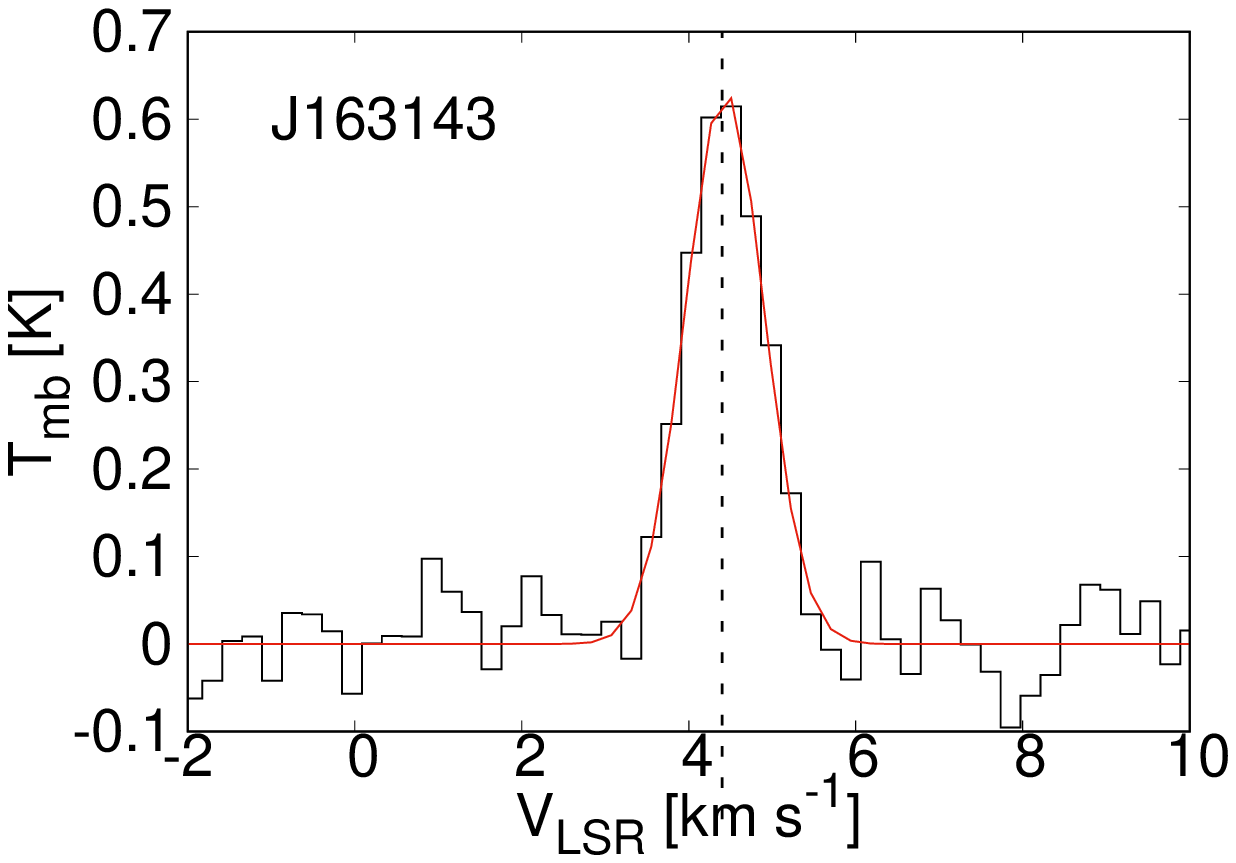}     
     \includegraphics[width=1.8in]{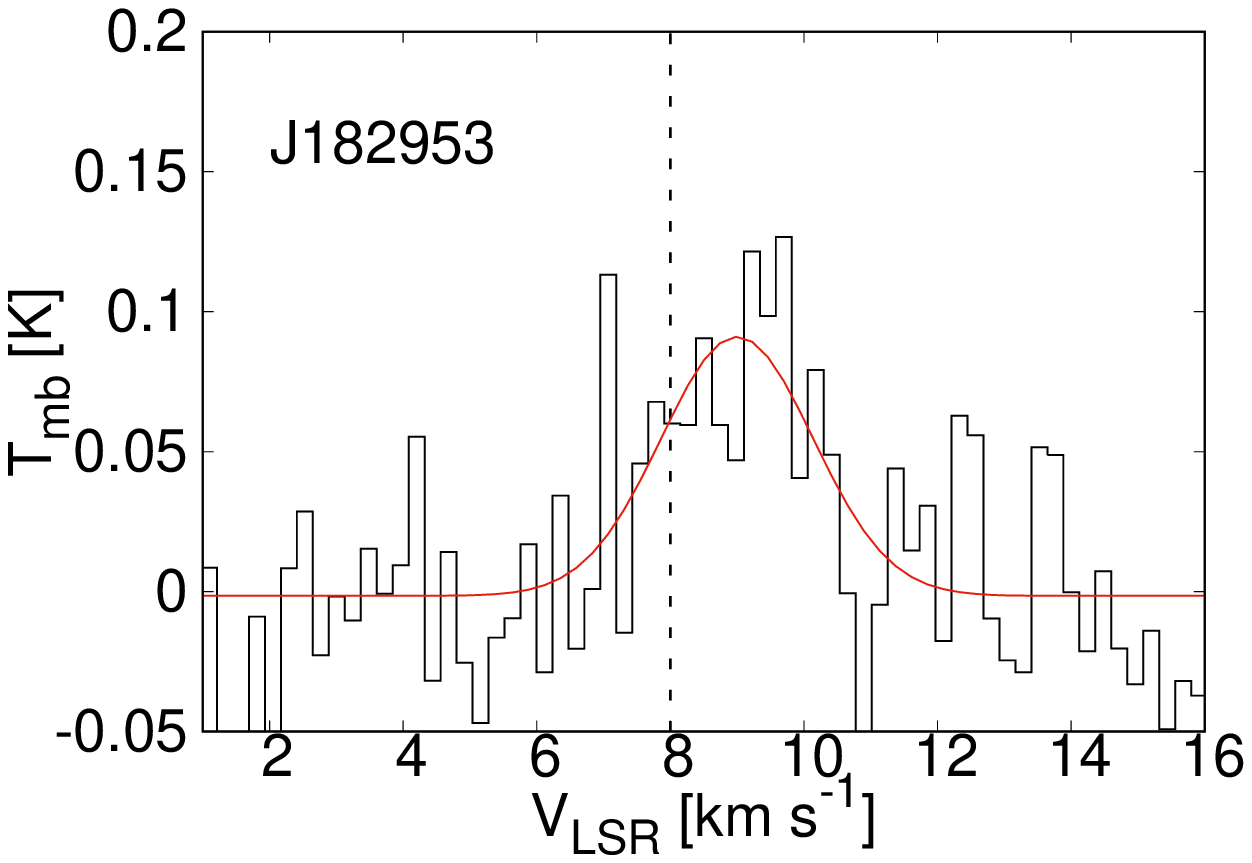}     
     \caption{The observed CS (5-4) spectra (black) with the Gaussian fits (red). Black dashed line marks the cloud systemic velocity of $\sim$8 km/s and $\sim$4.4 km/s in Serpens and Ophiuchus, respectively.  }
     \label{cs-figs}     
  \end{figure*}

Figures~\ref{hcop-figs},~\ref{o-h2co-figs},~\ref{p-h2co-figs}, and \ref{cs-figs} show the HCO$^{+}$ (3-2), ortho- and para-H$_{2}$CO (3-2), and CS (5-4) spectra, respectively. Also overplotted as red lines are the single- or multi-peaked Gaussian fits to the spectra (Sect.~3.1).

\clearpage

\section{Tables}

The line parameters derived from the HCO$^{+}$ (3-2), CS (5-4), ortho-H$_{2}$CO (3-2), para-H$_{2}$CO (3-2), C$^{17}$O (2-1), C$^{18}$O (2-1), $^{13}$CO (2-1), and $^{12}$CO (2-1) spectra are listed in Tables~\ref{hcop},~\ref{cs},~\ref{o-h2co},~\ref{p-h2co},~\ref{c17o},~\ref{c18o},~\ref{13co}, and ~\ref{12co}, respectively.

\begin{table*}
\centering
\caption{HCO$^{+}$ (3-2) Line Parameters\tnote{a}}
\label{hcop}
\begin{threeparttable}
\begin{tabular}{llllllll} 
\hline
Object   & Peak & V$_{\textrm{lsr}}$ (km s$^{-1}$) & T$_{\textrm{mb}}$ (K) & $\int{T_{\textrm{mb}} dv}$ (K km s$^{-1}$) &  $\Delta$v (km s$^{-1}$) \\
\hline

J182854	& 1	& 7.5 & 0.6 & 0.3 & 0.4 \\
		& 2	& 8.3 & 0.9 & 0.4 & 0.4 \\		
J182844	& 1	& 7.6 & 0.9 & 0.7 & 0.6 \\
		& 2	& 8.3 & 0.8 & 0.3 & 0.3 \\		
J183002	& 1	& 7.5 & 1.3 & 4.3 & 2.7 \\
		& 2	& 8.0 & 2.0 & 2.6 & 1.0 \\
		& 3	& 9.6 & 0.3 & 0.2 & 0.6 \\
J182959	& 1	& 7.1 & 0.4 & 0.3 & 0.6 \\
		& 2	& 7.9 & 1.6 & 1.0 & 0.5 \\
		& 3	& 8.7 & 0.6 & 0.5 & 0.7 \\
J163143	& 1	& 4.2 & 1.6 & 1.2 & 0.6 \\
		& 2	& 5.0 & 2.3 & 1.4 & 0.5 \\
J182953	& 1	& 9.1 & 0.8 & 0.6 & 0.6 \\
		& 2	& 10.0 & 0.7 & 0.6 & 0.7 \\
		& 3	& 9.2 & 0.4 & 1.4 & 2.7 \\
J163136	& 1	& 2.6 & 0.6 & 0.4 & 0.7 \\			
J182940	& --	& 8.0 & $<$0.09 & $<$0.09 & 1.0 \\
J182927	& --	& 8.0 & $<$0.1 & $<$0.1 & 1.0 \\
J182952	& 1	& 9.9 & 0.4 & 0.3 & 0.6 \\

\hline
\end{tabular}
\begin{tablenotes}
  \item[a] The uncertainty is estimated to be $\sim$15\%-20\% for the integrated intensity, $\sim$0.01-0.02 km s$^{-1}$ in V$_{\textrm{lsr}}$, $\sim$5\%-10\% on T$_{\textrm{mb}}$ and $\Delta$v.
\end{tablenotes}
\end{threeparttable}
\end{table*}

\begin{table*}
\centering
\caption{CS (5-4) Line Parameters\tnote{a}}
\label{cs}
\begin{threeparttable}
\begin{tabular}{llllllll} 
\hline
Object   & Peak & V$_{\textrm{lsr}}$ (km s$^{-1}$) & T$_{\textrm{mb}}$ (K) & $\int{T_{\textrm{mb}} dv}$ (K km s$^{-1}$) &  $\Delta$v (km s$^{-1}$) \\
\hline
J182854 & --	& 8.0		 & $<$0.15 & $<$0.15 & 1.0 \\
J182844 & 1	& 7.7 	 & 0.3 	  & 0.2 	  & 0.7 \\
J183002 & 1     & 7.8  	& 0.2 	& 0.3 	& 1.3  \\
	      & 2     & 9.4 	& 0.4 	& 0.4 	& 0.7 \\
J182959 & 1	& 8.3 	 & 0.5 	  & 0.5 	  & 0.8 \\
J163143 & 1	& 4.4 	 & 0.6 	  & 0.7 	  & 1.1 \\
J182953 & 1	& 8.9 	& 0.05 	& 0.1 	& 2.3   \\
J163136 & --	& 4.4		 & $<$0.12 & $<$0.12 & 1.0 \\	      
J182940 & --	& 8.0		 & $<$0.06 & $<$0.06 & 1.0 \\
J182927 & --	& 8.0		 & $<$0.09 & $<$0.09 & 1.0 \\
J182952 & --	& 8.0		 & $<$0.09 & $<$0.09 & 1.0   \\
\hline
\end{tabular}
\begin{tablenotes}
  \item[a] The uncertainty is estimated to be $\sim$15\%-20\% for the integrated intensity, $\sim$0.01-0.02 km s$^{-1}$ in V$_{\textrm{lsr}}$, $\sim$5\%-10\% on T$_{\textrm{mb}}$ and $\Delta$v.
\end{tablenotes}
\end{threeparttable}
\end{table*}

\begin{table*}
\centering
\caption{ortho-H$_{2}$CO (3-2) Line Parameters\tnote{a}}
\label{o-h2co}
\begin{threeparttable}
\begin{tabular}{lllllllll} 
\hline
Object   & Peak & V$_{\textrm{lsr}}$ (km s$^{-1}$) & T$_{\textrm{mb}}$ (K) & $\int{T_{\textrm{mb}} dv}$ (K km s$^{-1}$) &  $\Delta$v (km s$^{-1}$) \\
\hline
J182854 & --	& 8.0		& $<$0.06 	& $<$0.06 	& 1.0 	\\
J182844 & 1	& 7.7 	& 0.8 	 	& 0.5 	    	& 0.6		 \\
J183002 & 1	& 8.0		& 0.4			& 1.1			& 2.1		\\
	       & 2     & 9.4	&  0.4		& 0.3			& 0.5		\\
J182959 & 1	& 8.2 	& 1.3 	  	& 1.7 		& 1.2		 \\
J163143 & 1	& 4.7 	& 1.2 	  	& 1.6 	   	& 1.2 	\\
J182953 & 1	& 9.8 	& 0.2 		& 0.2 		& 0.7		    \\
J163136 & --	& 4.4		& $<$0.15 	& $<$0.15 	& 1.0		 \\	       
J182940 & --	& 8.0		& $<$0.07 	& $<$0.07 	& 1.0		 \\
J182927 & --	& 8.0		& $<$0.12 	& $<$0.12 	& 1.0		 \\
J182952 & 1 	& 10.0 	& 0.1 		& 0.07 		& 0.5		    \\	      
\hline
\end{tabular}
\begin{tablenotes}
  \item[a] The uncertainty is estimated to be $\sim$15\%-20\% for the integrated intensity, $\sim$0.01-0.02 km s$^{-1}$ in V$_{\textrm{lsr}}$, $\sim$5\%-10\% on T$_{\textrm{mb}}$ and $\Delta$v.
\end{tablenotes}
\end{threeparttable}
\end{table*}

\begin{table*}
\centering
\caption{para-H$_{2}$CO (3-2) Line Parameters\tnote{a}}
\label{p-h2co}
\begin{threeparttable}
\begin{tabular}{lllllllll} 
\hline
Object   & Peak & V$_{\textrm{lsr}}$ (km s$^{-1}$) & T$_{\textrm{mb}}$ (K) & $\int{T_{\textrm{mb}} dv}$ (K km s$^{-1}$) &  $\Delta$v (km s$^{-1}$) \\
\hline
J182854 & 1	& 7.9 	& 0.09 		& 0.07 		& 0.5		 \\
J182844 & 1	& 7.8 	& 0.5 		& 0.3 		& 0.5		 \\
J183002 & 1	& 5.1 	& 0.2 		& 0.3 		& 1.3		 \\
	       & 2 	& 7.8 	& 0.5 		& 1.1 		& 1.8	 	\\
	       & 3  	& 9.5 	& 0.2 		& 0.2 		& 0.6	 	\\
J182959 & 1	& 8.2 	& 1.2 		& 1.5 		& 0.9	 	\\
J163143 & 1	& 4.7 	& 0.9 		& 1.1 		& 0.9		 \\
J182953 & 1 	& 9.8 	& 0.2 		& 0.3 		& 1.2 	\\
J163136 & --	& 4.4   	& $<$0.06 	& $<$0.06 	& 1.0		\\	       
J182940 & --	& 8.0    	& $<$0.06 	& $<$0.06 	& 1.0		\\
J182927 & --	& 8.0    	& $<$0.06 	& $<$0.06 	& 1.0		\\
J182952 & 1 	& 10.0 	& 0.09 		& 0.08 		& 0.7	 	\\
\hline
\end{tabular}
\begin{tablenotes}
  \item[a] The uncertainty is estimated to be $\sim$15\%-20\% for the integrated intensity, $\sim$0.01-0.02 km s$^{-1}$ in V$_{\textrm{lsr}}$, $\sim$5\%-10\% on T$_{\textrm{mb}}$ and $\Delta$v.
\end{tablenotes}
\end{threeparttable}
\end{table*}

\begin{table*}
\centering
\caption{C$^{17}$O (2-1) Line Parameters\tnote{a}\tnote{b}}
\label{c17o}
\begin{threeparttable}
\begin{tabular}{llllllll} 
\hline
Object   & V$_{\textrm{lsr}}$ (km s$^{-1}$) & T$_{\textrm{mb}}$ (K) & $\int{T_{\textrm{mb}} dv}$ (K km s$^{-1}$) &  $\Delta$v (km s$^{-1}$) \\
\hline

J182854	& 7.9 & 0.4 & 0.6 & 1.0 \\
J182844	& 7.8 & 0.4 & 0.5 & 1.1 \\
J183002	& 8.6 & 0.5 & 1.4 & 1.1  \\
J182959	& 8.4 & 2.8 & 4.6 & 1.6 \\
J163143	& 4.6 & 1.1 & 1.7 & 1.4 \\
J182953	& 9.4 &  0.5 & 0.8   &  1.8  \\
J163136	& 2.6 & 1.3 & 1.3 & 1.0 \\
J182940	& 7.9 & 0.07 & 0.06 & 0.8 \\
J182927	& 8.4 & 0.08 & 0.1 & 1.8 \\
J182952	& 9.5 & 0.3 & 0.3 & 1.9   \\

\hline
\end{tabular}
\begin{tablenotes}
  \item[a] The uncertainty is estimated to be $\sim$15\%-20\% for the integrated intensity, $\sim$0.01-0.02 km s$^{-1}$ in V$_{\textrm{lsr}}$, $\sim$5\%-10\% on T$_{\textrm{mb}}$ and $\Delta$v.
 \item[b] The observed C$^{17}$O (2-1) emission is a composite of all 9 hyperfine components.
\end{tablenotes}
\end{threeparttable}
\end{table*}

\begin{table*}
\centering
\caption{C$^{18}$O (2-1) Line Parameters\tnote{a}}
\label{c18o}
\begin{threeparttable}
\begin{tabular}{llllllll} 
\hline
Object   & Peak & V$_{\textrm{lsr}}$ (km s$^{-1}$) & T$_{\textrm{mb}}$ (K) & $\int{T_{\textrm{mb}} dv}$ (K km s$^{-1}$) &  $\Delta$v (km s$^{-1}$) \\
\hline
J182854	& 1 & 8.0 & 2.0 & 1.2 & 0.5 \\
		& 2 & 9.0 & 0.7 & 0.4 & 0.4 \\
J182844	& 1 & 7.8 & 1.5 & 1.4 & 0.6 \\
J183002	& 1 & 7.8 & 1.8 & 4.9 & 2.2 \\
		& 2 & 8.1 & 2.2 & 1.5 & 0.5 \\
		& 3 & 9.1 & 1.3 & 1.4 & 0.8 \\
J182959	& 1 & 7.2 & 4.6 & 3.9 & 0.7 \\
		& 2 & 8.3 & 5.3 & 5.9 & 0.9 \\
J163143	& 1 & 4.6 & 4.5 & 4.7 & 0.8 \\
J182953	& 1 & 8.4 & 1.3 & 1.1 & 0.7 \\
		& 2 & 9.9 & 2.3 & 3.7 & 0.8 \\
J163136	& -- & -- 	   & -- 	& --	     & --    \\ 
J182940	& -- & -- 	   & -- 	& --	     & --    \\ 
J182927	& -- & -- 	   & -- 	& --	     & --    \\ 
J182952	& 1 & 8.1 & 1.0 & 0.7 & 0.6 \\
		& 2 & 9.9 & 2.5 & 1.6 & 0.5 \\
\hline
\end{tabular}
\begin{tablenotes}
  \item[a] The uncertainty is estimated to be $\sim$15\%-20\% for the integrated intensity, $\sim$0.01-0.02 km s$^{-1}$ in V$_{\textrm{lsr}}$, $\sim$5\%-10\% on T$_{\textrm{mb}}$ and $\Delta$v.
\end{tablenotes}
\end{threeparttable}
\end{table*}

\begin{table*}
\centering
\caption{$^{13}$CO (2-1) Line Parameters\tnote{a}}
\label{13co}
\begin{threeparttable}
\begin{tabular}{llllllll} 
\hline
Object   & Peak & V$_{\textrm{lsr}}$ (km s$^{-1}$) & T$_{\textrm{mb}}$ (K) & $\int{T_{\textrm{mb}} dv}$ (K km s$^{-1}$) &  $\Delta$v (km s$^{-1}$) \\
\hline
J182854	& 1 & 7.9 & 4.5 & 5.0 & 0.9 \\
		& 2 & 9.0 & 1.6 & 1.7 & 0.8 \\
J182844	& 1 & 7.7 & 2.1 & 4.2 & 1.5 \\
J183002	& 1 & 4.9 & 3.0 & 3.9 & 1.0 \\
		& 2 & 7.7 & 12.9 & 32.5 & 2.0 \\
		& 3 & 9.5 & 5.8 & 8.1 & 1.1 \\
J182959	& 1 & 6.9 & 10.4 & 14.3 & 1.1 \\  
		& 2 & 8.6 & 7.7 & 11.8 & 1.2 \\
J163143	& 1 & 2.7 & 2.9 & 2.7 & 0.7 \\
		& 2 & 4.6 & 13.6 & 20.4 & 1.2 \\
J182953	& 1 & 8.2 & 2.0 & 6.8 & 2.7 \\
		& 2 & 9.5 & 3.2 & 6.3 & 1.5 \\
		& 3 & 10.4 & 3.6 & 2.7 & 0.6 \\
J163136	& -- & -- 	   & -- 	& --	     & --    \\  
J182940	& -- & -- 	   & -- 	& --	     & --    \\  
J182927	& -- & -- 	   & -- 	& --	     & --    \\  
J182952	& 1 & 5.7 & 0.7 & 0.6 & 0.6 \\
		& 2 & 7.9 & 3.6 & 6.6 & 1.6 \\
		& 3 & 9.9 & 4.6 & 4.8 & 0.8 \\
\hline
\end{tabular}
\begin{tablenotes}
  \item[a] The uncertainty is estimated to be $\sim$15\%-20\% for the integrated intensity, $\sim$0.01-0.02 km s$^{-1}$ in V$_{\textrm{lsr}}$, $\sim$5\%-10\% on T$_{\textrm{mb}}$ and $\Delta$v.
\end{tablenotes}
\end{threeparttable}
\end{table*}

\begin{table*}
\centering
\caption{$^{12}$CO (2-1) Line Parameters\tnote{a}}
\label{12co}
\begin{threeparttable}
\begin{tabular}{llllllll} 
\hline
Object   & Peak & V$_{\textrm{lsr}}$ (km s$^{-1}$) & T$_{\textrm{mb}}$ (K) & $\int{T_{\textrm{mb}} dv}$ (K km s$^{-1}$) &  $\Delta$v (km s$^{-1}$) \\
\hline
J182854	& 1	& 1.2 & 0.4 & 0.5 & 1.0 \\
		& 2	& 3.5 & 0.5 & 1.2 & 1.9 \\
		& 3	& 5.2 & 2.0 & 2.7 & 1.1 \\
		& 4	& 7.5 & 2.5 & 6.5 & 2.1 \\
		& 5	& 10.2 & 1.4 & 1.8 & 1.1 \\
J182844	& 1	& 4.4 & 0.2 & 0.2 & 0.7 \\
		& 2	& 7.2 & 1.6 & 6.7 & 3.2 \\
		& 3	& 10.6 & 0.7 & 1.2 & 10.6 \\
		
J183002	& 1 	& 6.2  & 13.0 & 56.8 & 3.5 \\
		& 2 	& 10.7 & 7.0 & 25.5 & 3.0 \\

J182959 & 1	& 1.7 & 0.8 & 0.7 & 0.8 \\
		& 2	& 6.5 & 6.8 & 9.6 & 1.1 \\
		& 3	& 8.1 & 5.1 & 22.9 & 3.6 \\
		& 4	& 10.3 & 1.6 & 1.7 & 0.9 \\
J163143	& 1	& 2.9 & 12.1 & 27.3 & 1.8 \\
		& 2	& 5.1 & 10.4 & 14.1 & 1.1 \\

J182953  	& 1	& 6.2 & 4.3 & 10.2 & 1.8 \\
		& 2	& 9.6 & 2.3 & 6.6 & 2.4 \\
		
J163136	& 1	& 2.4 & 5.9 & 10.1 & 1.4 \\
		& 2	& 2.7 & 2.9 & 11.9 & 3.3 \\
J182940	& 1	& 6.8 & 2.5 & 5.4 & 1.7 \\
		& 2	& 9.1 & 1.2 & 1.6 & 1.1 \\
		& 3	&10.4 & 1.8 & 1.8 & 0.8 \\
J182927	& 1	& 6.3 & 1.8 & 9.1 & 3.8 \\
		& 2	& 6.5 & 2.5 & 2.9 & 0.9 \\
		& 3	& 9.0 & 1.6 & 1.1 & 0.6 \\
		& 4	& 10.9 & 1.6 & 3.1 & 1.5 \\

J182952  	& 1	& 6.2 & 3.0 & 9.4 & 2.6 \\
		& 2	& 9.8 & 2.0 & 6.0 & 2.3 \\

\hline
\end{tabular}
\begin{tablenotes}
  \item[a] The uncertainty is estimated to be $\sim$15\%-20\% for the integrated intensity, $\sim$0.01-0.02 km s$^{-1}$ in V$_{\textrm{lsr}}$, $\sim$5\%-10\% on T$_{\textrm{mb}}$ and $\Delta$v.
\end{tablenotes}
\end{threeparttable}
\end{table*}


\bsp	
\label{lastpage}
\end{document}